\documentclass[doublespacing]{elsart}

\usepackage{graphicx}

\usepackage{amssymb}

\begin{document}

\begin{frontmatter}



\title{Spatially self-organized resilient networks \\
by a distributed cooperative mechanism}


\author[JAIST]{Yukio Hayashi}, 

\address[JAIST]{Japan Advanced Institute of Science and Technology, 
Ishikawa 923-1292, Japan}

\begin{abstract}
The robustness of connectivity and the efficiency of paths 
are incompatible in many real networks. 
We propose a self-organization mechanism 
for incrementally generating onion-like networks 
with positive degree-degree correlations 
whose robustness is nearly optimal. 
As a spatial extension of the generation model 
based on cooperative copying and adding shortcut, we show that 
the growing networks become more robust and efficient
through enhancing the onion-like topological structure 
on a space.
The reasonable constraint for locating nodes on the perimeter 
in typical surface growth as a self-propagation 
does not affect these properties 
of the tolerance and the path length. 
Moreover, 
the robustness can be recovered in the random growth 
damaged by insistent sequential attacks even without any 
remedial measures.
\end{abstract}

\end{frontmatter}

\newpage
\section{Introduction}
In modern society, 
our daily activities depend on 
energy supply, communication, transportation, economic, and 
ecological networks, 
however their infrastructure systems are complex and 
not constructed by a central control.
Unfortunately, 
natural and man-made disasters occur at many locations in the world,
sometimes they bring to the crises of such network infrastructures. 
For the improvements with robust connectivity, 
it is expected to study the admirable self-organizations appeared 
in natural and social systems \cite{Dressler07}.
In particular, several fundamental mechanisms: 
preferential attachment \cite{Barabasi99}, 
copying \cite{Sole02,Satorras03}, 
survival \cite{Helbing97,Tero10,Hayashi12a}, 
subdivision 
(fragmentation) \cite{Doye05,Zhou05,Nagel08,Hayashi09a,Hayashi09b,Krapivsky10}, 
or aggregation \cite{Krapivsky10,Alava05}
are attractive 
for generating networks in the interdisciplinary research fields of 
physics, biology, sociology, and computer science.

With the break of complex network science 
in the beginning of 21st century, 
it is well known that in many social, technological, and biological 
networks there exists a common {\it scale-free} (SF)
structure whose power-law degree distribution is generated
by the preferential attachment
\cite{Barabasi99} likened to ``rich-get-richer'' rule.
The SF networks have the efficient {\it small-world} (SW) 
property \cite{Watts98} 
that the path length counted by the hops 
through minimum intermediate nodes 
between any two nodes is short even for a large network size, 
however they also have an extreme vulnerability against 
intentional attacks in spite of having the tolerance of connectivity 
against random failures \cite{Albert00}. 
The vulnerability comes from a few existing of 
large degree hub nodes in a power-law distribution. 
Thus, the 
efficiency of path and the robustness of connectivity are 
incompatible in many real networks, such as Internet, power-grids, 
airline networks, 
metabolic networks, and so on. 
If we do not persist the generation mechanism of real networks, 
we may find other more desirable mechanism to maintain both the 
robustness and the efficiency.

Since 
the natural design methods are not limited to only the selfish rule 
for a good network in efficiency, scalability, 
stability, adaptivity, or other criteria, 
we consider a random copying process to generate
complex networks as one of other candidates for the design of future 
network infrastructure. 
We take into account self-propagation 
in maintaining the robust 
network structure without degrading the communication or 
transportation performance in the growth.
Although the duplication process has been so far 
considered to be fundamental
in a model of protein-protein interaction networks
\cite{Sole02,Satorras03}, its generation mechanism with 
some modifications may be 
applied to a self-organized design of social and/or 
technological networks in urban planning, civil engineering, 
or information system science.

On the other hand, by numerical and theoretical analysis, 
it has been shown that 
\cite{Schneider11,Herrmann11,Tanizawa12} 
onion-like topological structure with positive degree-degree correlations 
gives the optimal robustness against targeted attacks to hub nodes
in an SF network.
For any degree distribution, 
the onion-like topology consists of a core of 
highly connected nodes hierarchically surrounded by rings of nodes 
with decreasing degree.
In a related generation method in a family of SF networks,
a deterministic model called as {\it mandala network} that consist
of recurrently expanded intra- and inner-connections of ringed nodes 
has been studied \cite{Filho15}; 
the robustness is improved by rewiring or adding links
applied to the outmost two rings of nodes from the core of 
connected hub nodes. 
Furthermore, an efficient rewiring algorithm 
has been developed for generating a network of 
onion-like topology
with the nearly optimal robustness 
under a given degree distribution \cite{Wu11}. 
However, these constructions are based on swapping endpoints of 
randomly chosen two links \cite{Schneider11,Herrmann11}, 
expanding the rings by simultaneously adding 
$n_{i} = 2 n_{i-1}$ nodes 
for the $i$-th iterations \cite{Filho15}, 
and entirely rewiring of links \cite{Wu11} 
like a configuration model \cite{Newman01}, 
an incremental generation method was not found 
for the networks with onion-like topology. 
Recently, an incrementally growing method 
of such networks with onion-like topology
has been proposed \cite{Hayashi14} 
as a modification of the {\it duplication-divergence} (D-D) model 
\cite{Sole02,Satorras03} in enhancing the degree-degree correlations.

In this paper, 
we focus on a self-organized design for 
infrastructural communication and transportation 
network systems on a space, 
rather than the detail technologies and facilities.
In particular, 
we consider a spatial growth of network with the robust onion-like 
topology.
Although there are many types of failures and attacks, e.g. 
locally spreading damages in a disaster, this paper treats 
typical random and targeted removal of nodes 
for investigating the fundamental property of robustness. 
Instead, we take notice of the uncertainty: 
whether the robust structure can be maintained or not in 
a spatial growing of network, 
because the embedding on a space imposes some kind of constraints 
on the construction of network topology.
The organization of this paper is as follows. 
In Sec. \ref{sec2}, we introduce a biologically inspired 
basic model \cite{Hayashi14} 
for generating onion-like topological structure 
with strong robustness. 
Without loss of the robust onion-like topology, 
we extend it to spatial networks according to 
surface growth.
Such growth gives a hint for finding plausible 
self-propagation mechanism in distributed network systems. 
In Sec. \ref{sec3}, 
we show good properties 
for the robustness of connectivity and the efficiency of path 
on the growing networks.
In particular, 
it is attractive that the reasonable constraint 
on a contact area of surface for locating nodes 
in the growth does not affect these properties of 
robustness and efficiency.
Moreover, we show the resilient connectivity. 
Even without any remedial measures, 
the robustness can be recovered in the growth damaged by 
sequential attacks. 
Resilience \cite{Zolli13} is an important concept 
to heal over, repair, and recover the performance from 
a damaged system, the recovery of robustness will give 
a first step to develop resilient network systems. 
Some strategies for the resilient networks are discussed 
particularly in interdependent networks 
\cite{Schneider13,Kertesz14} recently.
In Sec. \ref{sec4}, we summarize these results, 
and mention several issues to make a more resilient network 
system.

\section{Spatially growing onion-like networks} \label{sec2}

\subsection{Basic generation procedures}
This subsection introduces a basic model of incrementally growing 
networks with onion-like topology \cite{Hayashi14}.
In the next subsection, we extend it to a spatial model and 
explain how to locate a node on a space.

In the conventional 
D-D model \cite{Sole02,Satorras03}, as the duplication process, 
a new node added per time step links 
to connected neighbor nodes of a randomly chosen node.
Then deletion of the duplication links occurs with probability $\delta$. 
We modify the duplication process by non-trivial discovering ideas; 
the differences of our proposed network \cite{Hayashi14} 
from the D-D model are 
the adding of a mutual link between new and randomly chosen nodes, 
and the simultaneous progress of copying and adding shortcut links 
at a time-interval to enhance the onion-like topological structure.
We call the combination of a mutual link and duplication links 
as copying. 
Shortcut means an 
analogy to random connections in the SW model \cite{Watts98}.
The outline of network generation consists of 
(1) At each time step, a new node is added at a position on a square 
lattice. 
(2) As the copying process, 
the new node stochastically links to connected neighbor nodes 
of a chosen node 
limited on the perimeter (surface) of 
a spatially growing network. 
(3) In order to enhance the robustness of connectivity, 
we consider having positive degree-degree correlations in the 
copying and adding of shortcut links. 
These procedures are summarized as follows. 

\begin{figure}[tp]
 \begin{minipage}[htb]{0.47\textwidth} 
   \centering
   \includegraphics[height=50mm]{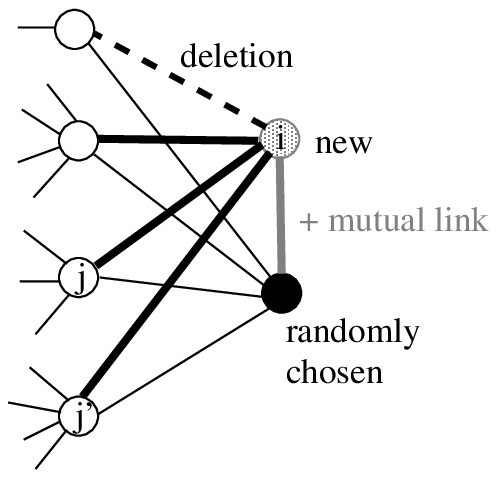}
     \begin{center} (a) Copying process \end{center}
 \end{minipage} 
 \hfill 
 \begin{minipage}[htb]{0.47\textwidth} 
   \centering
   \includegraphics[height=50mm]{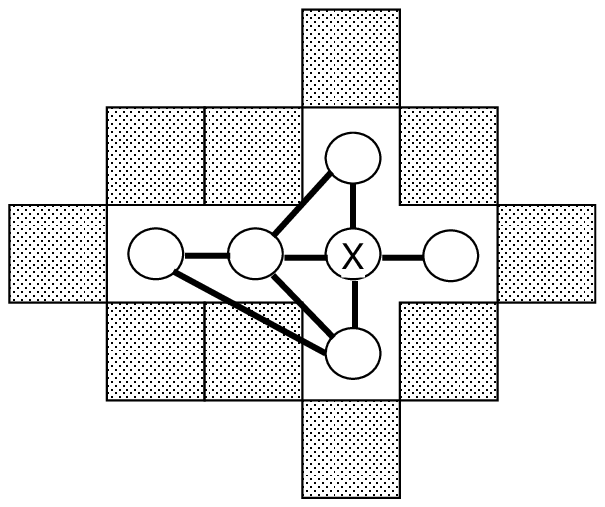}
     \begin{center} (b) Candidates of node's location \end{center}
 \end{minipage} 
\caption{Basic topological and spatial processes.
(a) Thick black and gray lines show the generated links by 
the copying process. Then deletion of dashed line occurs with 
probability $\delta \times p$.
Thin lines show the already existing links.
We remark that such a bypath of nodes j'-i-j is created 
for the path of nodes j'-chosen-j. 
(b) Open circles are selectable nodes, 
but one with cross mark is not.
Shaded sites show the perimeter as inserted locations of a new node.
}
\label{fig_copying}
\end{figure}

\begin{description}
  \item[Step 0] Set an initial configuration of connected $N_{0}$
    nodes. 
  \item[Step 1] At each time step $t = 1, 2, \ldots$, 
    a new node is added on a space.
    As shown to the topological structure in Fig. \ref{fig_copying}(a), 
    the new node $i$ connects to a randomly chosen node 
    and to the neighbor nodes $j$ 
    with a probability $(1 - \delta) \times p$ \cite{Wu11}, 
    \begin{equation}
      p \stackrel{\rm def}{=} \frac{1}{1 + a \mid k_{i} - k_{j} \mid},
      \label{eq_ass_link}
    \end{equation}
    where $\delta$ is a rate of link deletion, 
    $a \geq 0$ is a parameter, 
    and $k_{i}$ and $k_{j}$ denote the degrees of nodes $i$ and $j$. 
    Since the linking is a stochastic process for the new node $i$, 
    unknown $k_{i}$ in Eq.(\ref{eq_ass_link}) is anticipatorily set 
    as $(1 - \delta) \times$ the degree of the chosen node.
  \item[Step 2] 
    Moreover, at every time-interval $IT$, 
    shortcut links until the number $p_{sc} M(t)$ 
    are added between randomly chosen nodes 
    $i$ and $j$ according to the probability of Eq.(\ref{eq_ass_link})
    in prohibiting self-loops at a node 
    and duplicate connections between two nodes\footnote{In the 
    prohibitive case, selections of other nodes are tried in the same way.}.
    Here, $p_{sc}$ is a rate of adding shortcut links, 
    $M(t)$ denotes the number of links in the network at that time 
    $t = IT, 2 IT, 3 IT, \ldots$, and $IT$ is defined by time steps 
    greater than one.
  \item[Step 3] The above processes in Steps 1 and 2 
    are repeated up to a given size $N$ for $N(t) = t + N_{0}$. 
\end{description}

We emphasize the following 
effects of copying and shortcut on the robustness.
\begin{itemize}
  \item Local proxy function: by copying to make 
    bypaths via another access point of new node (see Fig. \ref{fig_copying}(a))
    as a duplexing and the accumulated multiplexing via other new nodes 
    in the growth
  \item Complementary function: by adding shortcut links especially 
    between small degree (homophily) nodes to enhance the robustness
\end{itemize}

Note that the local redundancy with bypaths 
is often used in distributed computer communication systems. 
Our generation method is based on a cooperation mechanism 
that consists of the above functions and the linking of similar 
degree nodes as homophily.
Each part constructed by the copying and adding shortcut links 
helps each other with the division of roles to be a robust network 
in taking into account positive degree-degree correlations. 
More precisely, we explain the roles as follows.
Since the double random selections for the neighbor nodes $j$ 
contribute to an equivalent effect of 
preferential attachment \cite{Yang13,Colman13}, large degree nodes 
$i$ and $j$ tend to be connected together
when the chosen node has a large degree. 
However, such positive 
correlations between small degree nodes are weak in 
the tree-like structure generated by only the copying process 
\cite{Hayashi14}.
Thus, in order to make an onion-like topological structure, 
we further consider addition of shortcut links 
\cite{Watts98} 
between randomly chosen nodes $i$ and $j$ with the probability
of Eq.(\ref{eq_ass_link}). 
The adding per a time-interval $IT$ instead of each time step 
is considered as non-dominant (complementary) 
but necessary process for generating an onion-like topology. 
It has already been shown that 
adding some shortcut links between randomly chosen nodes 
improves the robustness 
in the theory for the small-world model \cite{Newman00}
and also in the numerical simulations for geographical networks:
random Apollonian networks \cite{Hayashi06}, 
multi-scale quartered networks \cite{Hayashi10}, 
and link survival networks \cite{Hayashi12a}.
In addition, 
positive degree-degree correlations tend to appear 
in randomly growing networks \cite{Callaway01}.
Therefore, in our network grown by the simultaneous progress 
of the copying and adding shortcut links, it is expected that 
the robustness becomes stronger due to enhancing 
positive degree-degree correlations for emerging 
the onion-like topology.

\subsection{Spatial networks generated by surface growth with robustness}
In this subsection, 
we explain how to locate a new node in the network according to typical 
models of surface growth.

There exist many complex pattern-formations far from equilibrium 
in nature \cite{Meakin98} as living open systems. 
Several models of fractal growing random pattern have been studied 
for the growth of biological cell colonies, 
fluid displacing in a porous medium, dendritic solidification, 
dielectric breakdown, snowflake formation, and bacterial colonies 
\cite{Meakin98,Ben-Jacob97}. 
Some most important classes of surface growth include 
{\it Diffusion-Limited Aggregation} (DLA) \cite{Wilkinson83,Stanley94}, 
{\it Invasion Percolation} (IP) \cite{Witten83}, and 
{\it Eden growth} \cite{Eden61}, 
which can be used as a basis for understanding a wide range of 
pattern-formation phenomena with disorderly growth.

Thus, we consider 
diffusively growing networks on surface with both 
onion-like topological and fractal spatial structures.
In our network, 
the position of new node is determined by DLA, IP, 
and Eden models on a square lattice. 
Exactly, 
only DLA and IP models at the critical threshold 
generate a fractal structure. 
These models have the following processes \cite{Meakin98}
and explanations to give a new insight of the self-organized 
design of spatial network.

\begin{description}
  \item[\underline{DLA model}] As an idealization of 
    the irreversible aggregation, 
    the following process forms a growing diffusive cluster.
    It is motivated from several phenomena of biological interest
    for the surface growth with complicated shape, which corresponds
    to a distributed local extension of technological or social 
    network system.\\
    The initial configuration is a single occupied site on a lattice. 
    At each generation step, the cluster of occupied sites is grown by 
    launching a random walker from outside of the occupied region, 
    and allowing it to its random walk path 
    until it reaches a site that is adjacent to the occupied site. 
    The process is repeated with a new random walker.
  \item[\underline{IP model}] It is based on a transport process 
    by the slow displacement of a wetting fluid in a porous media.
    The percolation growth corresponds to an extension 
    in avoiding geographical obstacles engraved by 
    longstanding rains and winds
    for the network construction.\\
    Initially, random numbers or thresholds are assigned to 
    the sites of a lattice. They represent the capillary pressures 
    for penetrating through the porous medium or 
    the difficulties (low priorities) to install network connections
    on a geographical space. 
    At each generation step, 
    an unoccupied site with the lowest threshold
    (smallest random number, highest priority) is selected 
    as an injection point on the perimeter, and occupied.
    Such process is repeated. 
  \item[\underline{Eden model}] 
    It imitates a cell division process, 
    which corresponds to the most local extension to neighbor space
    in a simple way.\\
    Starting with an occupied cell (site), 
    an occupied cell on the perimeter of the cluster is randomly 
    selected with equal probability. 
    One of its nearest neighbor unoccupied perimeter cell is 
    selected and occupied. The cluster is grown by adding 
    the selected nearest neighbor at each generation step.
    Such process is repeated for a new selection of cell
    with the division.
\end{description}

As a constraint in the surface growth, 
the inserted position of new node is limited on the 
perimeter of connected cluster.
Figure \ref{fig_copying}(b) shows that 
the internal node(s) marked by $\times$ can't be selected, 
since the only neighbors marked by $\bigcirc$ 
of shaded positions are the candidates. 
In other words, 
the selection of a node is not 
uniformly at random (u.a.r) in the growing network.
However, the limitation is rather reasonable, 
since the perimeter is a contact area 
from the outer world into the network through the growth.
Moreover, although u.a.r selection is not equivalent to 
independent one, 
the perimeter sites can grow simultaneously 
in distributed processes on several places. 
These sites autonomously perform the processes initiated 
by diffusive growing, own local timers corresponded to 
thresholds\footnote{The threshold may be related to the amount 
of potential communication or transportation requests 
which are proportional to a population density.}, 
or random numbers.
It is an important discussion point in this paper 
whether the constraint on the surface growth 
hardly affect the emergence of an onion-like topological structure.

\begin{figure}[tp]
 \begin{minipage}{\textwidth} 
   \centering
   \includegraphics[height=26mm]{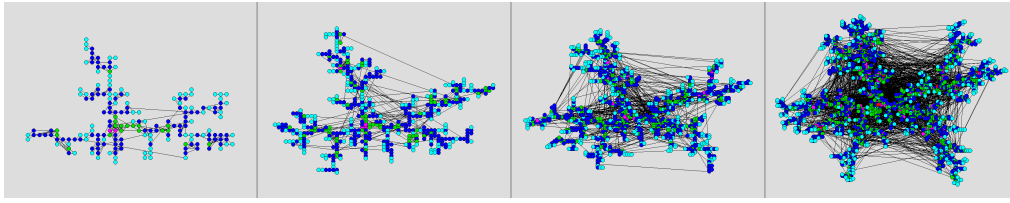}
     \begin{center} (a) DLA Model: $\delta=0.3$, $p_{sc}=0.015$ \end{center}
 \end{minipage} 
 \hfill 
 \begin{minipage}{\textwidth} 
   \centering
   \includegraphics[height=26mm]{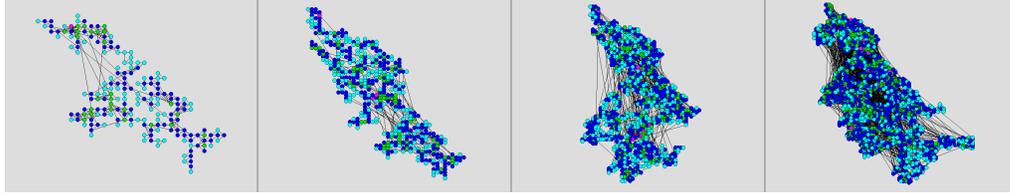}
     \begin{center} (b) IP Model: $\delta=0.3$, $p_{sc}=0.013$ \end{center}
 \end{minipage} 
 \hfill 
 \begin{minipage}{\textwidth} 
   \centering
   \includegraphics[height=26mm]{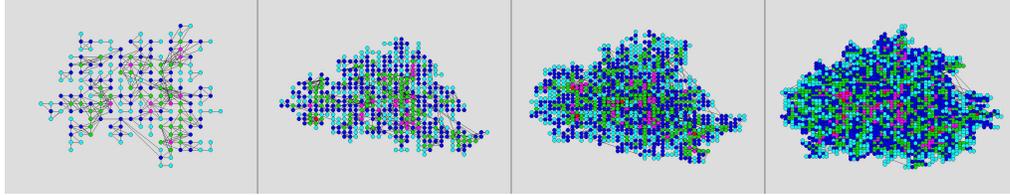}
     \begin{center} (c) Eden Model: $\delta=0.3$, $p_{sc}=0.009$ \end{center}
 \end{minipage} 
\caption{(Color online) 
Growing spatial networks for $N(t) = 200, 500, 1000$ and $2000$ 
from left to right with arranged scales. 
Gradually colored node is according to its degree: 
1-3(cyan), 4-6(blue), 7-9(green), 10-12(magenta), 
and the larger(red).} \label{fig_vis_growth}
\end{figure}

Typical shapes of the growing spatial networks
are shown in Fig. \ref{fig_vis_growth}.
These models form (a) dendritic, (b) porous, 
and (c) compact patterns, respectively, on a square lattice.
Large degree (magenta and red)
nodes are spontaneously interspersed 
in constructing densely connected parts.  
The rate $p_{sc}$ of adding shortcut links 
is regulated to be a same condition of the connection density as 
$\langle k \rangle \approx 5.6$ at $N=2000$ in all models. 
We also set initial complete graph of $N_{0}=4$, $IT=50$, and 
$a=0.3$. 
Since we want to investigate the effect of the copying process 
with adding shortcut links on the robustness and the efficiency, 
we use a small value of $a$ which is different from $3.0$ 
in the rewiring algorithm \cite{Wu11}. 
Too large value of $a$ restricts linking 
in the copying process, only the mutual link may be remained
at each time step in the growing.
This point will be mentioned again in subsection 3.1.

\begin{figure}[tp]
 \begin{minipage}{0.47\textwidth} 
   \centering
   \includegraphics[height=67mm,angle=-90]{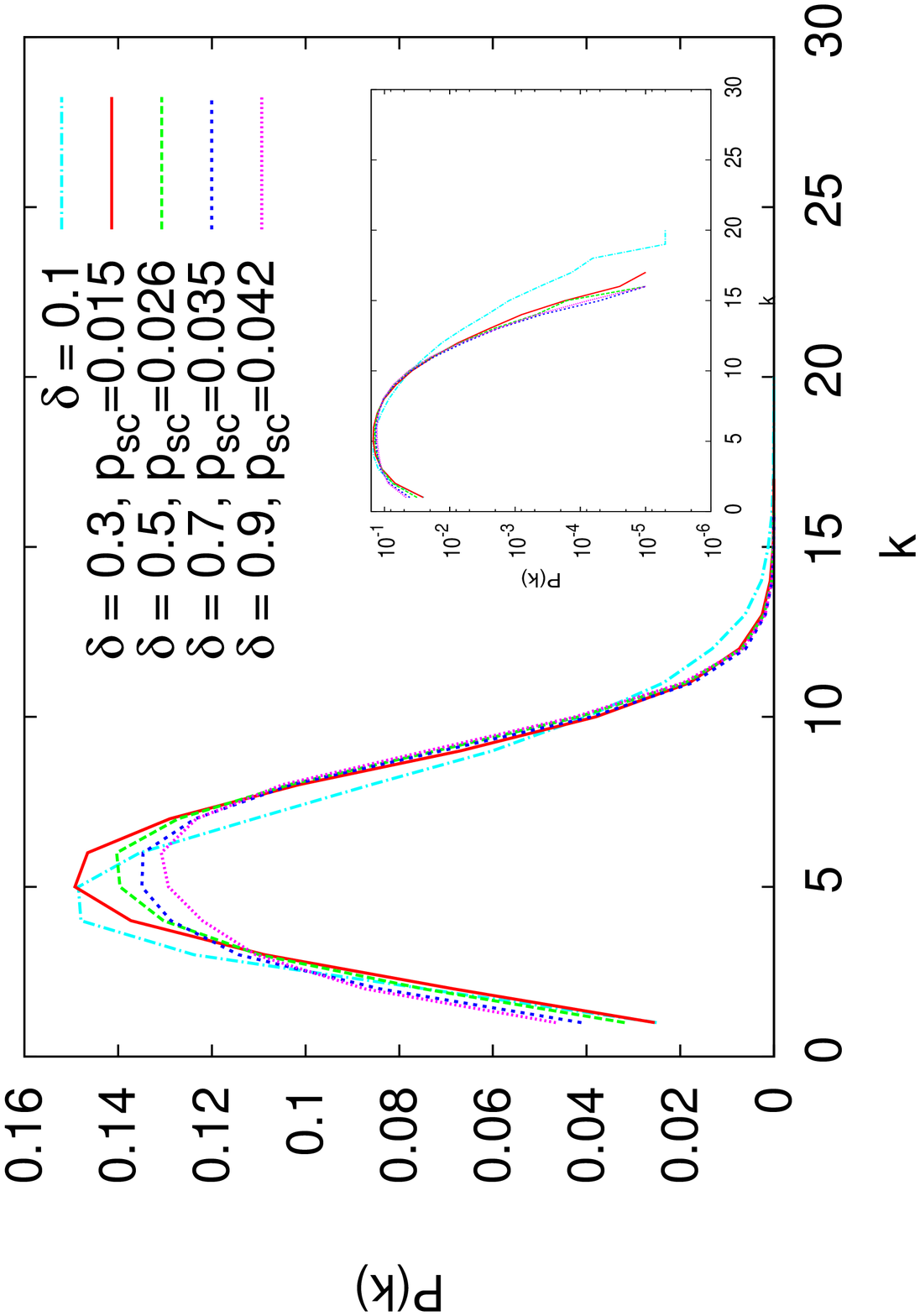}
     \begin{center} (a) DLA model \end{center}
 \end{minipage} 
 \hfill 
 \begin{minipage}{0.47\textwidth} 
   \centering
   \includegraphics[height=67mm,angle=-90]{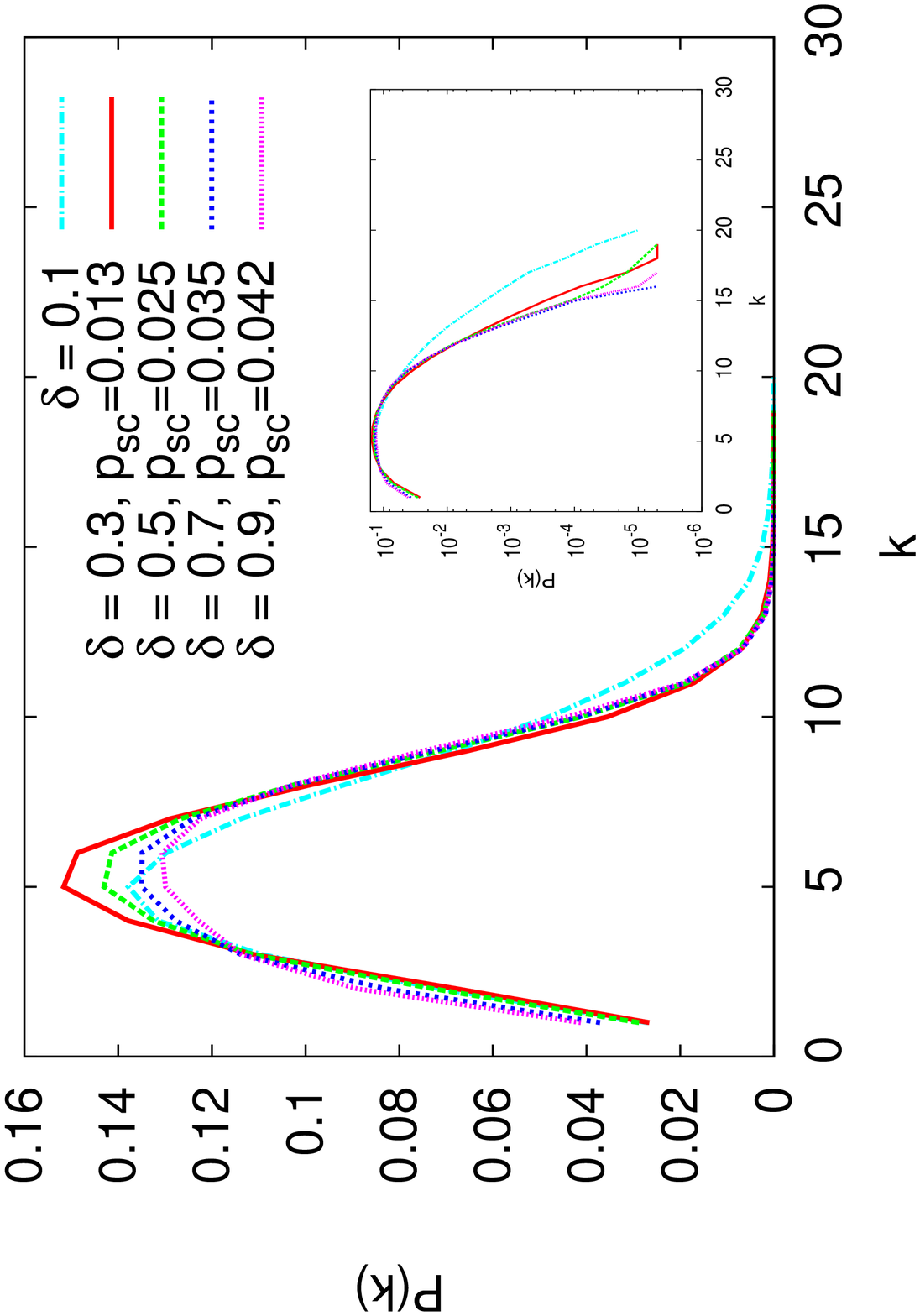}
       \begin{center} (b) IP model \end{center}
 \end{minipage} 
 \hfill 
 \begin{minipage}{0.47\textwidth} 
   \centering
   \includegraphics[height=67mm,angle=-90]{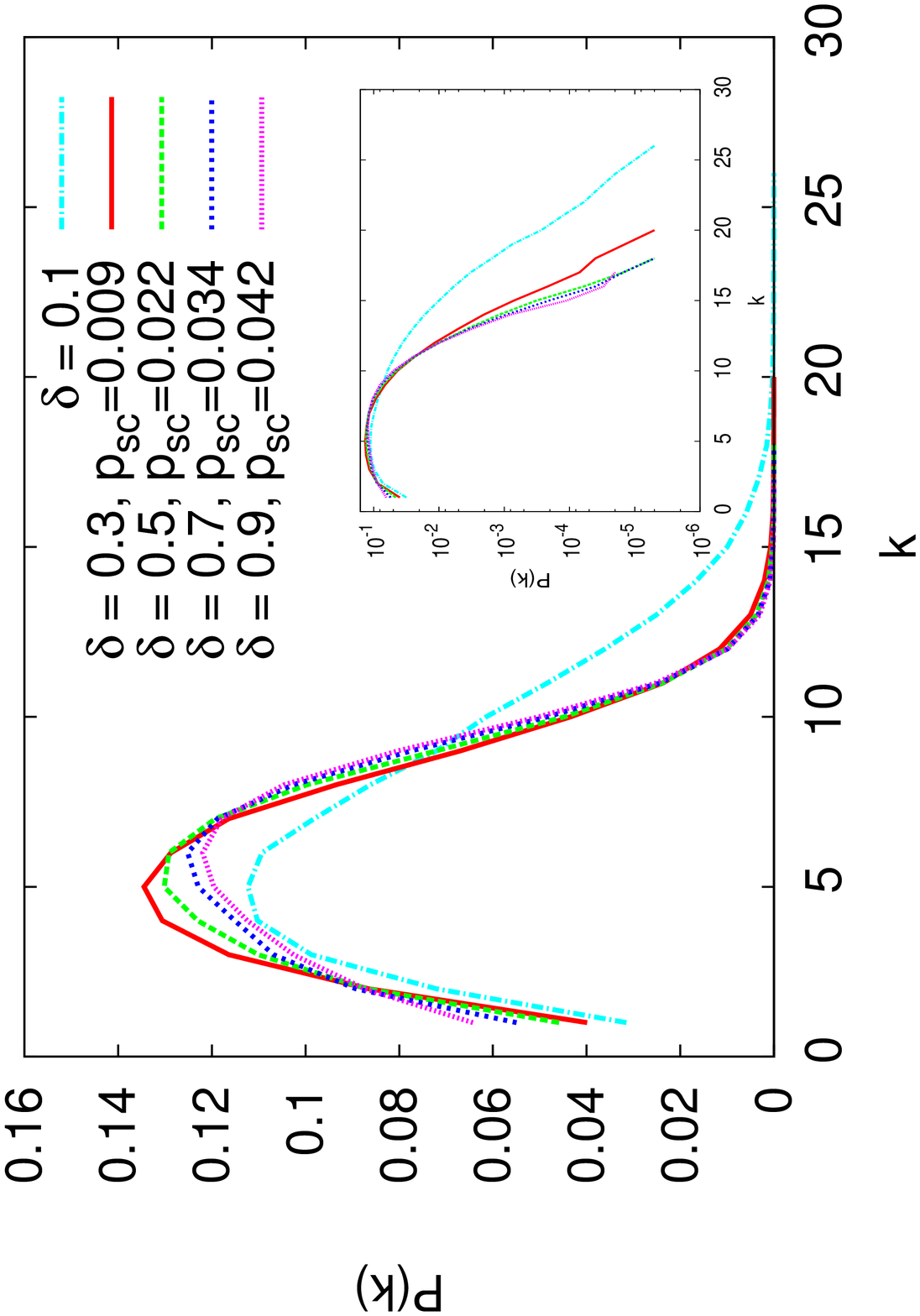}
      \begin{center} (c) Eden model \end{center}
 \end{minipage} 
 \hfill 
 \begin{minipage}{0.47\textwidth} 
   \centering
   \includegraphics[height=67mm,angle=-90]{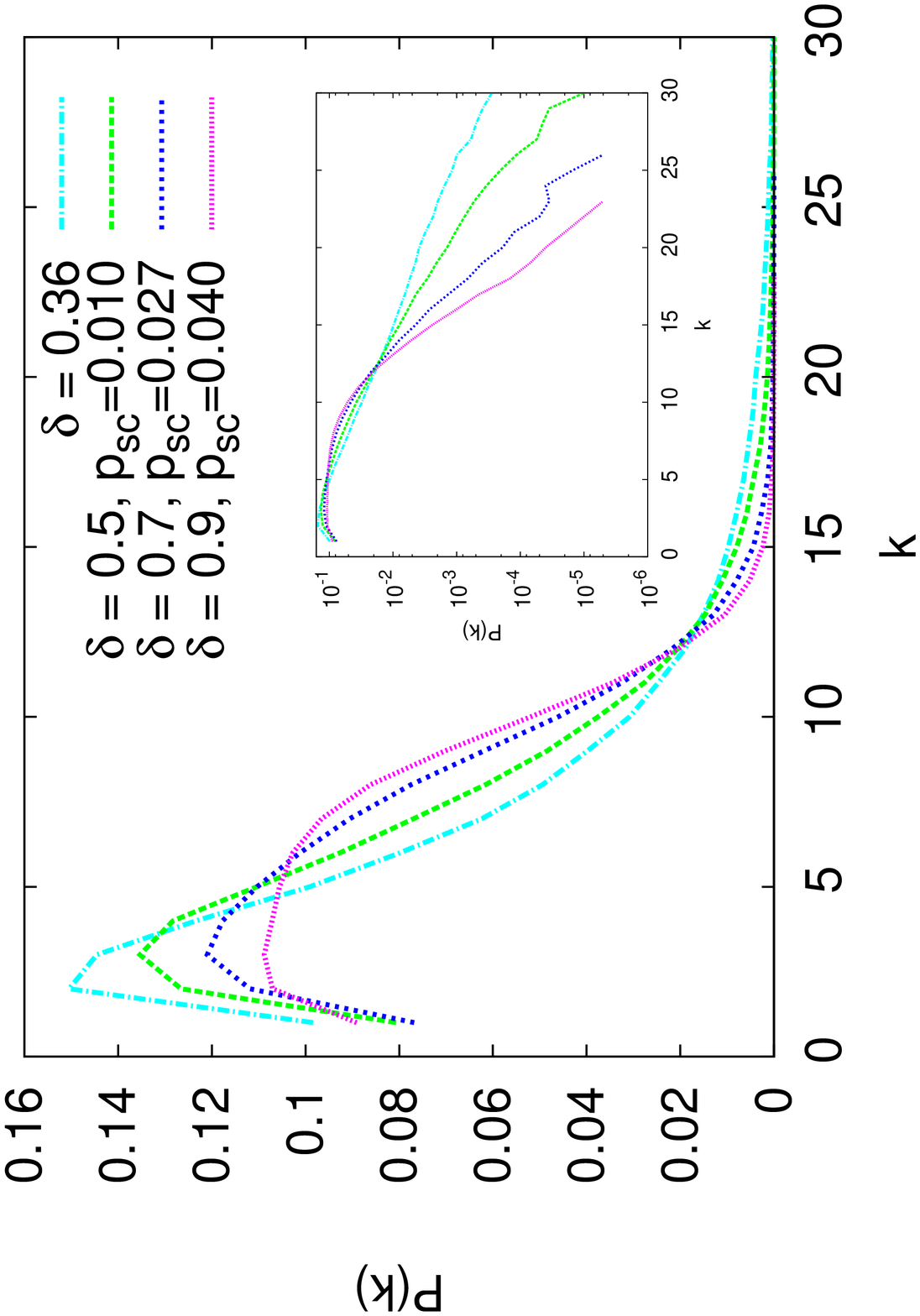}
      \begin{center} (d) Spatial growing model \end{center}
 \end{minipage} 
\caption{(Color online) Degree distribution in the networks 
according to (a) DLA, (b) IP, (c) Eden, 
and (d) Spatial growing models 
for $N = 2000$ with $\langle k \rangle \approx 5.6$.
Inset shows the exponential decay of tail part 
approximated by a straight line in semi-log plot.
These results are averaged over 100 samples.}
\label{fig_pk}
\end{figure}

Figures \ref{fig_pk}(a)-(c) 
show the degree distributions in the onion-like networks 
according to DLA, IP, and Eden models. 
We remark that the largest degree $k_{max}$ is bounded around 
$20$ with the exponential tail for $N=2000$ (see Inset).
The load at a node for maintaining links becomes smaller than 
that in SF networks, 
since there is no huge hubs.
We compare Figs. \ref{fig_pk}(a)-(c) 
with the result for the previous generation method of 
spatial growing networks \cite{Hayashi14}
without the constraint in the surface growth, 
in which a new node is located on random radius between 
$r_{min}$ and $r_{max}$ with a random 
direction \cite{Brunet07} from the position of a uniformly 
randomly chosen node 
in order to make proximity connections heuristically.
Figure \ref{fig_pk}(d) shows the distribution with slightly larger 
$k_{max}$ but the exponential tail is more clear in the 
spatial growing networks without the constraint in the surface growth.
The smaller $k_{max}$ in Fig \ref{fig_pk}(a)-(c) is probably caused 
by the constraint.

\begin{figure}[htp]
 \begin{minipage}{0.31\textwidth} 
   \centering
   \includegraphics[height=35mm]{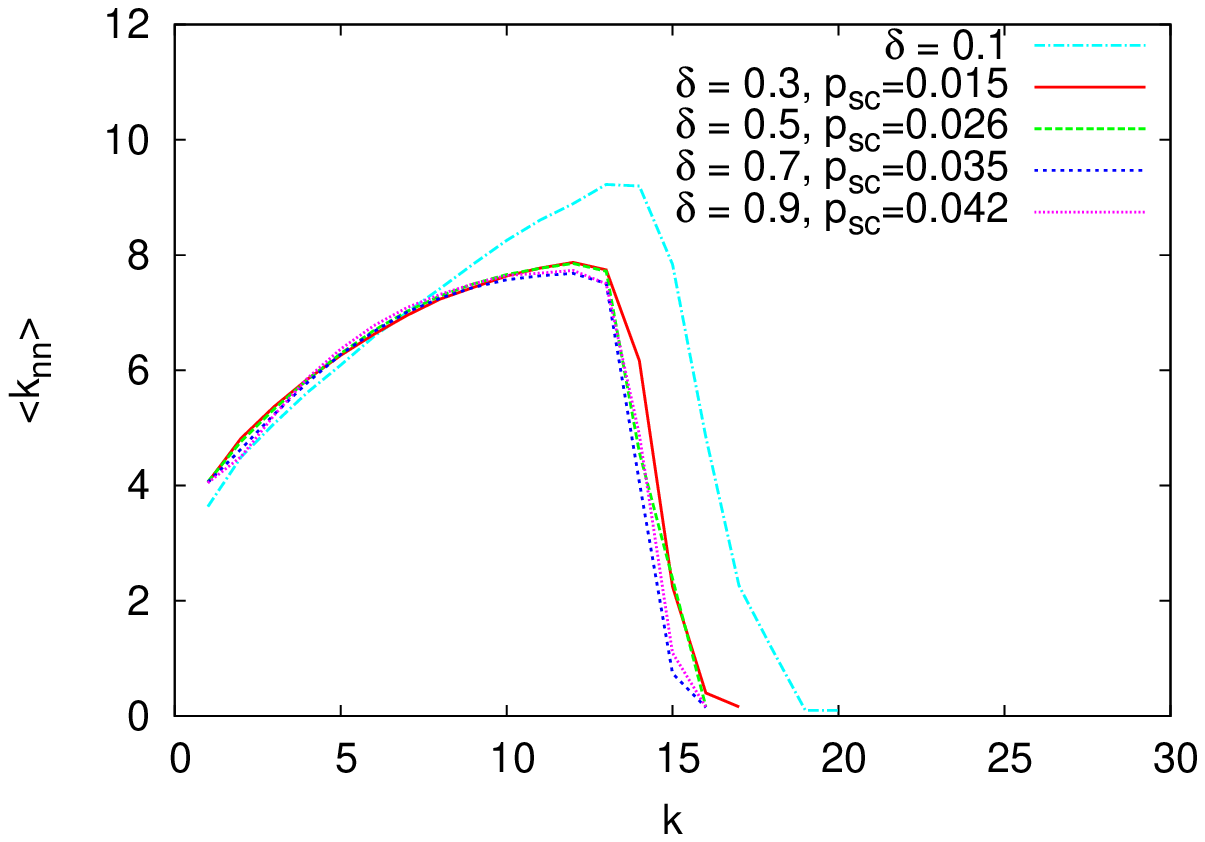}
     \begin{center} (a) DLA model \end{center}
 \end{minipage} 
 \hfill 
 \begin{minipage}{0.31\textwidth} 
   \centering
   \includegraphics[height=35mm]{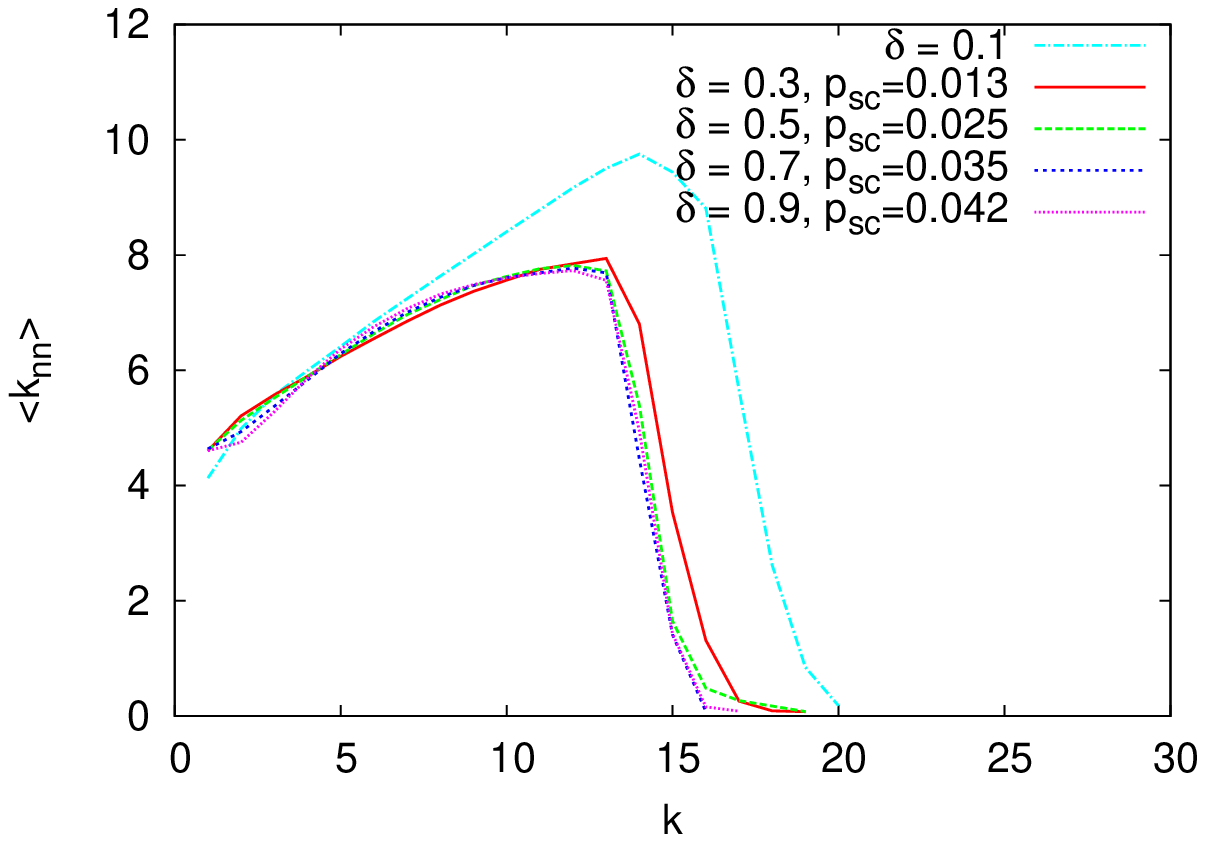}
     \begin{center} (b) IP model \end{center}
 \end{minipage} 
 \hfill 
 \begin{minipage}{0.31\textwidth} 
   \centering
   \includegraphics[height=35mm]{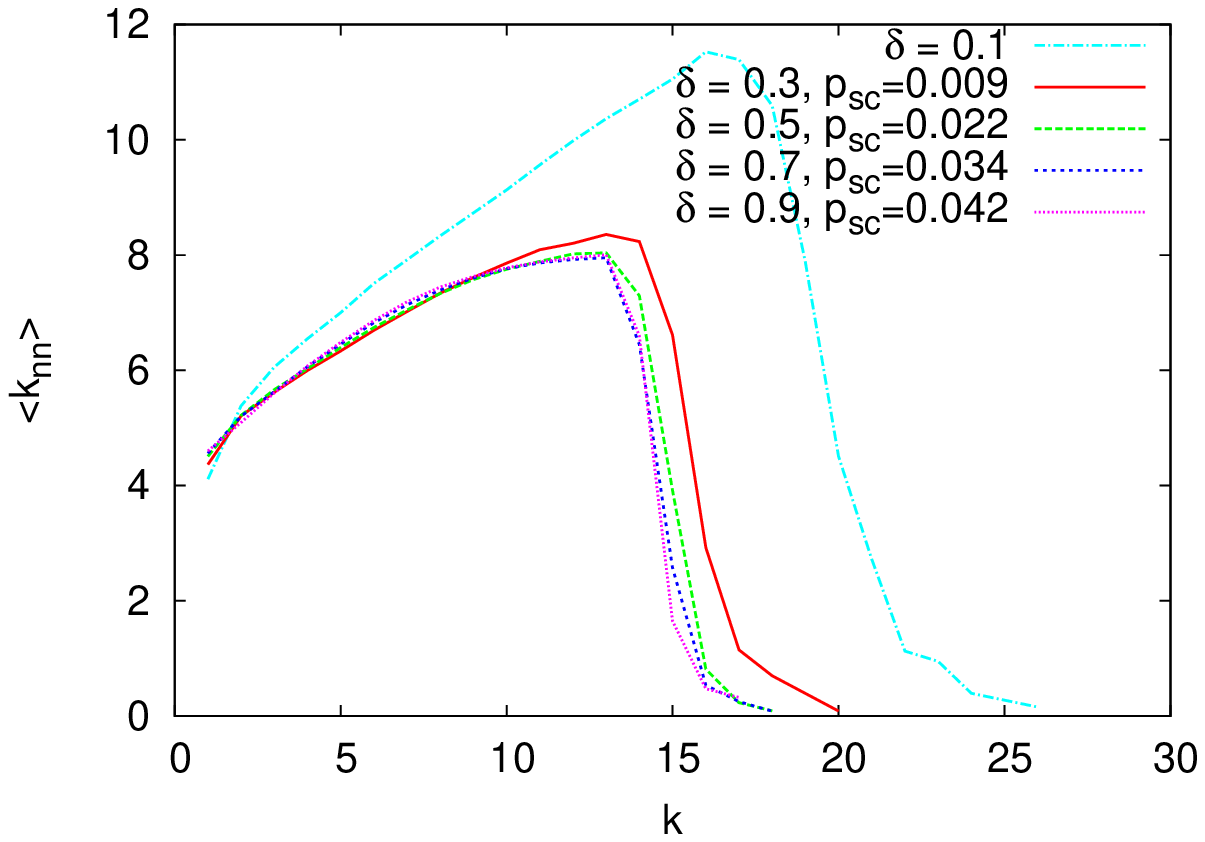}
     \begin{center} (c) Eden model \end{center}
 \end{minipage} 
\caption{(Color online) Distribution of the average degree 
$\langle k_{nn} \rangle$ of the nearest neighbor nodes of node with 
degree $k$
in the networks according to (a) DLA, (b) IP, and (c) Eden models 
from left to right for $N=2000$ with $\langle k \rangle \approx 5.6$.
Positive degree-degree correlations appear except the tails with 
finite-size effect in all cases. 
These results are averaged over 100 samples.}
\label{fig_knn}
\end{figure}

\begin{figure}[htp]
 \begin{minipage}{0.31\textwidth} 
   \centering
   \includegraphics[height=155mm]{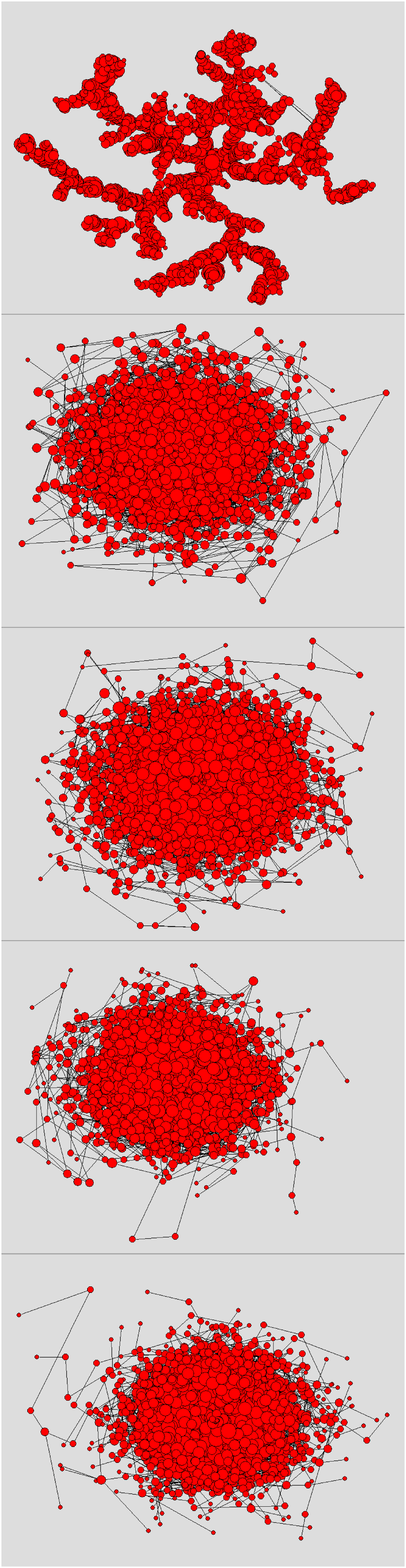}
     \begin{center} (a) DLA model \end{center}
 \end{minipage} 
 \hfill 
 \begin{minipage}{0.31\textwidth} 
   \centering
   \includegraphics[height=155mm]{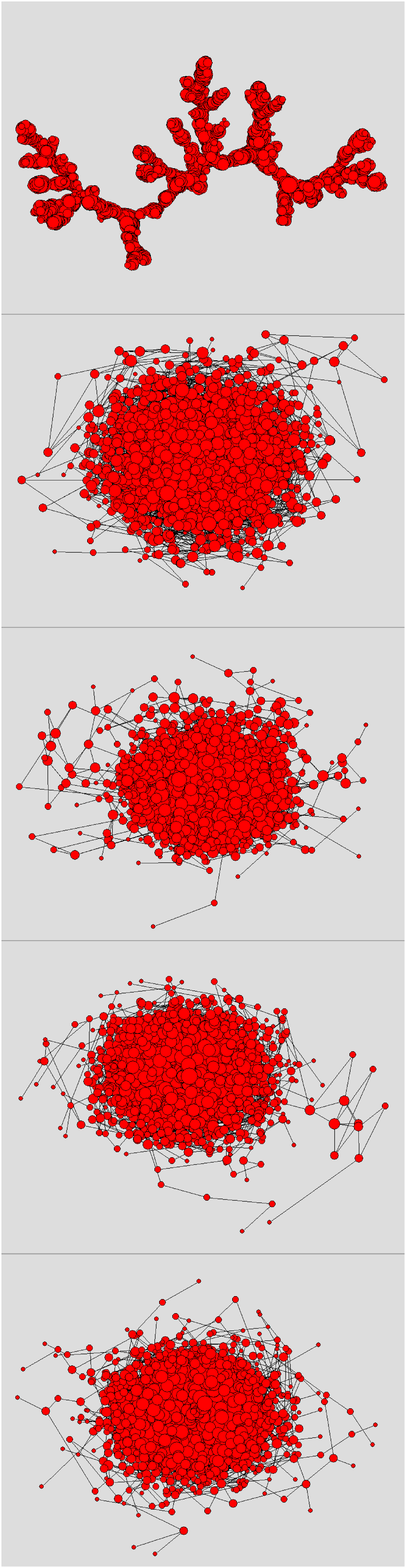}
     \begin{center} (b) IP model \end{center}
 \end{minipage} 
 \hfill 
 \begin{minipage}{0.31\textwidth} 
   \centering
   \includegraphics[height=155mm]{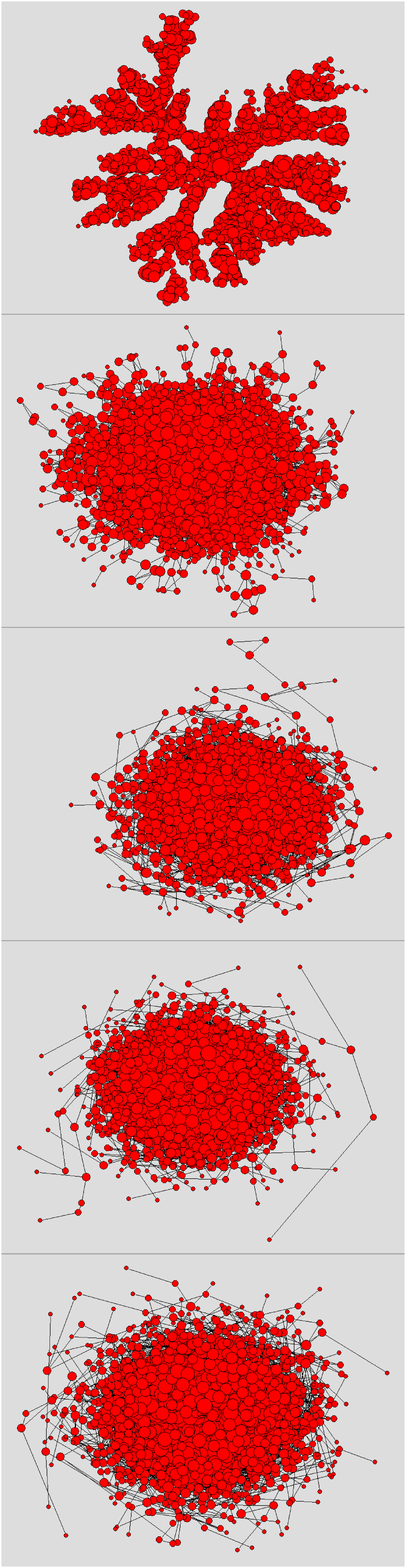}
     \begin{center} (c) Eden model \end{center}
 \end{minipage} 
\caption{Visualization of the topological structures of the networks 
according to (a) DLA, (b) IP, and (c) Eden models from left to right.
A tree-like structure appears at the top for $\delta=0.1$, 
in otherwise onion-like structures with a core area of high degree nodes 
surrounded by low degree nodes appear from the 2nd row to the bottom for 
$\delta = 0.3, 0.5, 0.7$ and $0.9$. 
The node size is proportional to its degree.}
\label{fig_vis}
\end{figure}

We shows 
the average degree $\langle k_{nn} \rangle$ of the nearest neighbor 
nodes of node with degree $k$ in Fig. \ref{fig_knn}.
The increasing slope represents 
positive degree-degree correlations which are necessary to be 
onion-like topology. 
The drops in $k \geq 15$ are ignorable because of finite-size effect 
(see the tails of $p(k)$ in Fig. \ref{fig_pk}).
Note that 
the case of $\delta = 0.1$ has also positive degree-degree 
correlations in spite of the tree-like structure 
without onion-like topology. 
From top to bottom in Fig. \ref{fig_vis}, 
we show examples of tree-like and onion-like topological structures 
visualized by Pajek \cite{Pajek} in ignoring the spatial positions 
of nodes on the surface growth. 
In particular, from the 2nd row to the bottom for 
$\delta = 0.3, 0.5, 0.7,$ and $0.9$, high degree nodes concentrate on 
the center area while low degree nodes surround them.

\section{Robustness and efficiency on the growing networks} \label{sec3}
We investigate the robustness against random failures and malicious 
attacks in subsection 3.1, the growing behavior in subsection 3.2, 
and the efficiency of path in subsection 3.3. 
In the malicious attack, 
nodes are removed in decreasing order of the current 
degrees through the recalculations \cite{Holme02}.
We also discuss the resilience against sequential attacks 
in subsection 3.4.

\subsection{Robustness of connectivity on the growing networks}
We consider following two measures for the robustness of 
connectivity and the degree-degree correlation in a network.
For the robustness, 
we investigate an index \cite{Schneider11,Herrmann11,Wu11}:
\begin{equation}
  R = \frac{1}{N} \sum_{q = 1/N}^{1} S(q), 
\label{eq_def_R}
\end{equation}
where 
$S(q)$ denotes the number of nodes in the giant component 
(GC: largest connected cluster) after removing $q N$ nodes, 
$q$ is a fraction of removed nodes
by random failures or malicious attacks.
The range of $R$ is $[0, 0.5]$, where $R=0$ corresponds to 
a completely disconnected network consisting of isolated nodes, 
and $R=0.5$ corresponds to the most robust network. 
As a measure of degree-degree correlation, 
we investigate the assortativity \cite{Newman03a,Newman10}
\[
  r = \frac{S_{1} S_{e} - S_{2}^{2}}{S_{1} S_{3} - S_{2}^{2}}, 
\]
where $S_{1} = \sum_{i} k_{i}$, $S_{2} = \sum_{i} k_{i}^{2}$
$S_{3} = \sum_{i} k_{i}^{3}$, $S_{e} = \sum_{ij} A_{ij} k_{i} k_{j}$, 
$A_{ij}$ denotes the $i$-$j$ element of the adjacency matrix. 
The range of $r$ is $[-1, 1]$ as the Pearson correlation 
coefficient for degrees. 
Nodes with similar degrees tend to be connected as 
$r > 0$ is larger. 
Note that 
onion structure and assortativeness with a large $r > 0$ 
are distinct properties  \cite{Schneider11}: 
{\it Not all assortative networks have onion structure
but all onion networks are assortative} \cite{Wu11}.
Therefore, the value of $r > 0$ is relatively 
large in an onion-like network.

Table \ref{table_DLA} shows the average values over 100 
samples for our networks with $a=0.3$ in the 4-6th columns 
and the corresponding rewired version reconstructed by 
Wu and Holme's algorithm \cite{Wu11} with $a=3.0$ 
in the 7-9th columns.
$R$:failures and $R$:attacks denote the robustness index 
defined by Eq. 
(\ref{eq_def_R}) against random failures and malicious 
attacks, respectively. Note that the rate $p_{sc}$ of adding 
shortcut links is regulated to be 
$\langle k \rangle \approx 5.6$ with a same connection density. 
We remark that, with higher values of $R$, 
the robustness is improved from the tree-like network 
generated by only the copying process for $\delta = 0.1$ 
to the networks for $\delta = 0.3 \sim 0.9$. 
These networks have an onion-like topological structure 
with high assortativity $r$, 
since the values of $R$ are slightly smaller but 
almost coincide with 
$R = 0.436 \sim 0.444$ against random failures and 
$R = 0.307 \sim 0.326$ against malicious attacks 
in the rewired version \cite{Wu11}
to be an onion-like network with the nearly optimal robustness. 
Although 
the assortativity $r$ in the rewired version becomes larger 
in all cases, the improvement on the robustness from our 
networks is small.
Table \ref{table_compare} shows the results for comparison 
between the cases of $a=0.3$ and $a=3.0$. 
In order to be $\langle k \rangle \approx 5.6$ in the case of $a=3.0$, 
larger rates $p_{sc}$ are necessary especially for large rates 
$\delta$ of deletion. The setting of $p_{sc}$ 
means that the effect of the 
copying process becomes very weak within limited strong degree-degree 
correlations, because almost all 
of duplication links are failed as deletion 
at each time step as similar to the case of $\delta=0.9$.
Not surprisingly, 
a larger assortativity $r$ is obtained in Table \ref{table_compare}
because of stronger degree-degree correlations by larger $a$ than 
that for each corresponding case in Table \ref{table_DLA}.
However, 
the value of robustness index $R$ in Table \ref{table_compare}
is at a same level as that 
for each corresponding case in Table \ref{table_DLA}.
There are no notable differences among these results for 
DLA, IP, Eden models from top to bottom in both Tables.
Thus, the setting of $a=0.3$ does not lose the intrinsic 
property for the robustness in our networks, moreover 
the effect of the copying process remains.

\begin{figure}[tp]
\centering
 \begin{minipage}{0.47\textwidth} 
   \includegraphics[height=70mm,angle=-90]{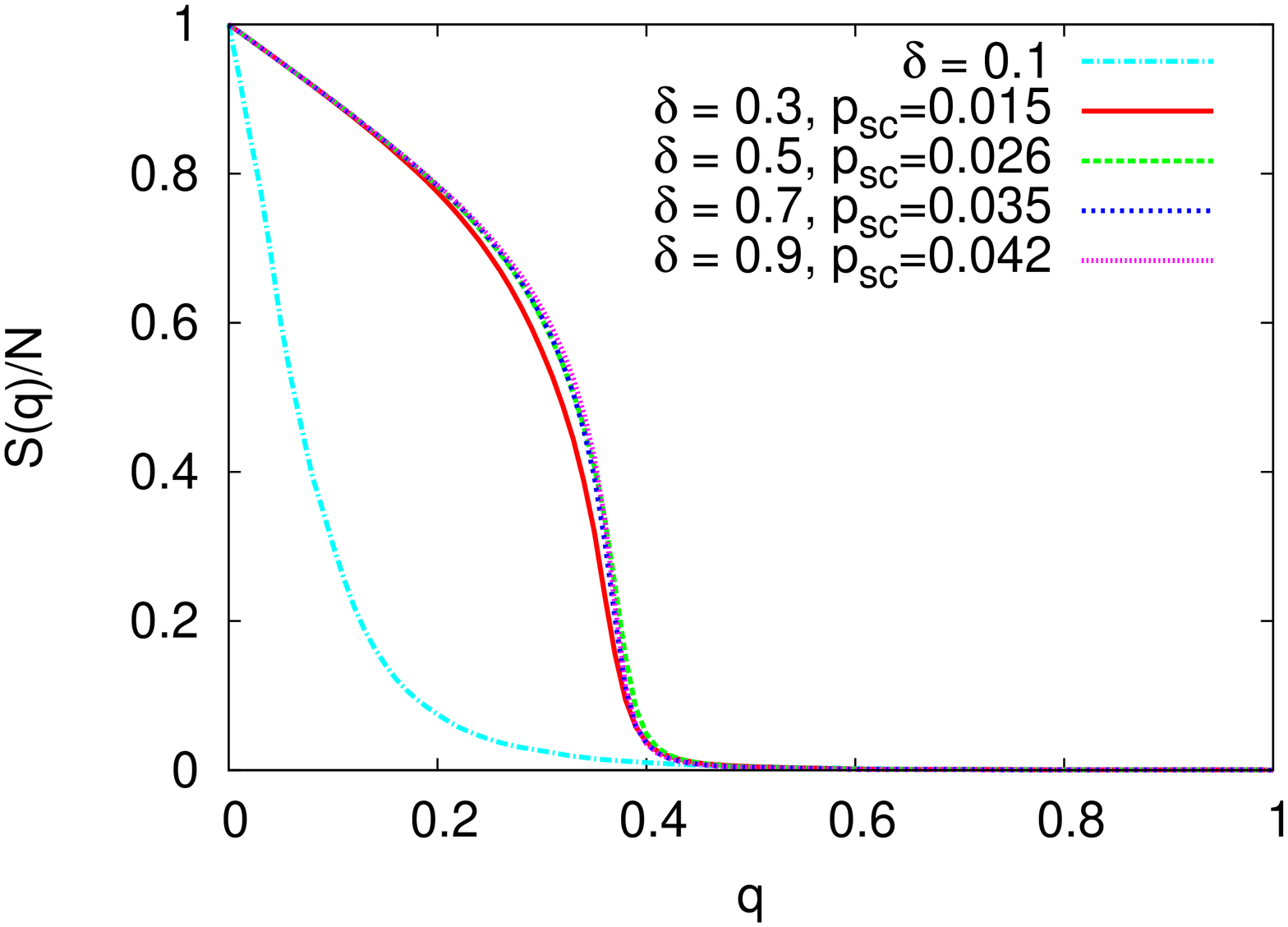}
    \begin{center} (a) Relative size of GC for attacks \end{center}
 \end{minipage} 
 \hfill 
 \begin{minipage}{0.47\textwidth} 
   \includegraphics[height=70mm,angle=-90]{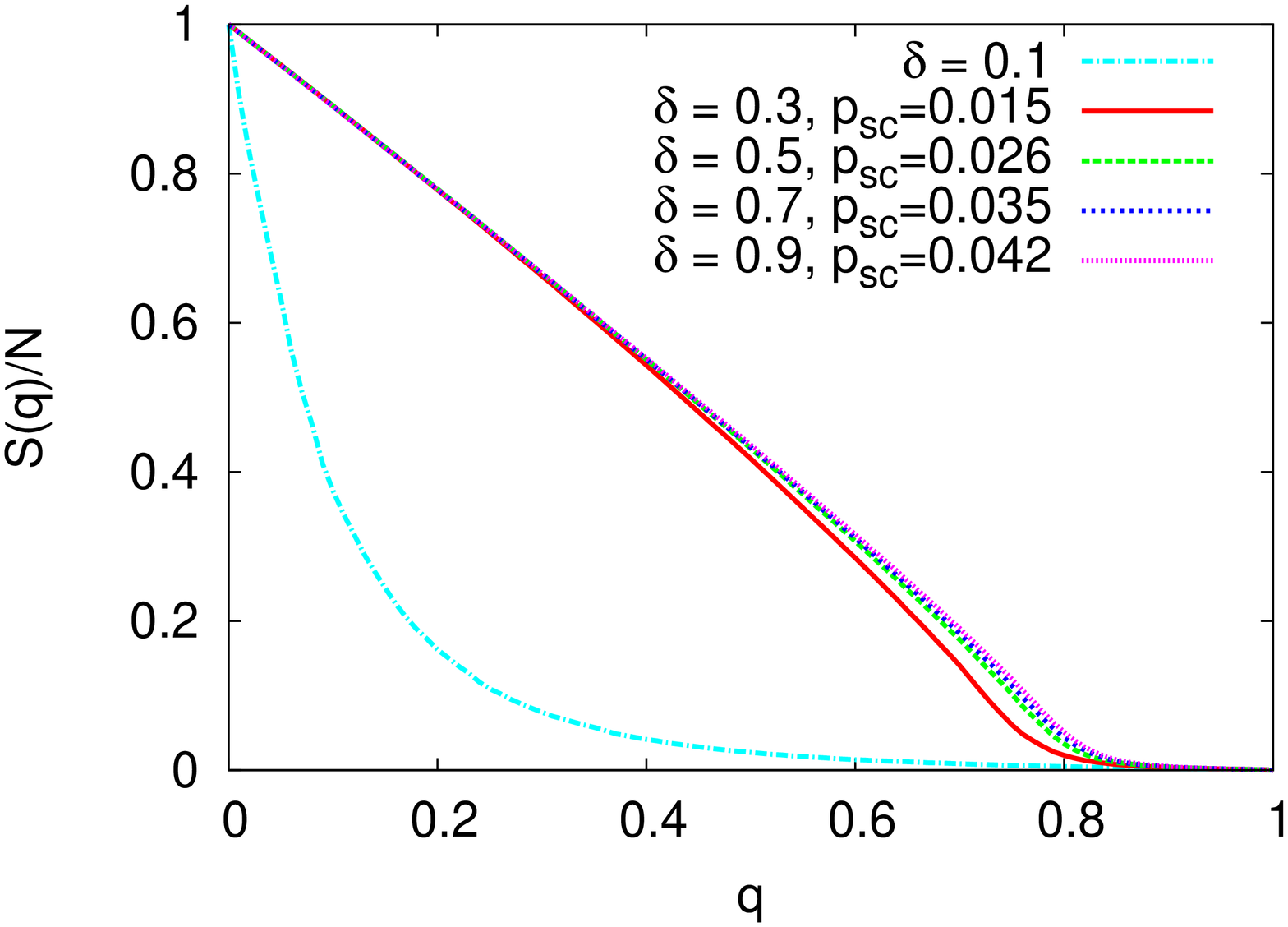}
    \begin{center} (b) Relative size of GC for failures \end{center}
 \end{minipage} 
 \hfill
 \begin{minipage}{0.47\textwidth} 
   \includegraphics[height=70mm,angle=-90]{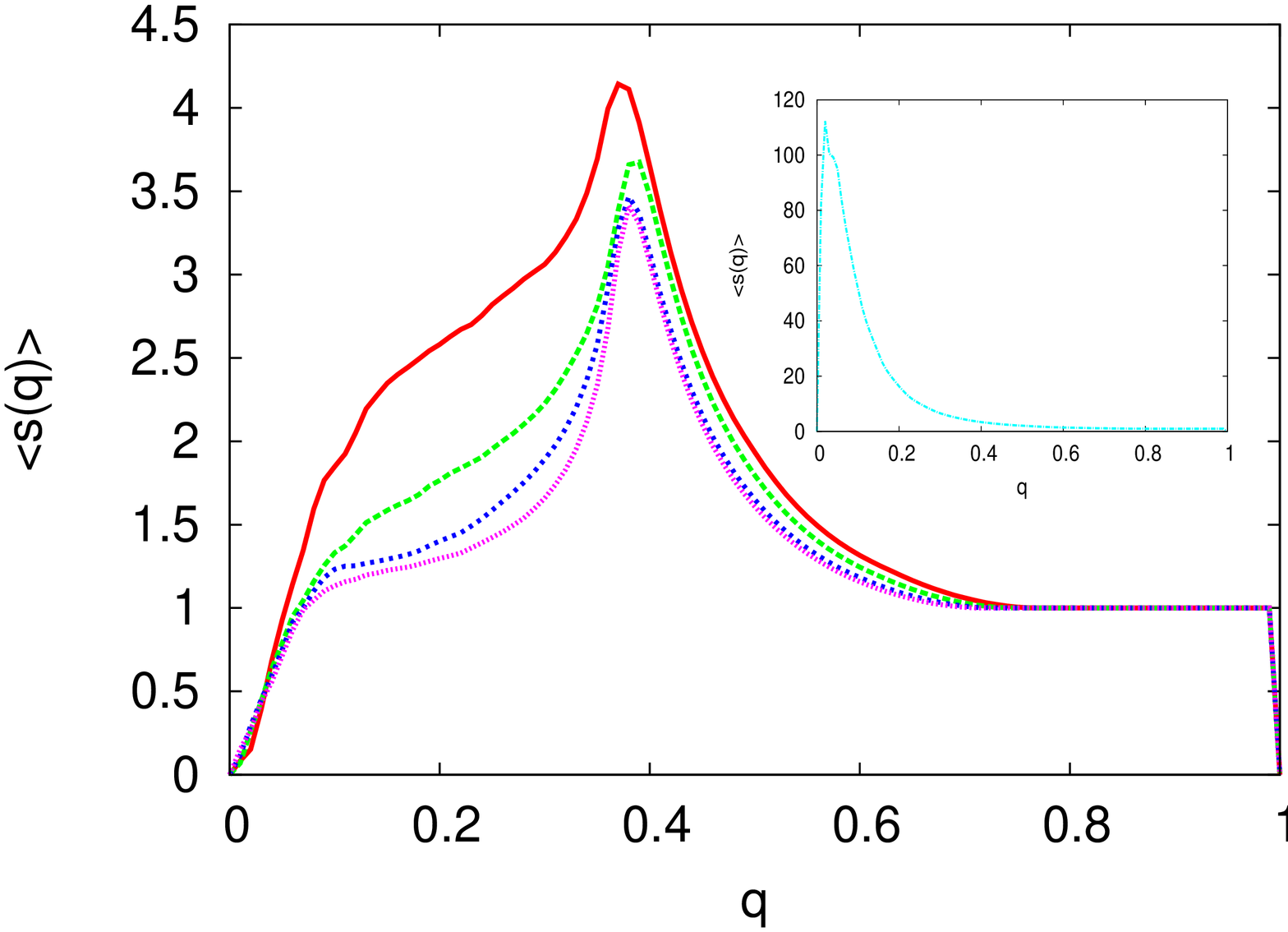}
    \begin{center} (c) Ave. size of clusters for attacks \end{center}
 \end{minipage} 
 \hfill 
 \begin{minipage}{0.47\textwidth} 
   \includegraphics[height=70mm,angle=-90]{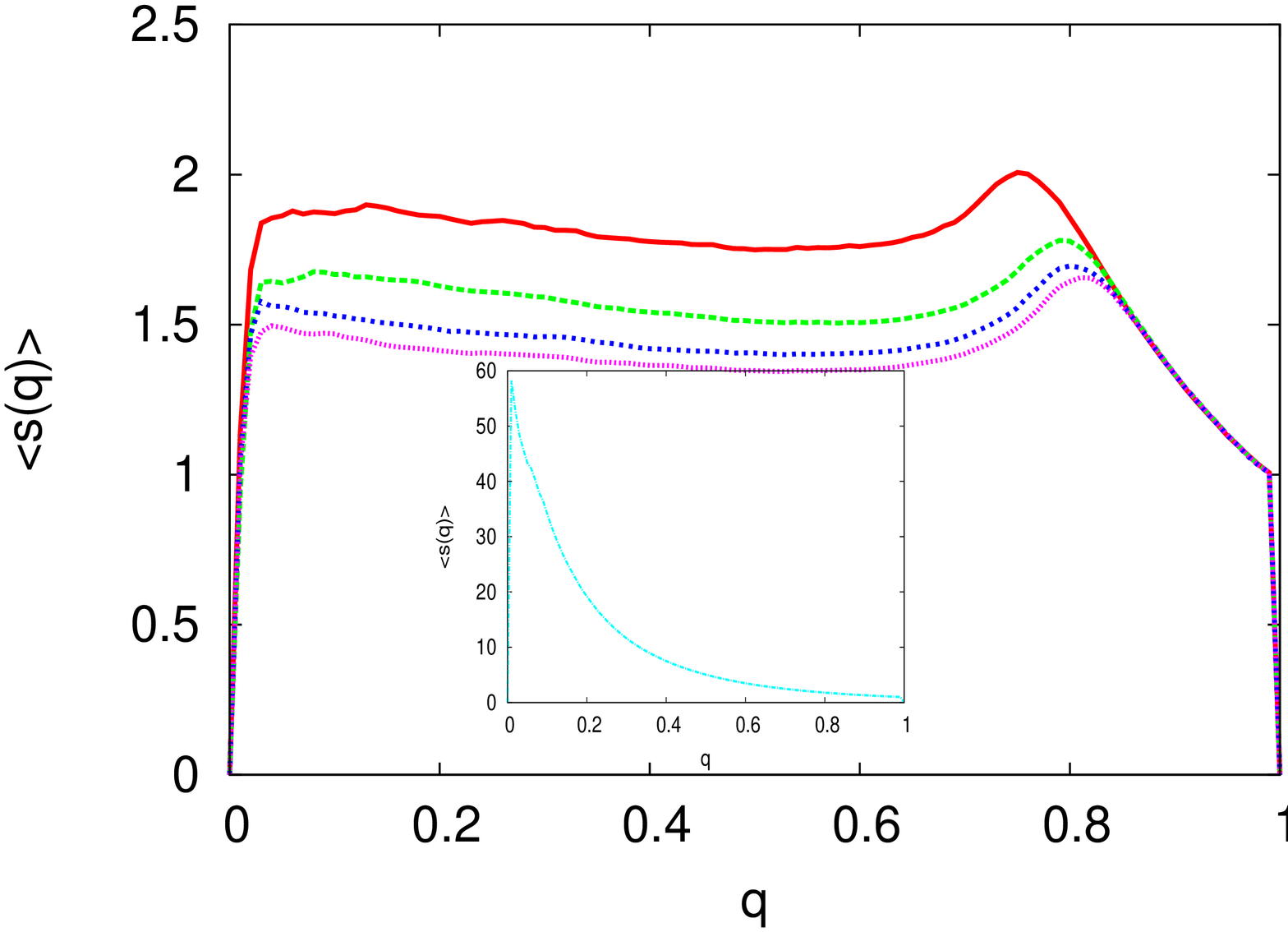}
    \begin{center} (d) Ave. size of clusters for failures \end{center}
 \end{minipage} 
\caption{(Color online) Robustness against 
(a)(c) malicious attacks and 
(b)(d) random failures in the networks according to DLA model 
for $N = 2000$ with $\langle k \rangle \approx 5.6$. 
The top:(a)(b)
shows the relative size $S(q)/N$ of 
the giant component (GC) versus fraction $q$ of removed nodes, 
and the bottom:(c)(d) 
shows the average size $\langle s(q) \rangle$ 
of isolated clusters except the GC.
Similar graphs are obtained for IP and Eden models.
}
\label{fig_SN_s}
\end{figure}

In the following until the end of next section, 
we set $a=0.3$ unless otherwise noted.
Figure \ref{fig_SN_s} shows typical results of the robustness. 
Inset (cyan line) shows the vulnerability  
in the case of tree-like networks for $\delta = 0.1$.  
Note that 
the value of $R$ defined by Eq.(\ref{eq_def_R}) corresponds  
to the area under the line of $S(q)/N$. 
The peak of $\langle s \rangle$ corresponds to the critical 
point $q_{c}$ at the breaking of the GC (dropping point of 
$S(q)/N$).
Although a larger $q_{c}$ indicates a more robust network, 
$R$ is generally a more precise index than $q_{c}$.
Because different values of $R$ can exist for a same value of $q_{c}$
depending on the steepness of dropping curve of $S(q)/N$.

\begin{figure}[tp]
 \begin{minipage}{0.31\textwidth} 
   \centering
   \includegraphics[height=50mm,angle=-90]{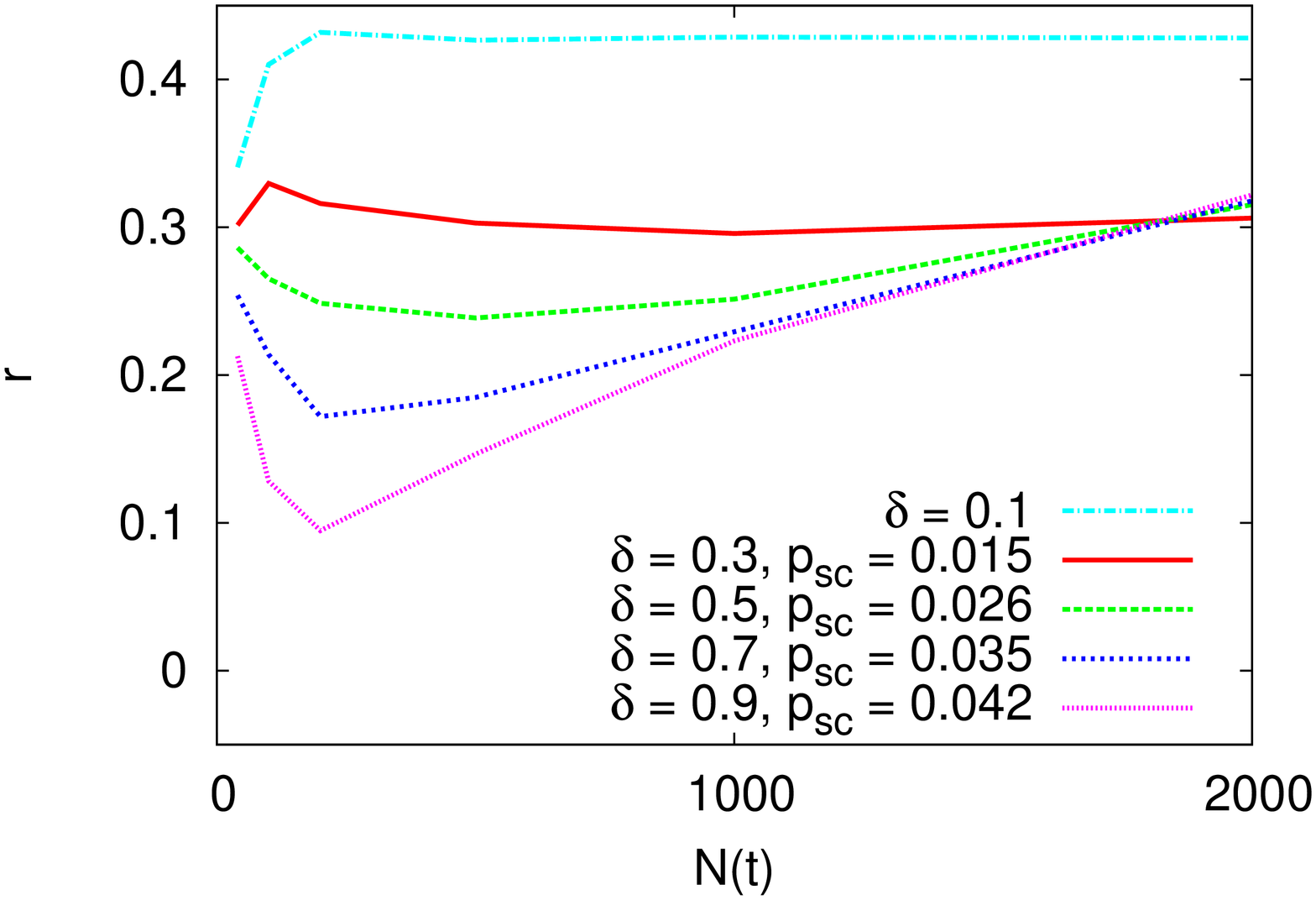}
 \end{minipage} 
 \hfill 
 \begin{minipage}{0.31\textwidth} 
   \centering
   \includegraphics[height=50mm,angle=-90]{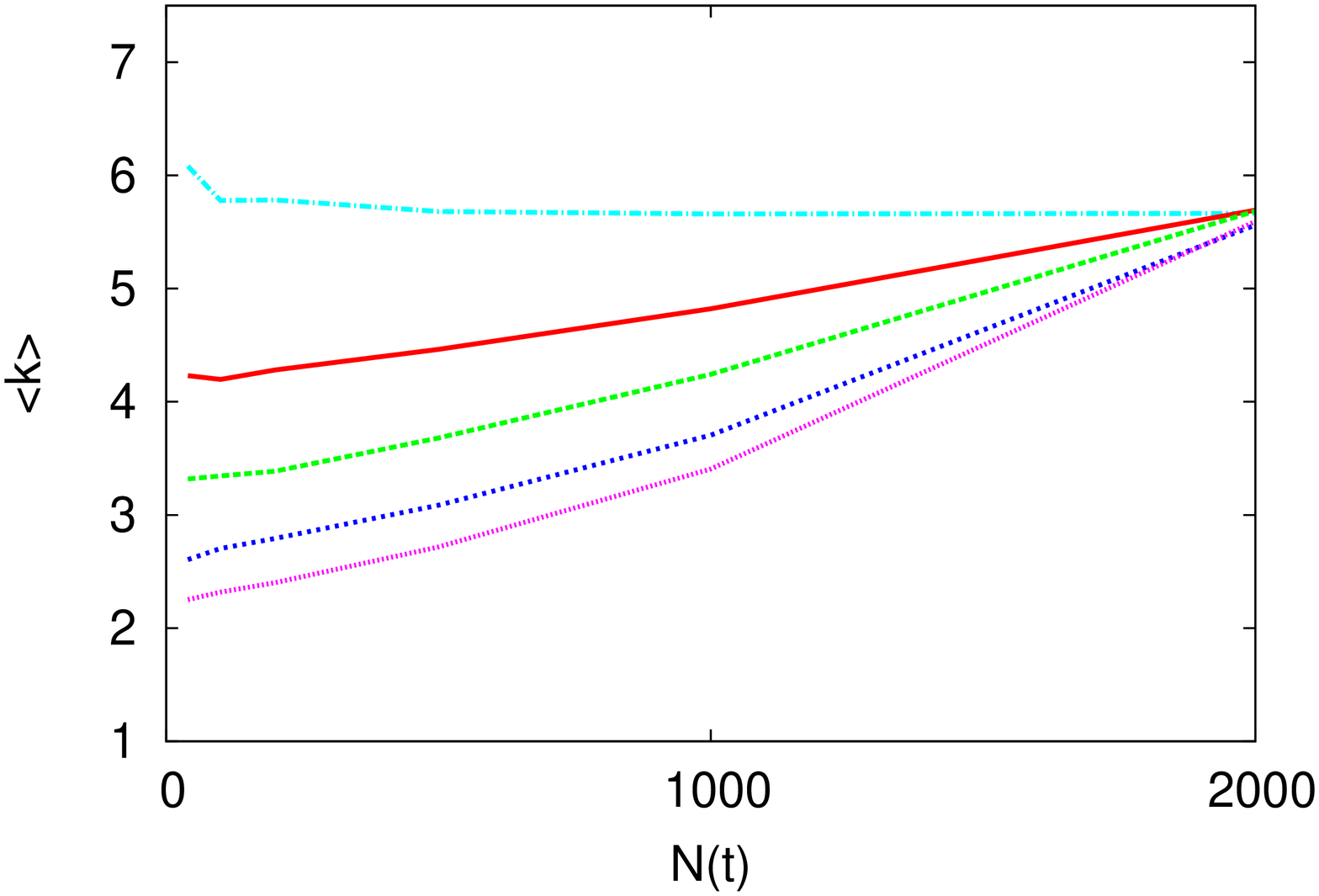}
 \end{minipage} 
 \hfill 
 \begin{minipage}{0.31\textwidth} 
   \centering
   \includegraphics[height=50mm,angle=-90]{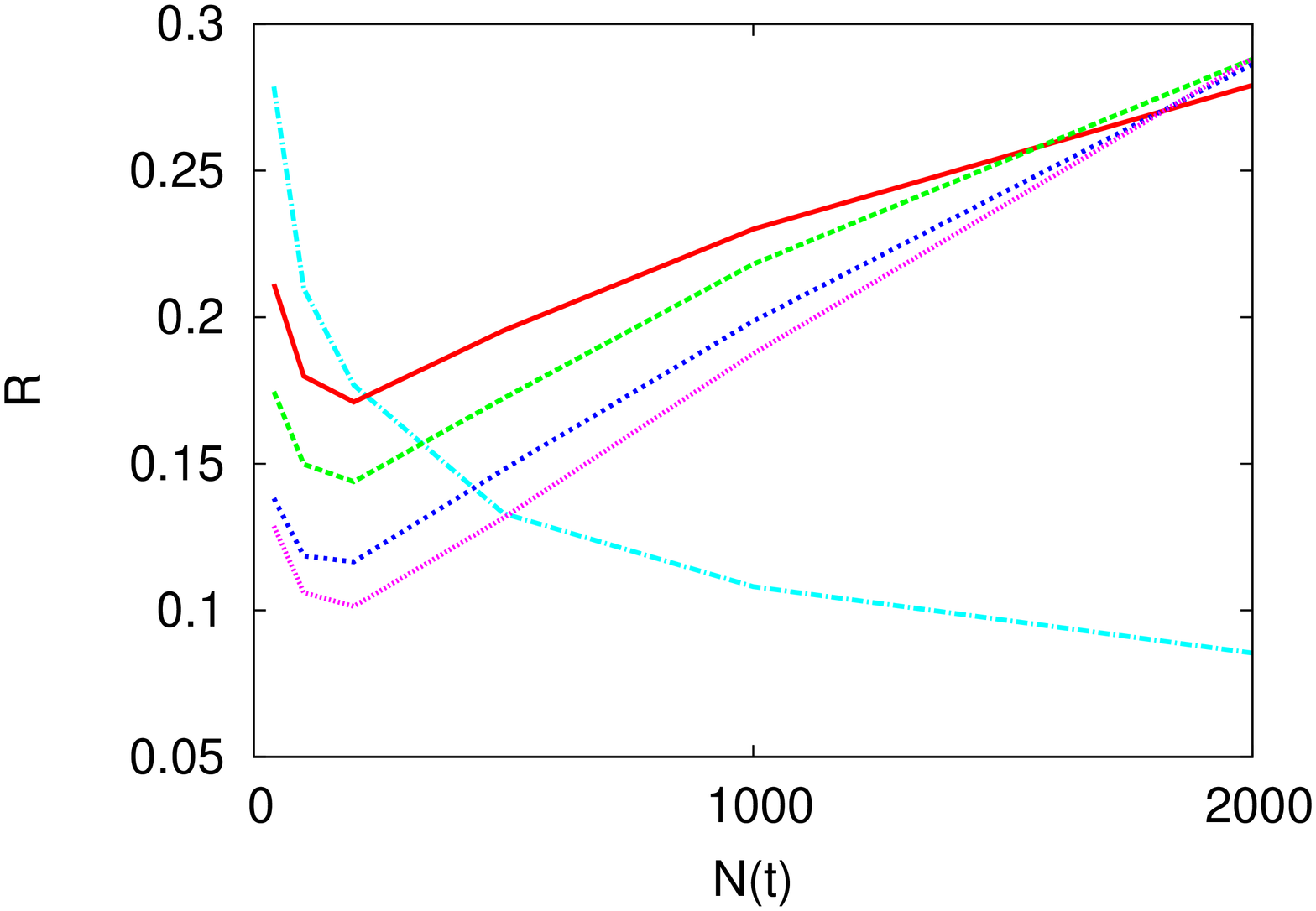}
 \end{minipage} 
 \hfill 
 \begin{minipage}{0.31\textwidth} 
   \centering
   \includegraphics[height=50mm,angle=-90]{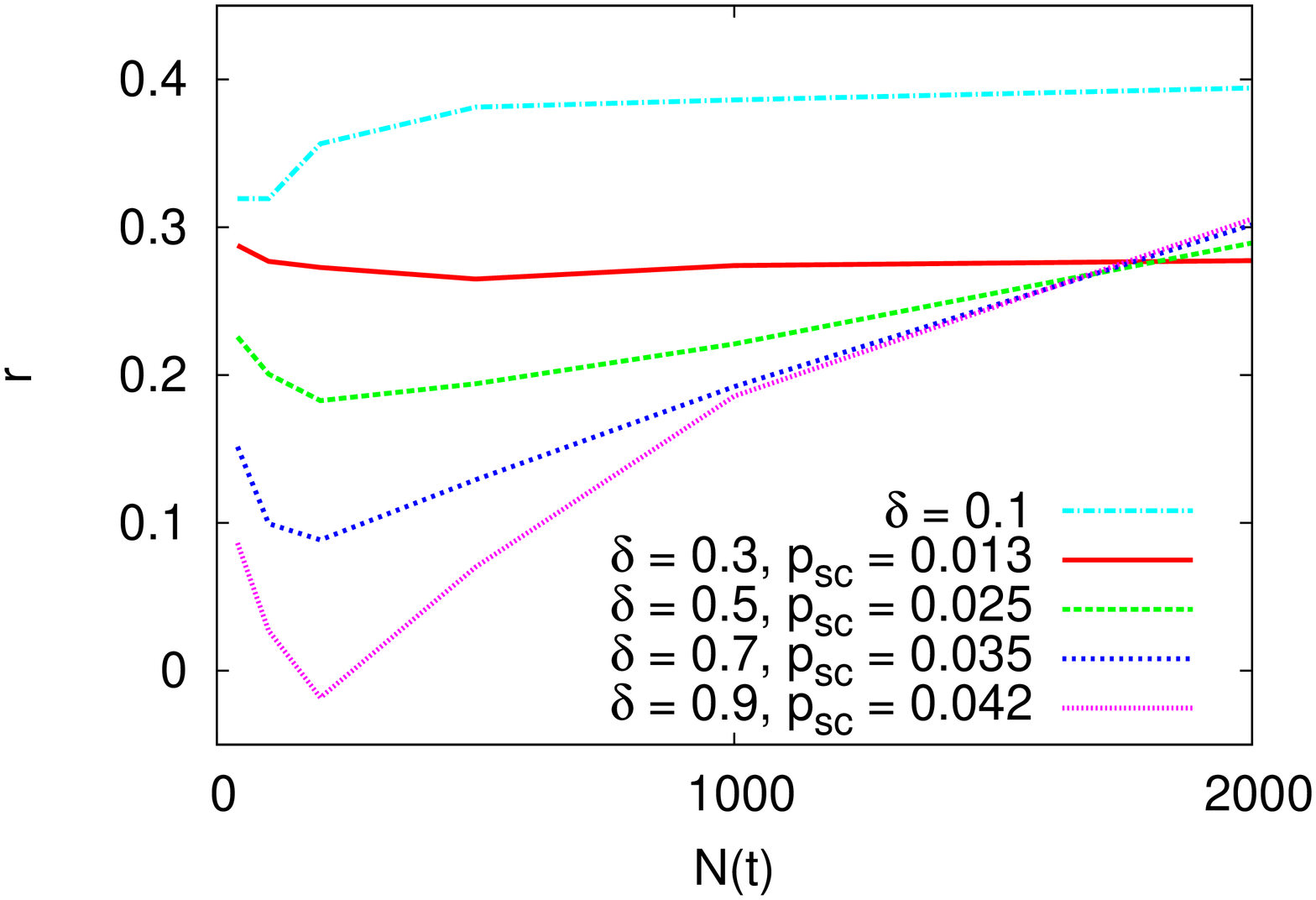}
 \end{minipage} 
 \hfill 
 \begin{minipage}{0.31\textwidth} 
   \centering
   \includegraphics[height=50mm,angle=-90]{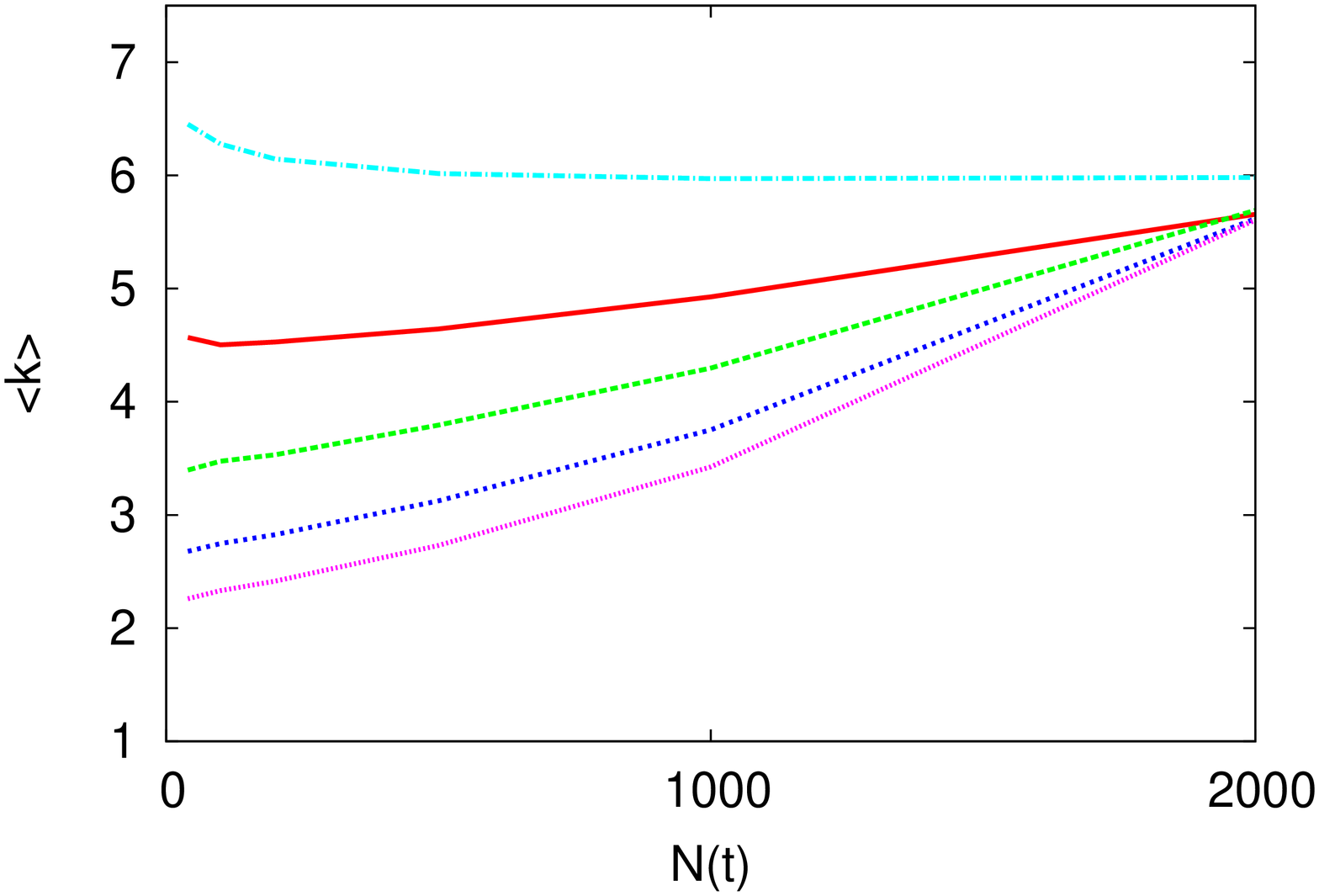}
 \end{minipage} 
 \hfill 
 \begin{minipage}{0.31\textwidth} 
   \centering
   \includegraphics[height=50mm,angle=-90]{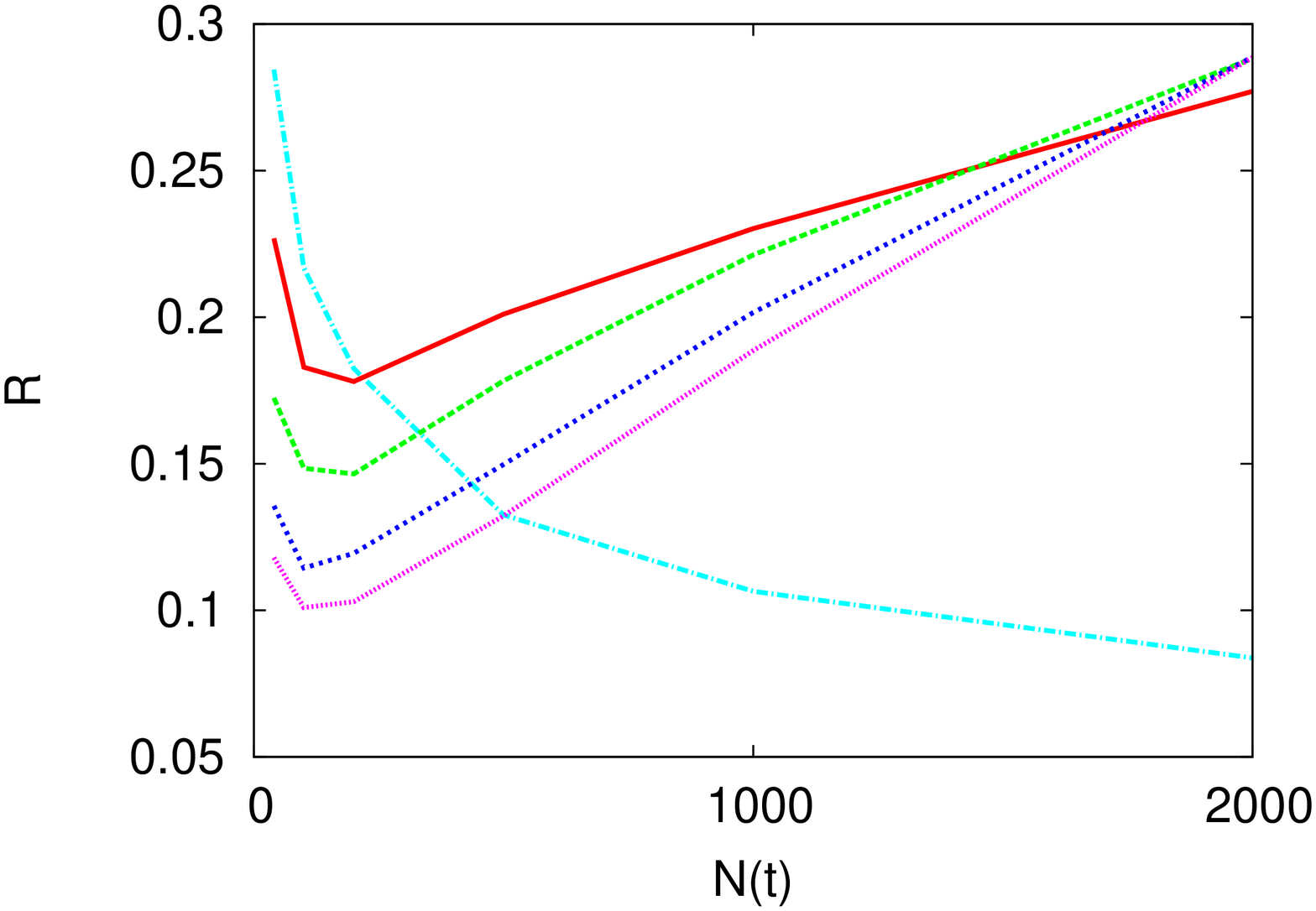}
 \end{minipage} 
 \hfill 
\begin{minipage}[htb]{0.31\textwidth} 
   \centering
   \includegraphics[height=50mm,angle=-90]{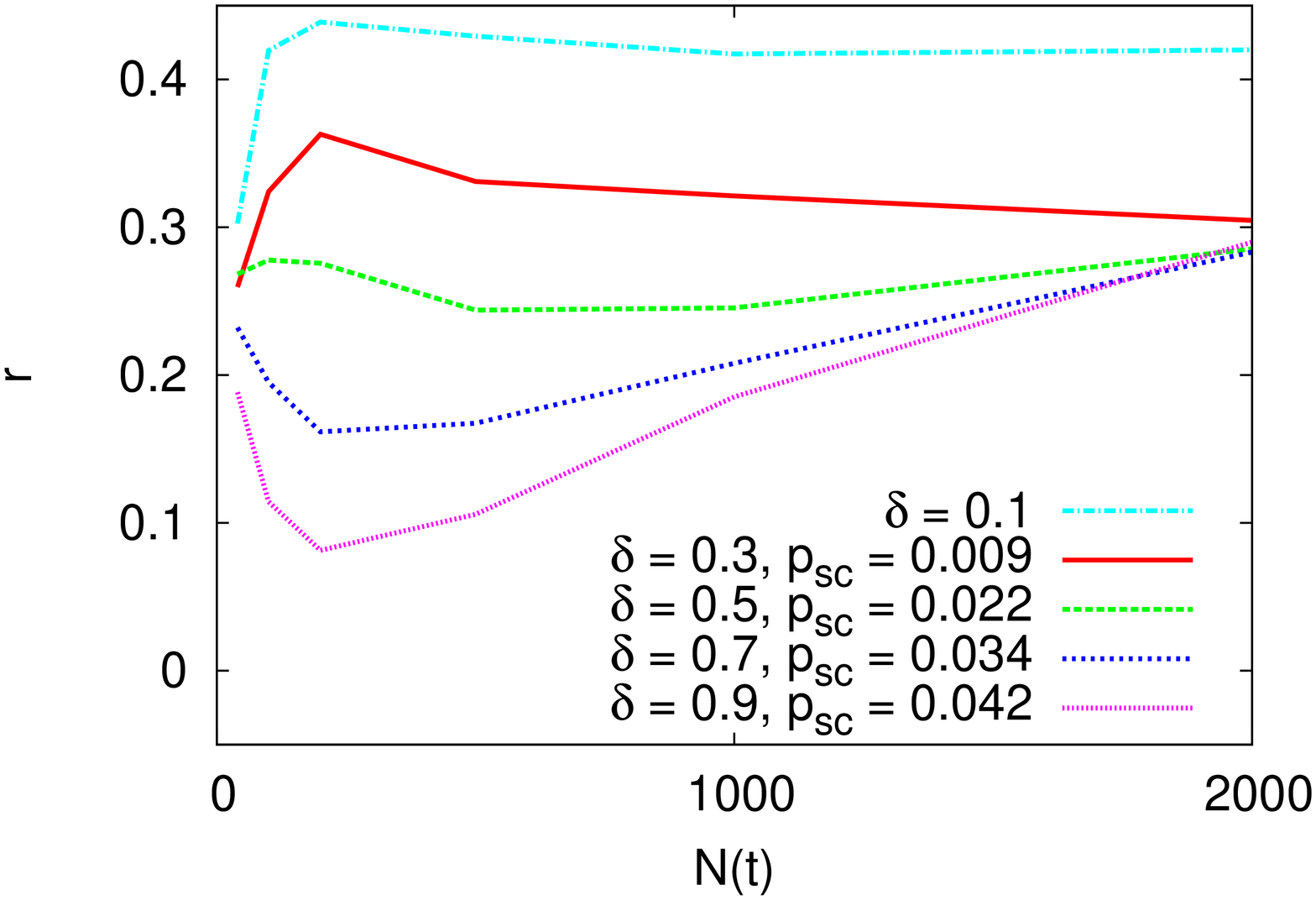}
 \end{minipage} 
 \hfill 
 \begin{minipage}[htb]{0.31\textwidth} 
   \centering
   \includegraphics[height=50mm,angle=-90]{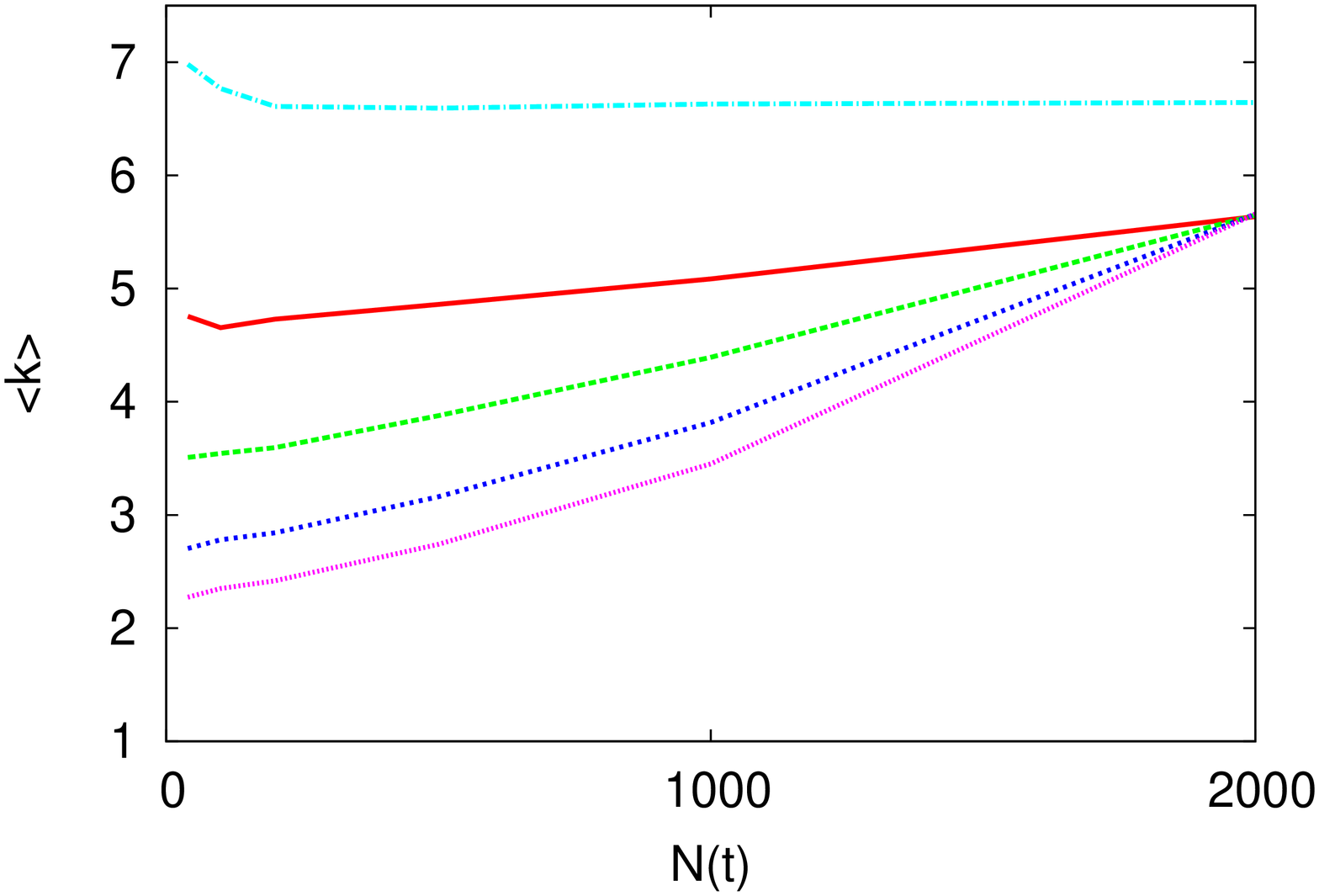}
 \end{minipage} 
 \hfill 
 \begin{minipage}[htb]{0.31\textwidth} 
   \centering
   \includegraphics[height=50mm,angle=-90]{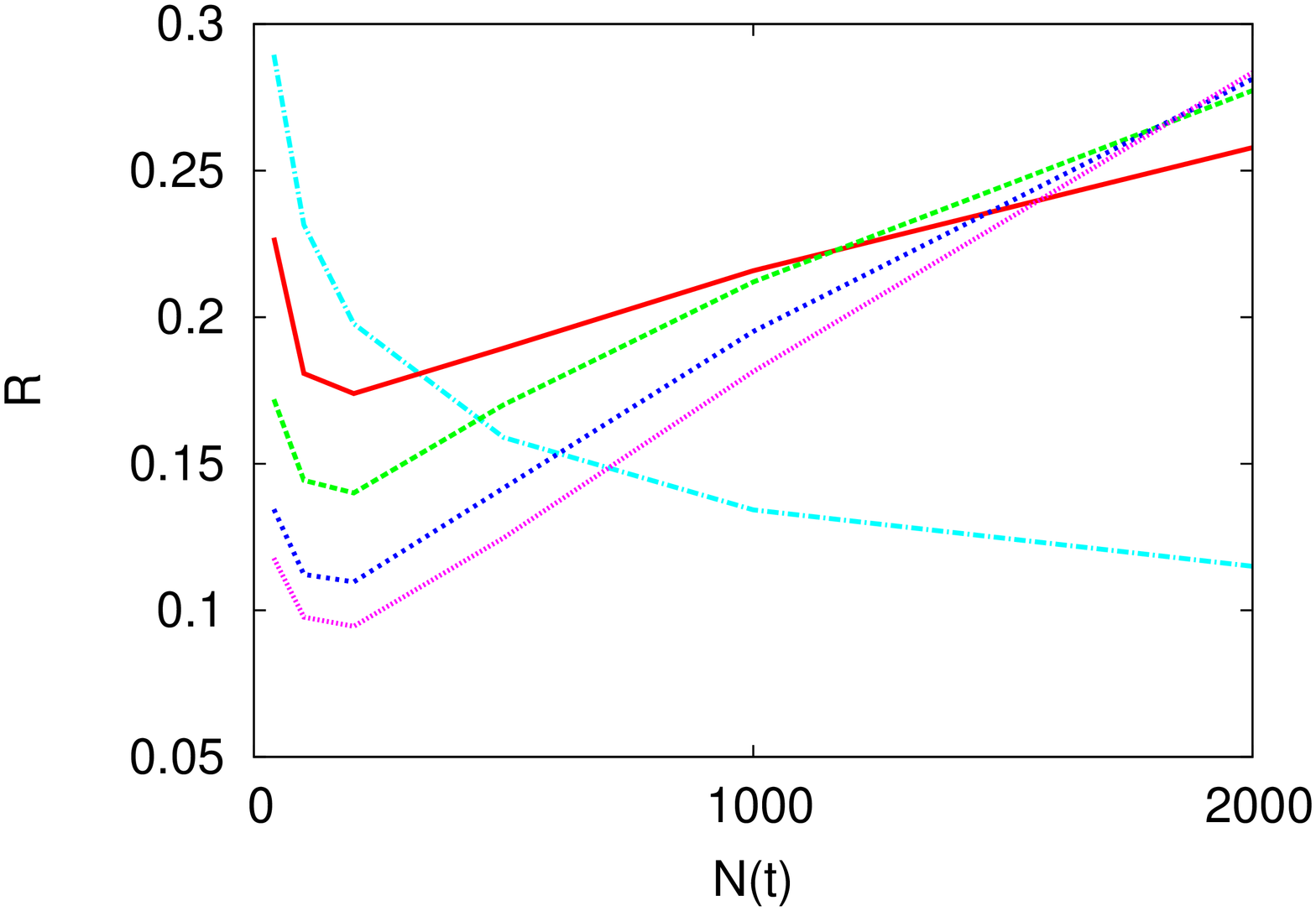}
 \end{minipage} 
\caption{(Color online) Time-courses of assortativity $r$, 
average degree $\langle k \rangle$, robustness index $R$ 
against the malicious attacks 
in the growing networks according to (top) DLA, (middle) IP, 
(bottom) Eden models for the size $N(t)$.
These results are averaged over 100 samples.} 
\label{fig_dla_r-k-R}
\end{figure}

\begin{table}[htb]
\caption{Average values over 100 samples 
in the networks for $N = 2000$ on the surface growth 
according to DLA, IP, and Eden models from top to bottom.
The 4-6th columns show the results for our networks with $a=0.3$, 
and the 7-9th columns show the results for the corresponding 
rewired version \cite{Wu11} with $a=3.0$ from our networks 
with $a=0.3$.
}\label{table_DLA}
\scalebox{0.75}[0.75]{
\begin{tabular}{ccc|ccc|ccc} \hline
DLA & & & our & networks & with $a=0.3$ & rewired & version & with $a=3.0$\\
$\delta$ & $p_{sc}$ & $\langle k \rangle$ & $r$ & $R$:failures & $R$:attacks 
& $r$ & $R$:failures & $R$:attacks \\ \hline
0.1 & 0.0   & 5.663790 & 0.428085 & 0.11414846 & 0.08546072 & 0.570401 & 0.44257568 & 0.32097051 \\ \hline
0.3 & 0.015 & 5.693130 & 0.306182 & 0.42286182 & 0.27909878 & 0.563234 & 0.44446583 & 0.32624905 \\
0.5 & 0.026 & 5.683880 & 0.315288 & 0.43330392 & 0.28798839 & 0.566574 & 0.44266511 & 0.32195528 \\
0.7 & 0.035 & 5.560350 & 0.317886 & 0.43543737 & 0.28620190 & 0.568082 & 0.43890241 & 0.31470397 \\
0.9 & 0.042 & 5.592120 & 0.321879 & 0.43776842 & 0.28822588 & 0.573602 & 0.43836274 & 0.31299371 \\ \hline\hline
IP & & & our & networks & with $a=0.3$ & rewired & version & with $a=3.0$\\
$\delta$ & $p_{sc}$ & $\langle k \rangle$ & $r$ & $R$:failures & $R$:attacks 
& $r$ & $R$:failures & $R$:attacks \\ \hline
0.1 & 0.0   & 5.980630 & 0.394308 & 0.11528832 & 0.08374447 & 0.577101 & 0.44597737 & 0.32535835 \\ \hline
0.3 & 0.013 & 5.656150 & 0.277479 & 0.41909362 & 0.27702885 & 0.561372 & 0.44345529 & 0.32521399 \\
0.5 & 0.025 & 5.691770 & 0.289450 & 0.4337521  & 0.28823022 & 0.569636 & 0.44294242 & 0.32367072 \\
0.7 & 0.035 & 5.616860 & 0.301741 & 0.43641043 & 0.28870629 & 0.574130 & 0.44029968 & 0.31788086 \\
0.9 & 0.042 & 5.606670 & 0.305765 & 0.43796751 & 0.28855921 & 0.571790 & 0.43886147 & 0.31492539 \\ \hline\hline
Eden & & & our & networks & with $a=0.3$ & rewired & version & with $a=3.0$\\
$\delta$ & $p_{sc}$ & $\langle k \rangle$ & $r$ & $R$:failures & $R$:attacks 
& $r$ & $R$:failures & $R$:attacks \\ \hline
0.1 & 0.0   & 6.644020 & 0.420114 & 0.19868989 & 0.11501878 & 0.600332 & 0.45091269 & 0.33193373 \\ \hline
0.3 & 0.009 & 5.635930 & 0.304668 & 0.41034536 & 0.25781734 & 0.567774 & 0.439060744 & 0.31314427 \\
0.5 & 0.022 & 5.648480 & 0.285373 & 0.43086621 & 0.27727353 & 0.568454 & 0.43895619 & 0.31237012 \\
0.7 & 0.034 & 5.657810 & 0.283205 & 0.43680778 & 0.28117685 & 0.572352 & 0.43777281 & 0.30982738 \\
0.9 & 0.042 & 5.658680 & 0.290042 & 0.43891585 & 0.28344713 & 0.566763 & 0.43673314 & 0.30796406 \\ \hline\hline
\end{tabular}
}
\end{table}

\begin{table}[htb]
\caption{Average values over 100 samples 
in the networks for $N = 2000$ on the surface growth 
according to DLA, IP, and Eden models from top to bottom.
The 4-6th columns show the results for comparison between the 
cases of $a=0.3$ and $a=3.0$ 
that is different parameter setting form Table \ref{table_DLA}, 
and the 7-9th columns show the results for the corresponding 
rewired version \cite{Wu11} with $a=3.0$ from our networks
with $a=3.0$.
}\label{table_compare}
\scalebox{0.75}[0.75]{
\begin{tabular}{ccc|ccc|ccc} \hline
DLA & & & our & networks & with $a=3.0$ & rewired & version & with $a=3.0$\\
$\delta$ & $p_{sc}$ & $\langle k \rangle$ & $r$ & $R$:failures & $R$:attacks 
& $r$ & $R$:failures & $R$:attacks \\ \hline
0.1 & 0.0   & 2.834040 & 0.437167 & 0.04345627 & 0.03607689 & 0.437008 & 0.32199987 & 0.18107685  \\ \hline
0.3 & 0.035 & 5.650160 & 0.457448 & 0.43313566 & 0.2979080  & 0.583822 & 0.43888272 & 0.31681199\\ \hline
0.5 & 0.039 & 5.629190 & 0.457572 & 0.43389278 & 0.29740136 & 0.579910 & 0.43849948 & 0.31413023\\ \hline
0.7 & 0.043 & 5.594100 & 0.453034 & 0.43468667 & 0.29602458 & 0.585048 & 0.43675333 & 0.31164826\\ \hline
0.9 & 0.045 & 5.651210 & 0.454004 & 0.43586855 & 0.29712904 & 0.582588 & 0.43772334 & 0.31261318\\ \hline\hline
IP & & & our & networks & with $a=3.0$ & rewired & version & with $a=3.0$\\
$\delta$ & $p_{sc}$ & $\langle k \rangle$ & $r$ & $R$:failures & $R$:attacks 
& $r$ & $R$:failures & $R$:attacks \\ \hline
0.1 & 0.0   & 2.970500 & 0.400235 & 0.04522621 & 0.03565373 & 0.435406 & 0.33601715 & 0.19365666\\ \hline
0.3 & 0.034 & 5.635630 & 0.436601 & 0.43351504 & 0.29754534 & 0.581092 & 0.43953115 & 0.31804051\\ \hline
0.5 & 0.038 & 5.561670 & 0.430161 & 0.43346665 & 0.29522308 & 0.582158 & 0.43758882 & 0.3138613\\ \hline
0.7 & 0.043 & 5.601620 & 0.430950 & 0.4351622  & 0.29589068 & 0.582668 & 0.43698282 & 0.31191742\\ \hline
0.9 & 0.045 & 5.662410 & 0.437168 & 0.43656923 & 0.2982953  & 0.585400 & 0.43781393 & 0.31344812\\ \hline\hline
Eden & & & our & networks & with $a=3.0$ & rewired & version & with $a=3.0$\\
$\delta$ & $p_{sc}$ & $\langle k \rangle$ & $r$ & $R$:failures & $R$:attacks 
& $r$ & $R$:failures & $R$:attacks \\ \hline
0.1 & 0.0   & 3.167770 & 0.384913 & 0.07280454 & 0.0402646  & 0.445969 & 0.35275171 & 0.19681491\\ \hline
0.3 & 0.032 & 5.666700 & 0.425713 & 0.43319376 & 0.29233634 & 0.580907 & 0.43759013 & 0.31193571 \\ \hline
0.5 & 0.037 & 5.629740 & 0.422828 & 0.4342548  & 0.29071317 & 0.582505 & 0.43608627 & 0.3089480\\ \hline
0.7 & 0.043 & 5.663520 & 0.419765 & 0.43617703 & 0.29168745 & 0.582820 & 0.43608627 & 0.30800115\\ \hline
0.9 & 0.045 & 5.664820 & 0.420572 & 0.43666387 & 0.2916347  & 0.579751 & 0.43535101 & 0.30689578\\ \hline\hline
\end{tabular}
}
\end{table}

\subsection{Growing behavior}
We investigate the growing behavior of our proposed networks. 
In Fig. \ref{fig_dla_r-k-R} left, middle, and right, 
for the time-courses of assortativity $r$, 
average degree $\langle k \rangle$, and robustness index $R$, 
we obtain similar results 
in the networks according to DLA, IP, and Eden models.
They are consistent with the behavior 
in the spatially growing model \cite{Hayashi14} 
without the constraint in the surface growth. 
As shown at the left of 
Fig. \ref{fig_dla_r-k-R} top, middle, and bottom, 
$r$ is almost constant through the growing 
except the cases of $\delta = 0.7$ and $0.9$ 
denoted by dashed (blue) and dotted (magenta) lines.
At the middle column of 
Fig. \ref{fig_dla_r-k-R} top, middle, and bottom, 
$\langle k \rangle$ increases for the 
the growing size $N(t) = t+4$.
The slope (increasing rate) becomes 
steeper as the deletion rate $\delta$ is larger. 
At the right of 
Fig. \ref{fig_dla_r-k-R} top, middle, and bottom, 
while $R$ also increases in the simultaneous progress of the 
copying and adding shortcut links, it decreases in the cases of 
$\delta = 0.1$ without shortcuts in the tree-like network
denoted by the dashed (cyan) lines. 
In general, 
larger $\langle k \rangle$ with more links 
tends to lead to higher robustness. 
Indeed, 
the increasing of $\langle k \rangle$ corresponds to 
the increasing of $R$ in Fig. \ref{fig_dla_r-k-R}
top, middle, and bottom.
Note that 
the decrease of $R$ in the first stage of small $N(t)$ 
is due to the tree-like structure before shaping an onion-like 
topological structure. 
Thus, 
our proposed networks 
become more robust in the growth with the enhancement 
of onion-like topology. 
We remark that 
the case of $\delta = 0.1$ seems to be good
with high $R$ in the first stage of small $N(t)$
while the cases of $\delta = 0.7$ and $0.9$ (blue and magenta lines)
seem to be good with larger increasing 
around the last stage of large $N(t)$. 
The case of $\delta = 0.3$ (red line)
is comparatively stable with smaller fluctuation 
in the entire stage.

\begin{figure}[htp]
 \begin{minipage}{0.47\textwidth} 
   \centering
   \includegraphics[height=67mm,angle=-90]{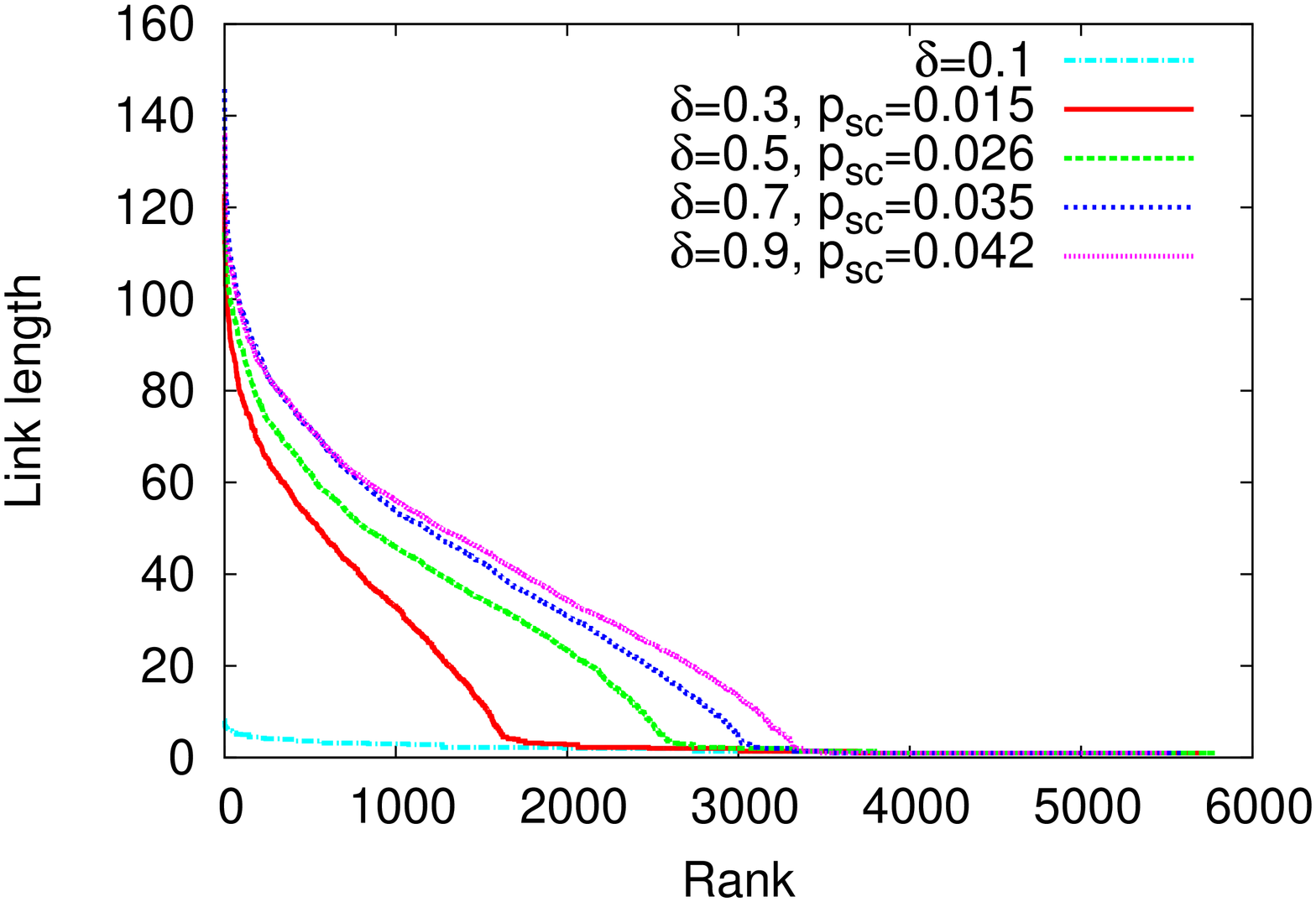}
     \begin{center} (a) DLA model \end{center}
 \end{minipage} 
 \hfill 
 \begin{minipage}{0.47\textwidth} 
   \centering
   \includegraphics[height=67mm,angle=-90]{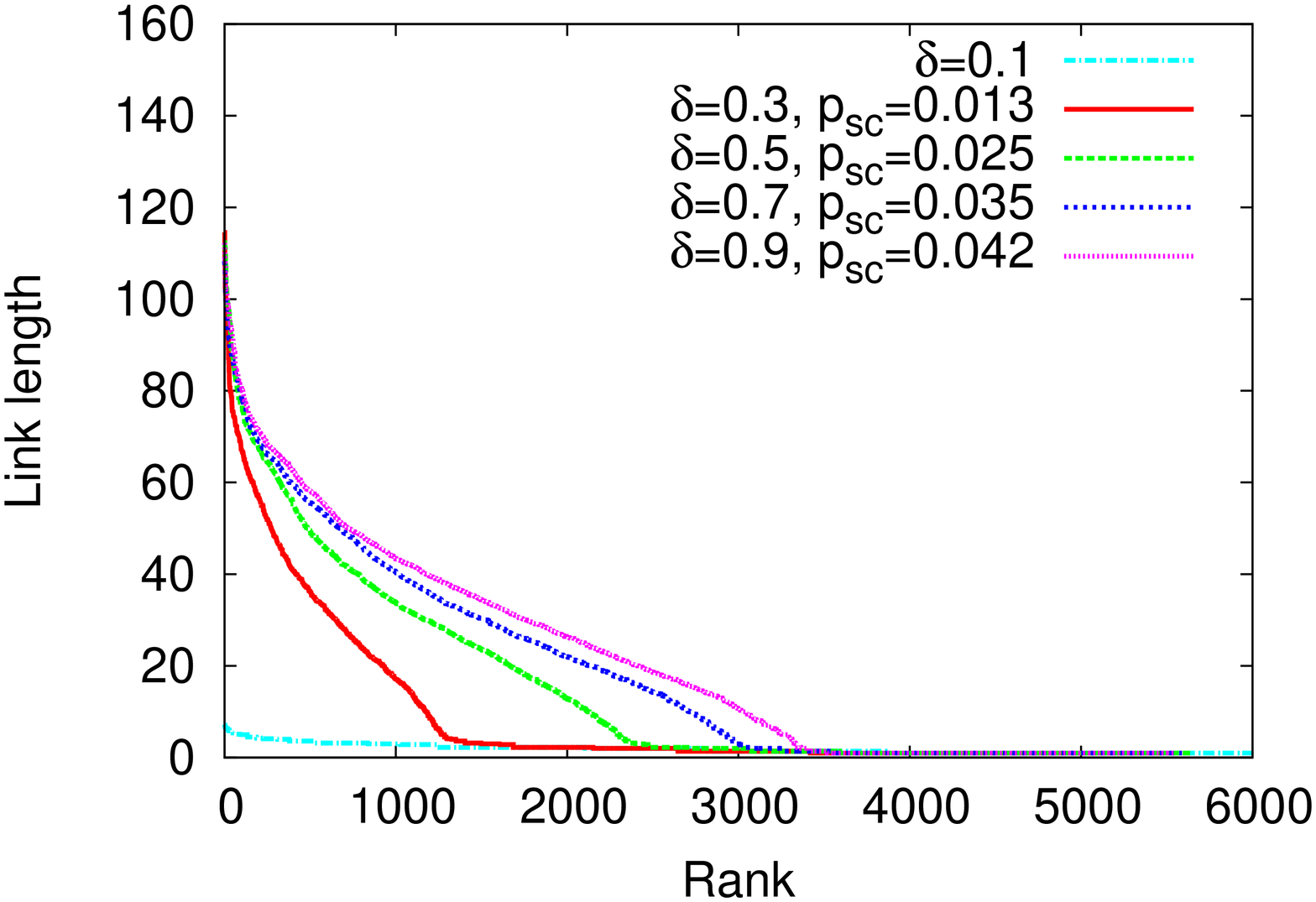}
       \begin{center} (b) IP model \end{center}
 \end{minipage} 
 \hfill 
 \begin{minipage}{0.47\textwidth} 
   \centering
   \includegraphics[height=67mm,angle=-90]{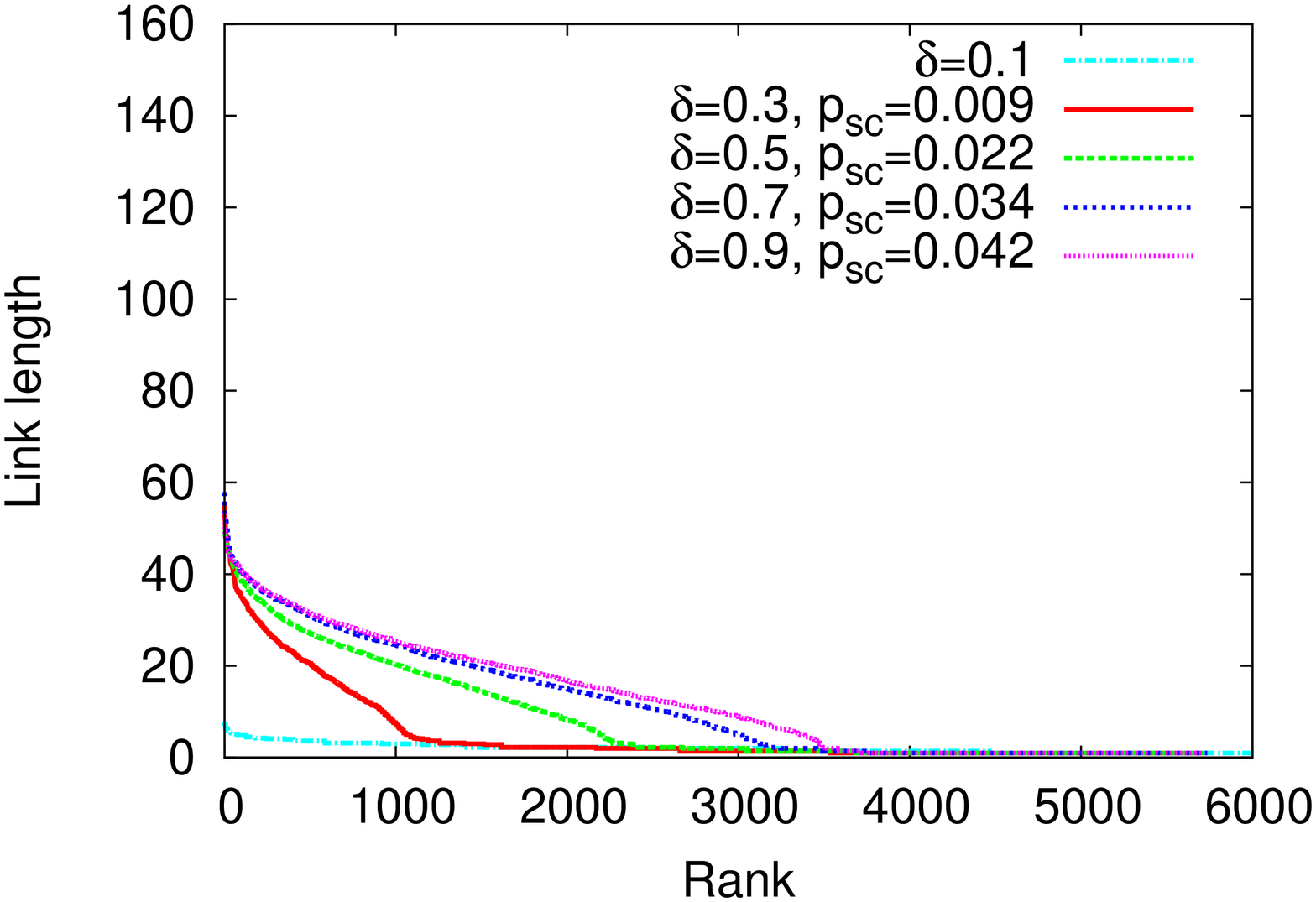}
      \begin{center} (c) Eden model \end{center}
 \end{minipage} 
 \hfill 
 \begin{minipage}{0.47\textwidth} 
   \centering
   \includegraphics[height=67mm,angle=-90]{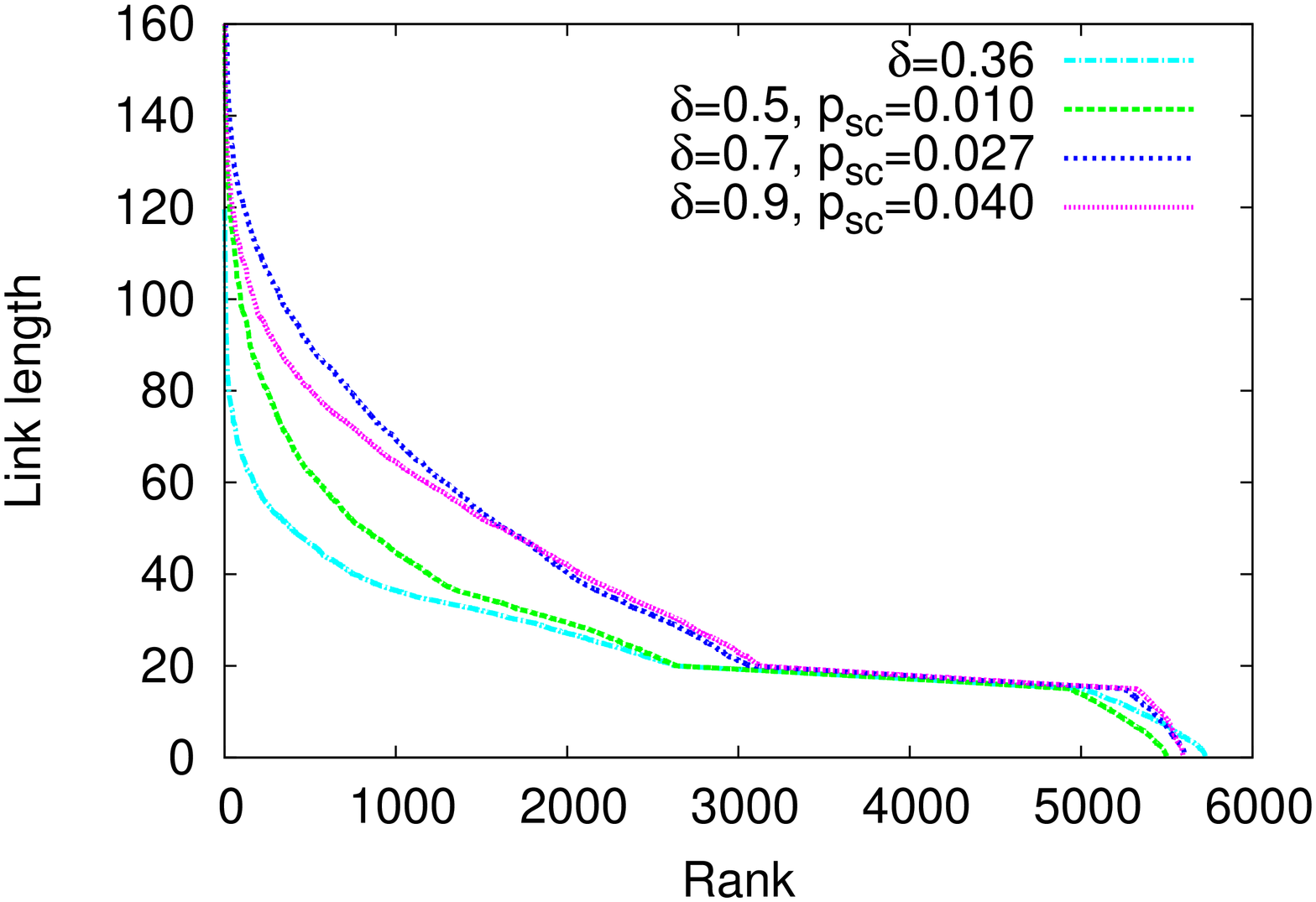}
      \begin{center} (d) Spatial growing model \end{center}
 \end{minipage} 
\caption{(Color online) Rank plot of link lengths measured by the 
Euclidean distance in the networks for $N = 2000$ 
with $\langle k \rangle \approx 5.6$.
These results are averaged over 100 samples.}
\label{fig_pl}
\end{figure}

\begin{figure}[htp]
\centering
 \begin{minipage}{0.47\textwidth} 
   \includegraphics[height=67mm,angle=-90]{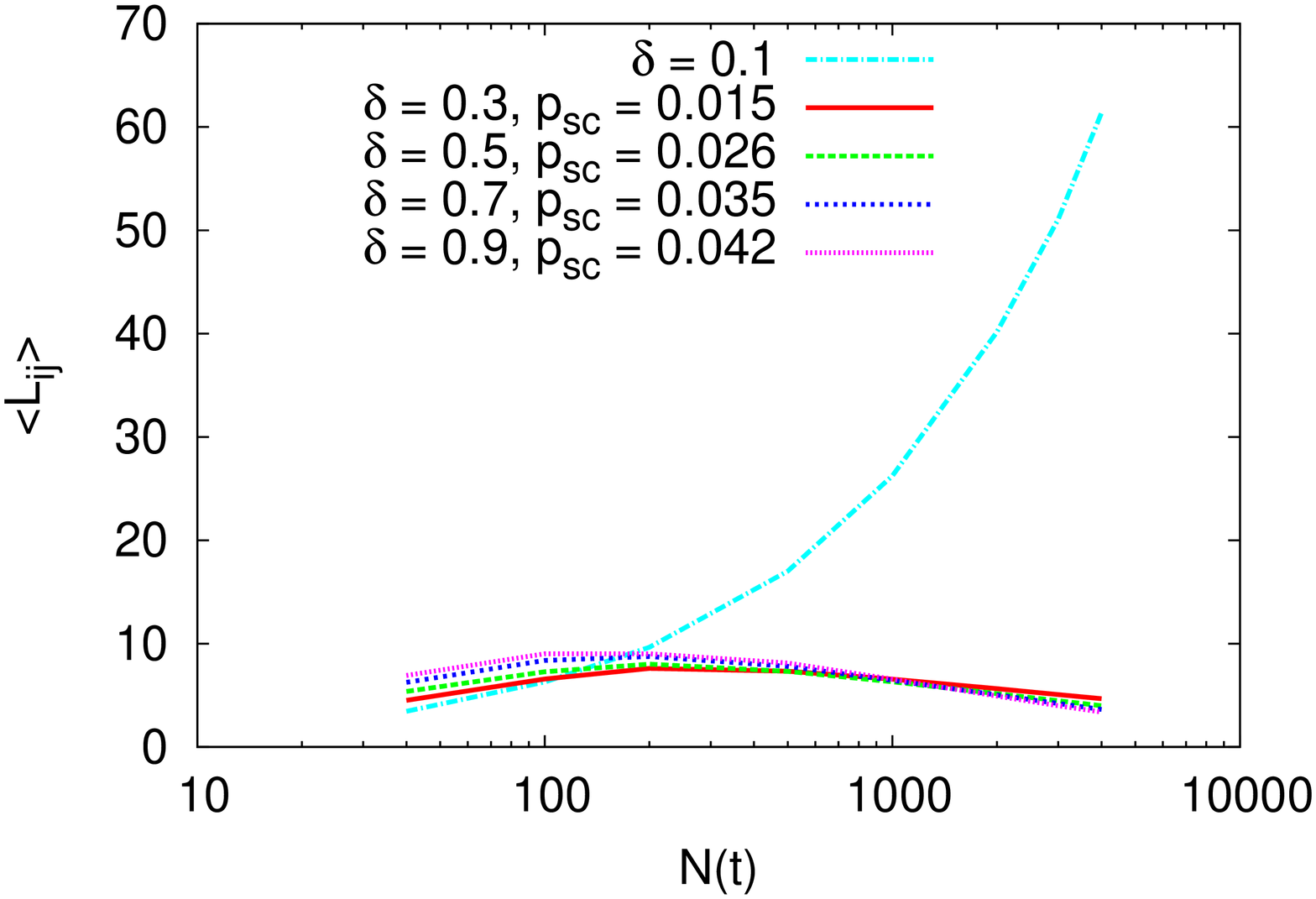}
    \begin{center} (a) Ave. hops on min. hop paths 
    on the nets according to DLA model \end{center}
 \end{minipage} 
 \hfill 
 \begin{minipage}{0.47\textwidth} 
   \includegraphics[height=67mm,angle=-90]{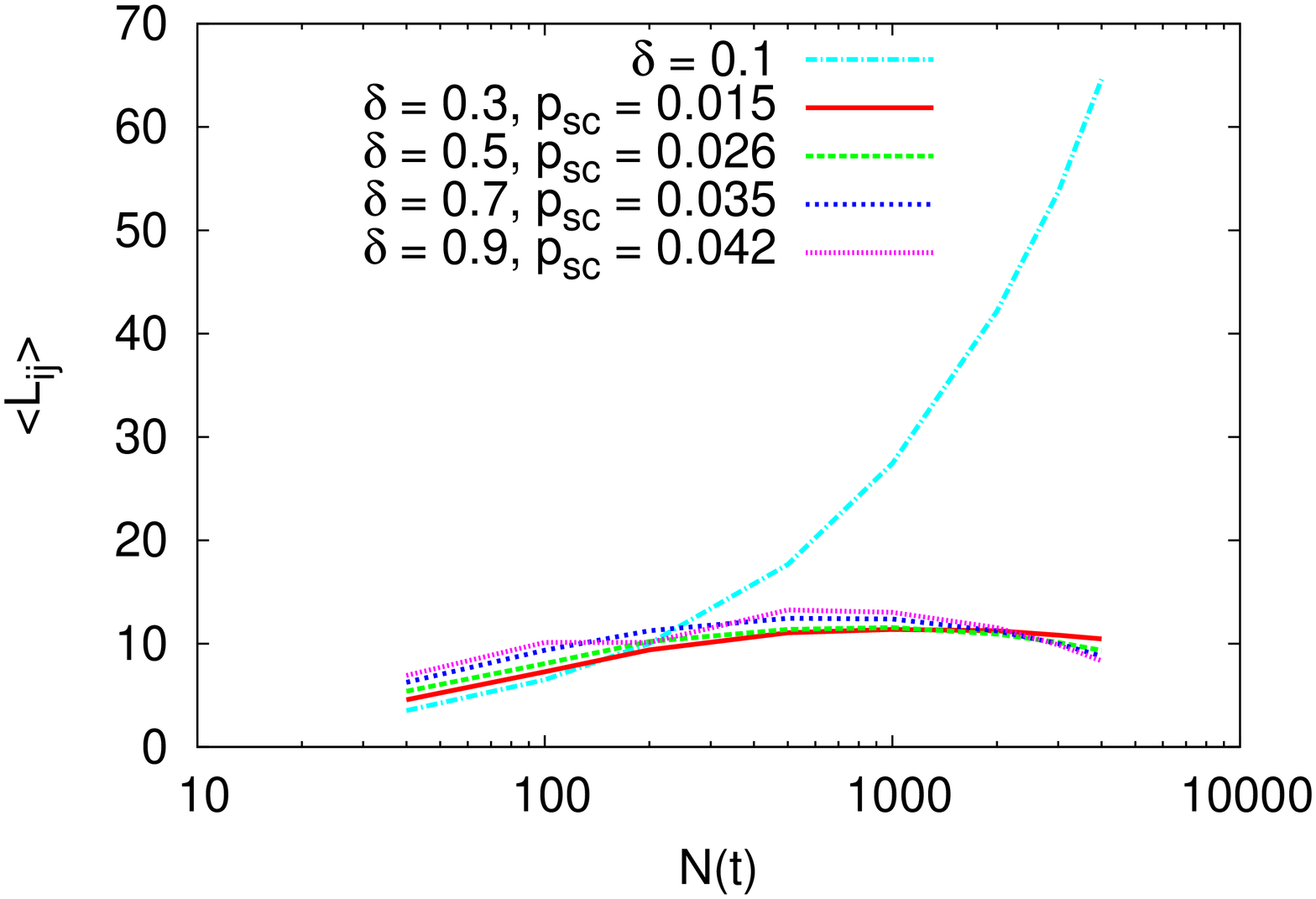}
    \begin{center} (b) Ave. hops on shortest paths 
    on the nets according to DLA model \end{center}
 \end{minipage} 
 \hfill
 \begin{minipage}{0.47\textwidth} 
   \includegraphics[height=67mm,angle=-90]{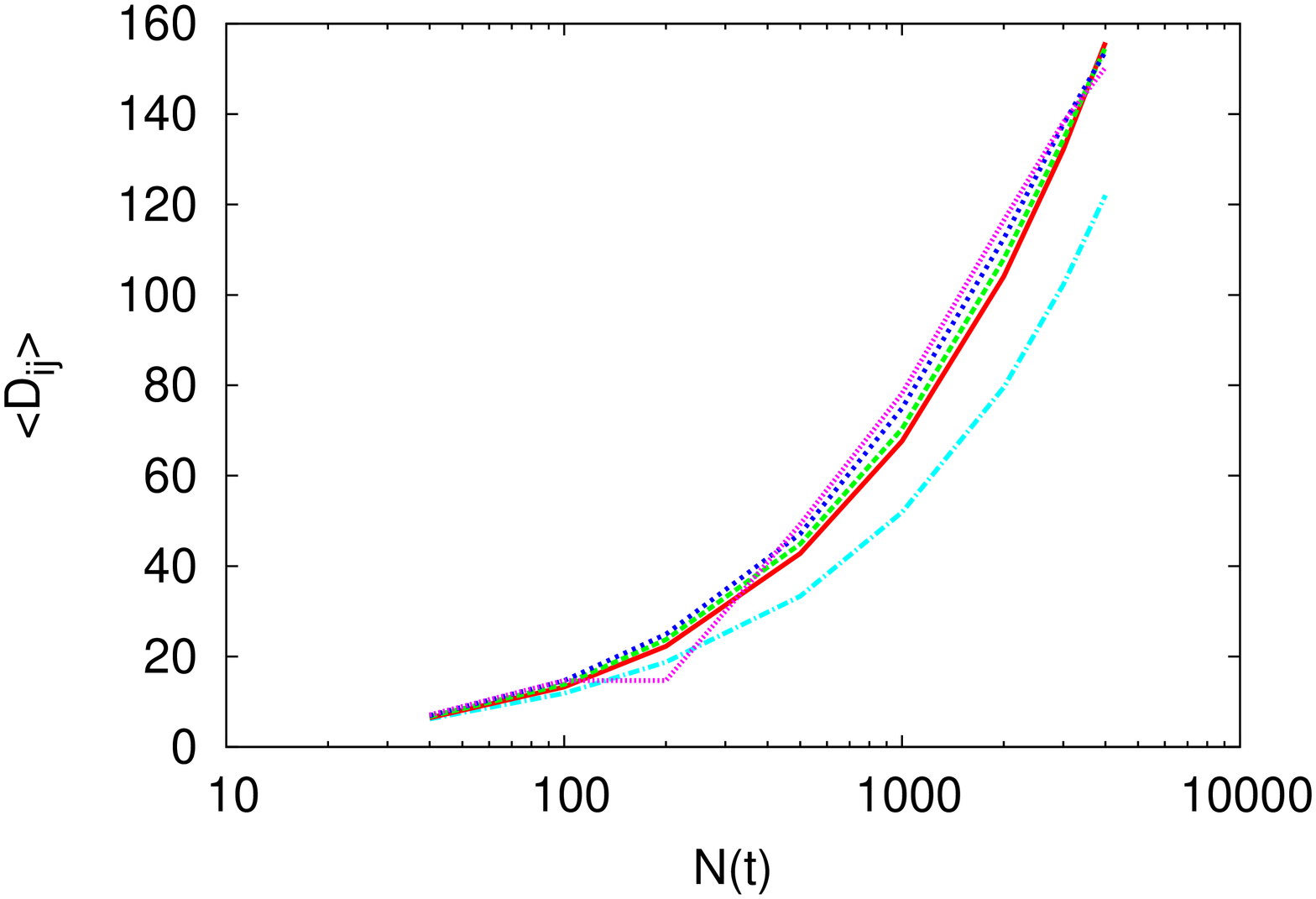}
    \begin{center} (c) Ave. dist. on min. hop paths 
    on the nets according to DLA model \end{center}
 \end{minipage} 
 \hfill 
 \begin{minipage}{0.47\textwidth} 
   \includegraphics[height=67mm,angle=-90]{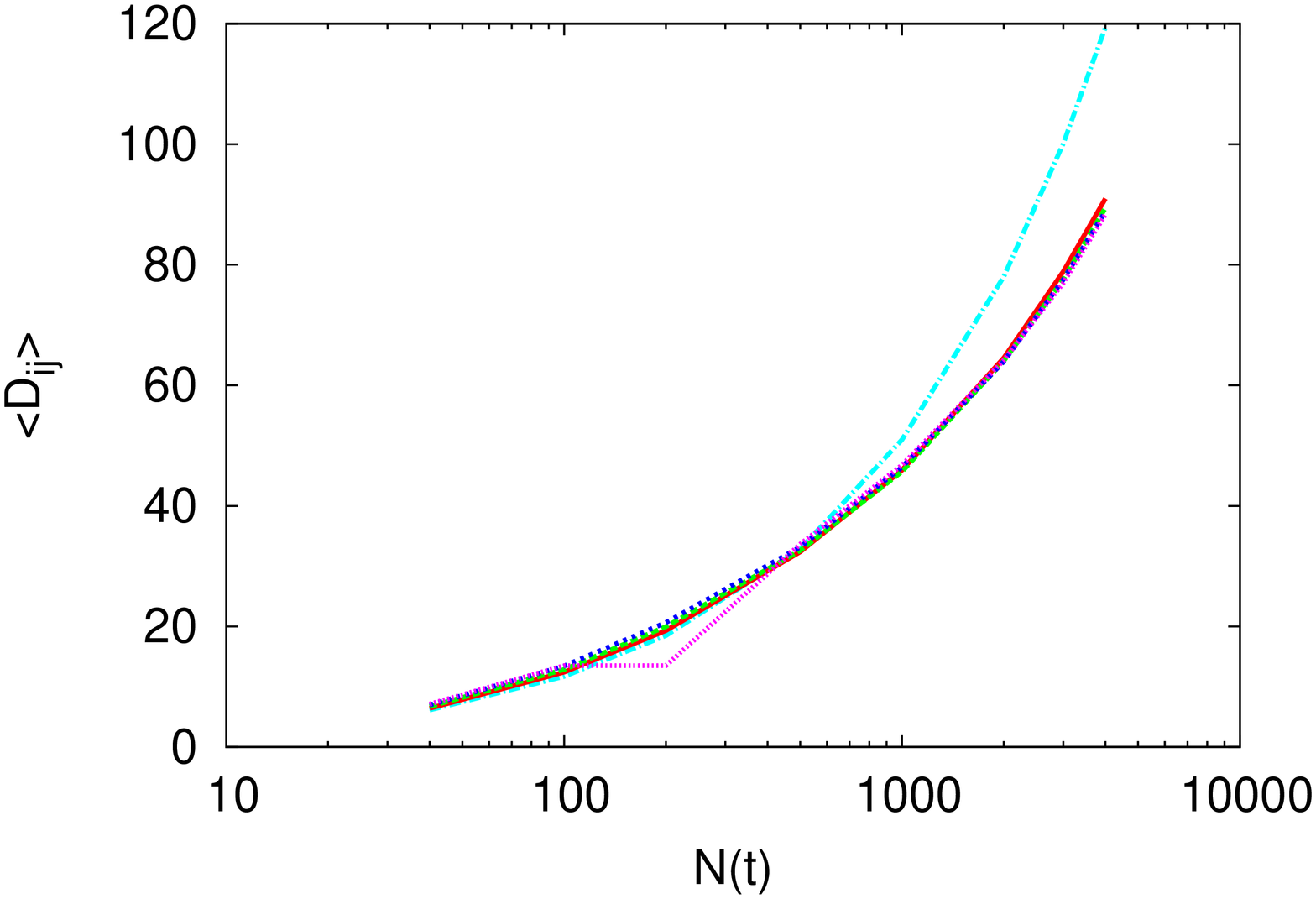}
    \begin{center} (d) Ave. dist. on shortest paths 
    on the nets according to DLA model \end{center}
 \end{minipage} 
\caption{(Color online) Average path length 
$\langle L_{ij} \rangle$ counted by hops (top:(a)(b))
and the average distance $\langle D_{ij} \rangle$ measured by 
Euclidean distance (bottom:(c)(d))
in the growing networks according to DLA model. 
The left:(a)(c) and right:(b)(d)
show the results for the minimum hop paths 
and the shortest distance paths between two nodes, respectively.
These results are averaged over 100 samples.}
\label{fig_dla_path}
\end{figure}

\begin{figure}[htp]
\centering
 \begin{minipage}{0.47\textwidth} 
   \includegraphics[height=67mm,angle=-90]{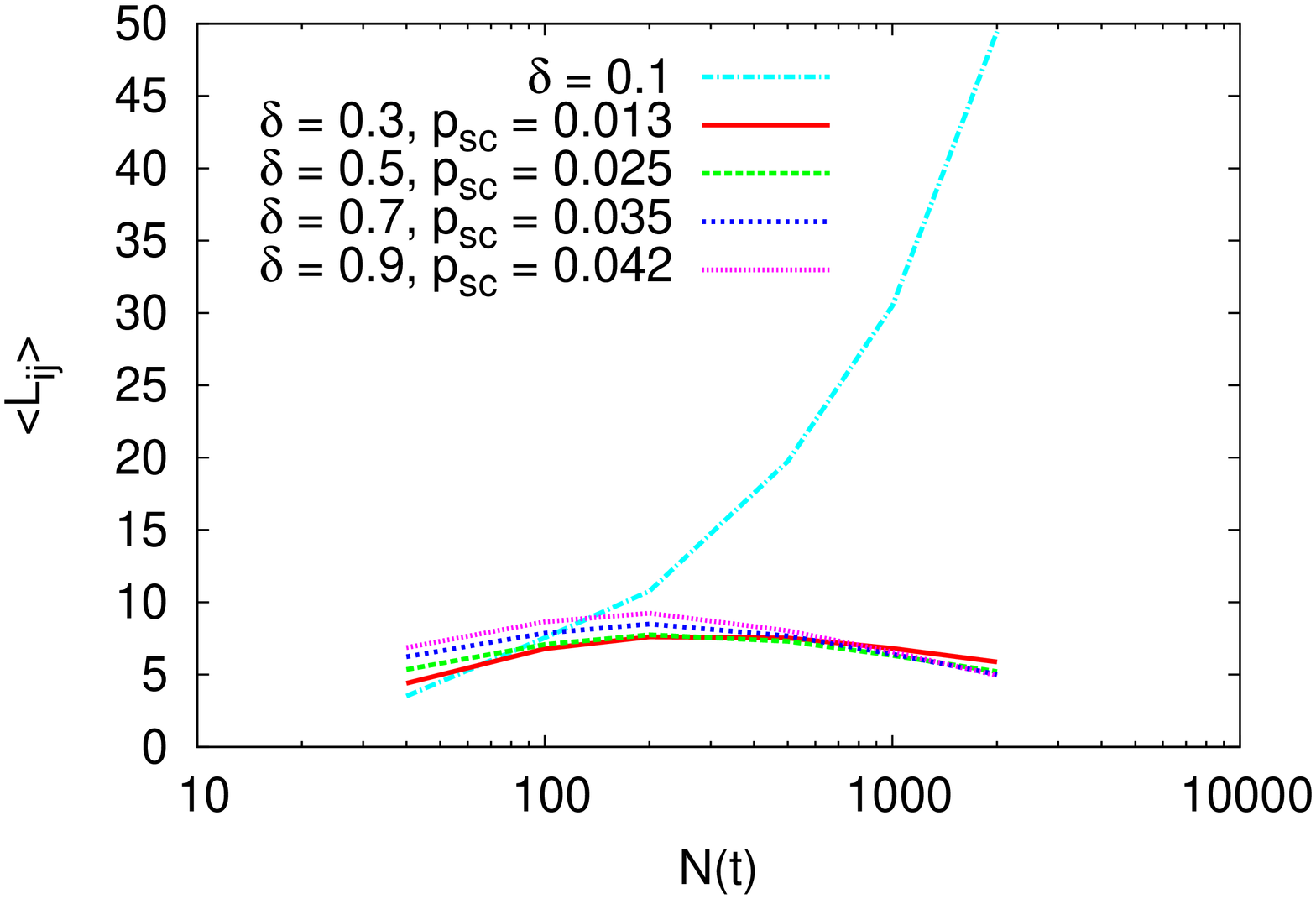}
    \begin{center} (a) Ave. hops on min. hop paths 
    on the nets according to IP model \end{center}
 \end{minipage} 
 \hfill 
 \begin{minipage}{0.47\textwidth} 
   \includegraphics[height=67mm,angle=-90]{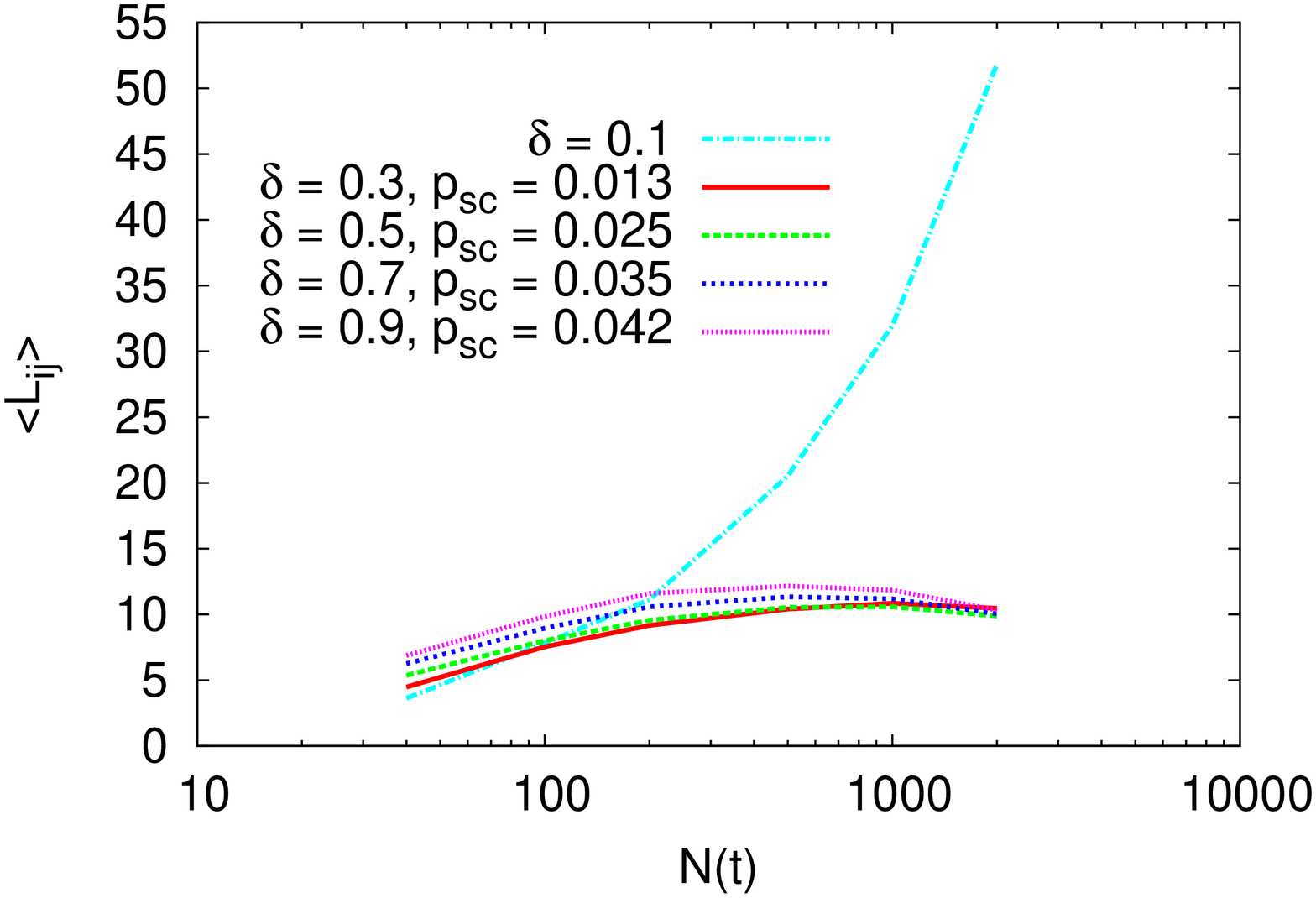}
    \begin{center} (b) Ave. hops on shortest paths 
    on the nets according to IP model \end{center}
 \end{minipage} 
 \hfill
 \begin{minipage}{0.47\textwidth} 
   \includegraphics[height=67mm,angle=-90]{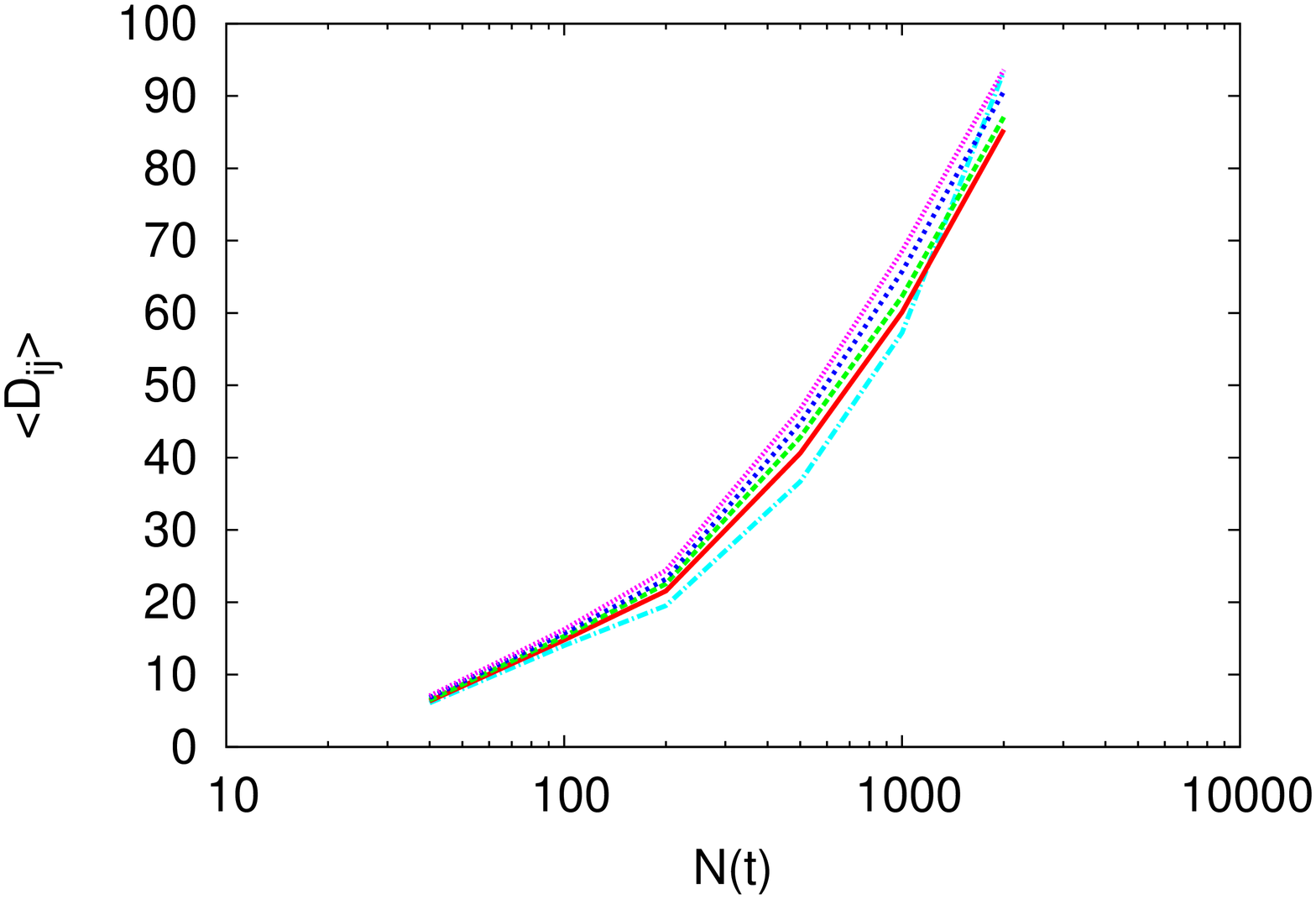}
    \begin{center} (c) Ave. dist. on min. hop paths 
    on the nets according to IP model \end{center}
 \end{minipage} 
 \hfill 
 \begin{minipage}{0.47\textwidth} 
   \includegraphics[height=67mm,angle=-90]{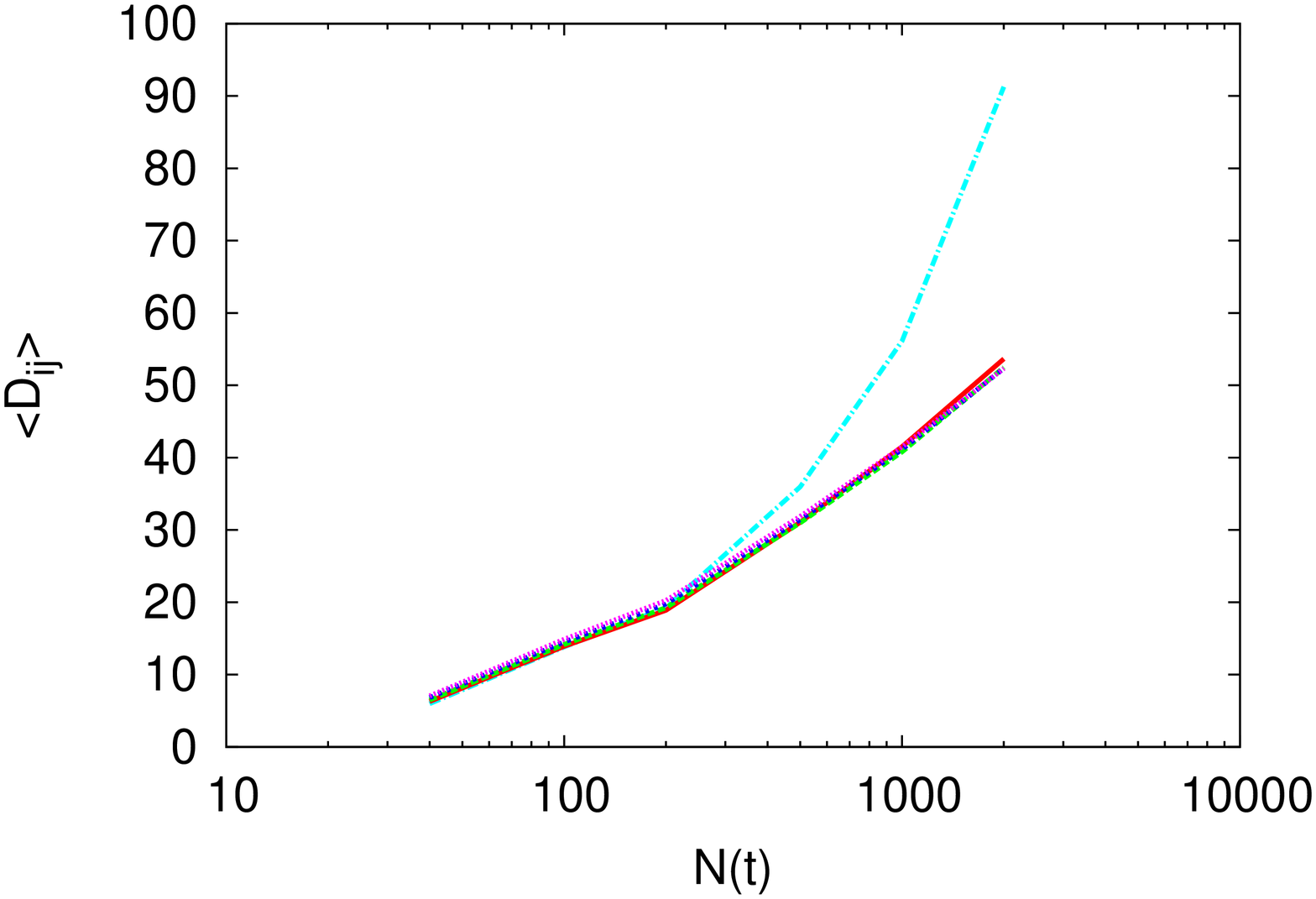}
    \begin{center} (d) Ave. dist. on shortest paths 
    on the nets according to IP model \end{center}
 \end{minipage} 
\caption{(Color online) Average path length 
$\langle L_{ij} \rangle$ counted by hops (top:(a)(b))
and the distance $\langle D_{ij} \rangle$ measured by 
Euclidean distance (bottom:(c)(d))
in the growing networks according to IP model. 
The left:(a)(c) and right:(b)(d) 
show the results for the minimum hop paths 
and the shortest distance paths between two nodes, respectively. 
These results are averaged over 100 samples.}
\label{fig_invasion_path}
\end{figure}

\begin{figure}[htp]
\centering
 \begin{minipage}{0.47\textwidth} 
   \includegraphics[height=67mm,angle=-90]{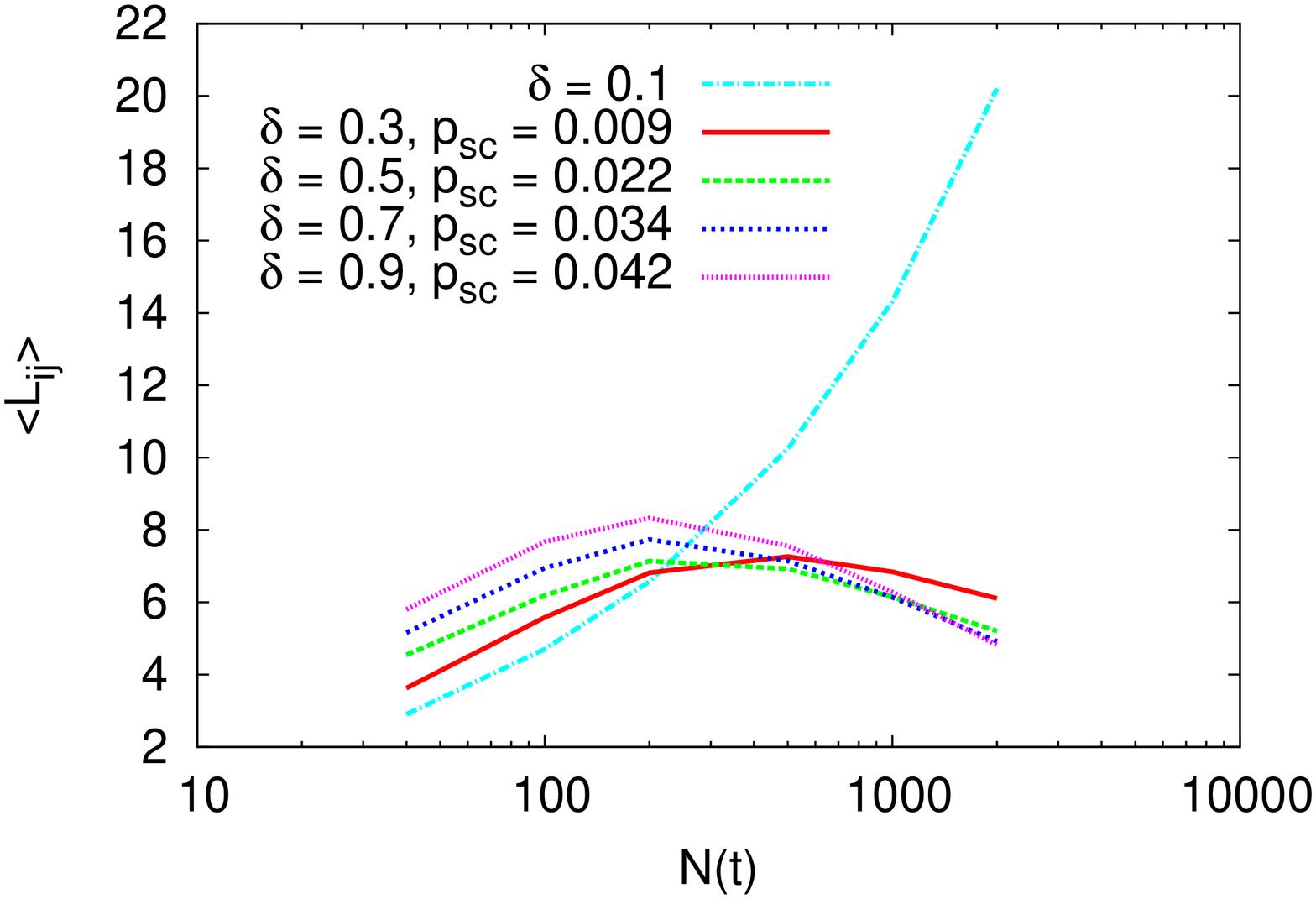}
   \begin{center} (a) Ave. hops on min. hop paths 
   on the nets according to Eden model \end{center}
 \end{minipage} 
 \hfill 
 \begin{minipage}{0.47\textwidth} 
   \includegraphics[height=67mm,angle=-90]{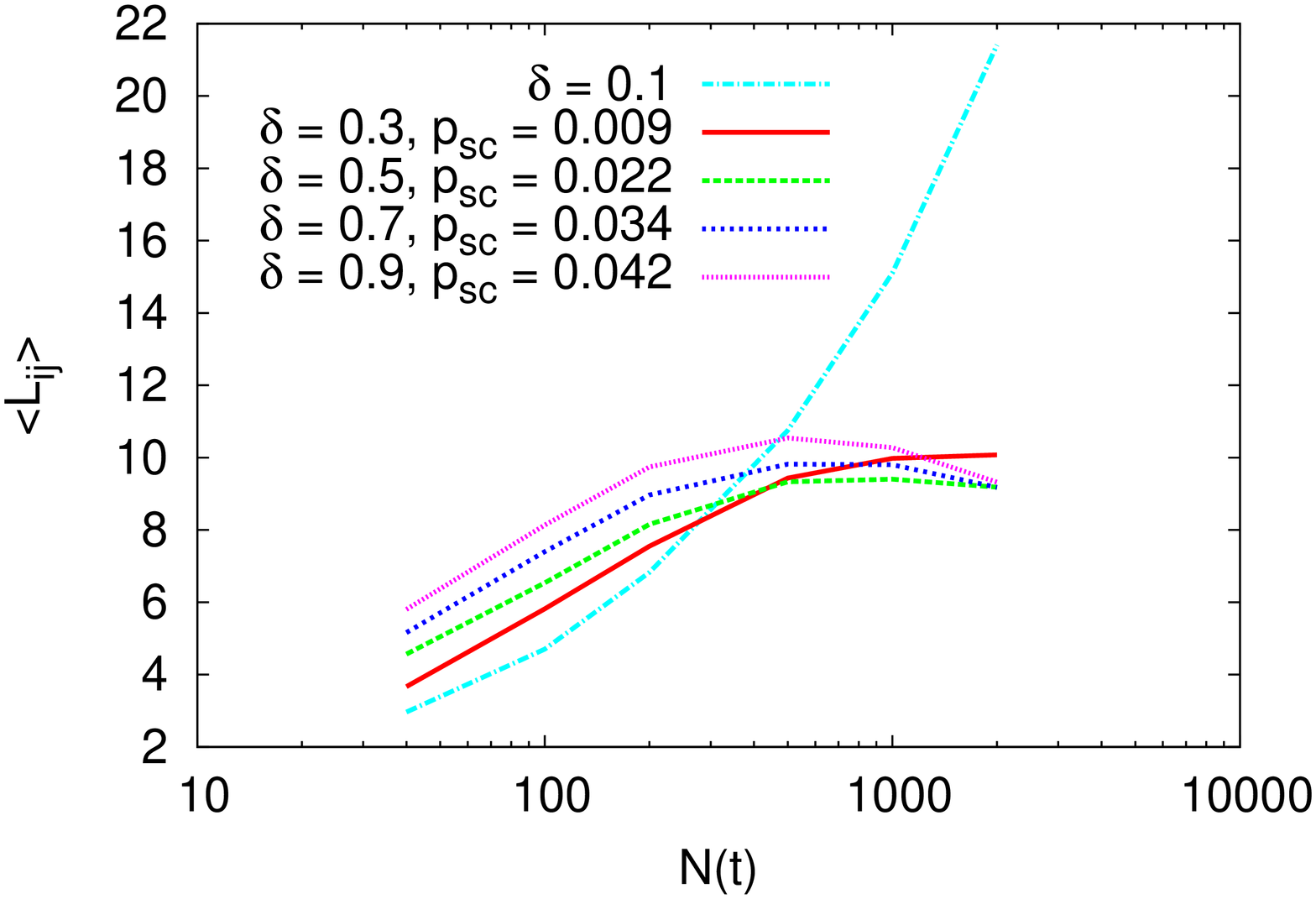}
   \begin{center} (b) Ave. hops on shortest paths 
   on the nets according to Eden model \end{center}
 \end{minipage} 
 \hfill
 \begin{minipage}{0.47\textwidth} 
   \includegraphics[height=67mm,angle=-90]{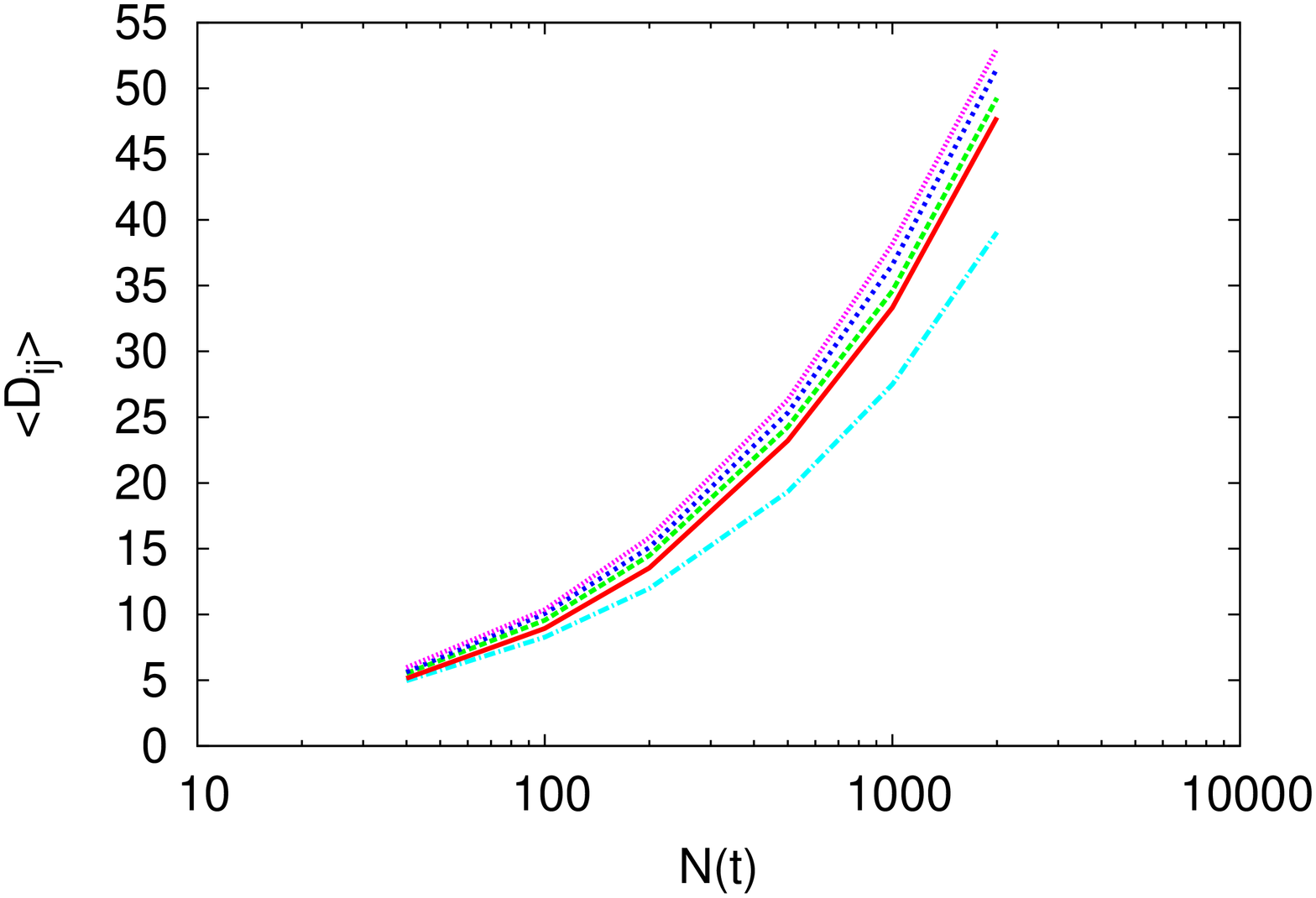}
   \begin{center} (c) Ave. dist. on min. hop paths 
   on the nets according to Eden model \end{center}
 \end{minipage} 
 \hfill 
 \begin{minipage}{0.47\textwidth} 
   \includegraphics[height=67mm,angle=-90]{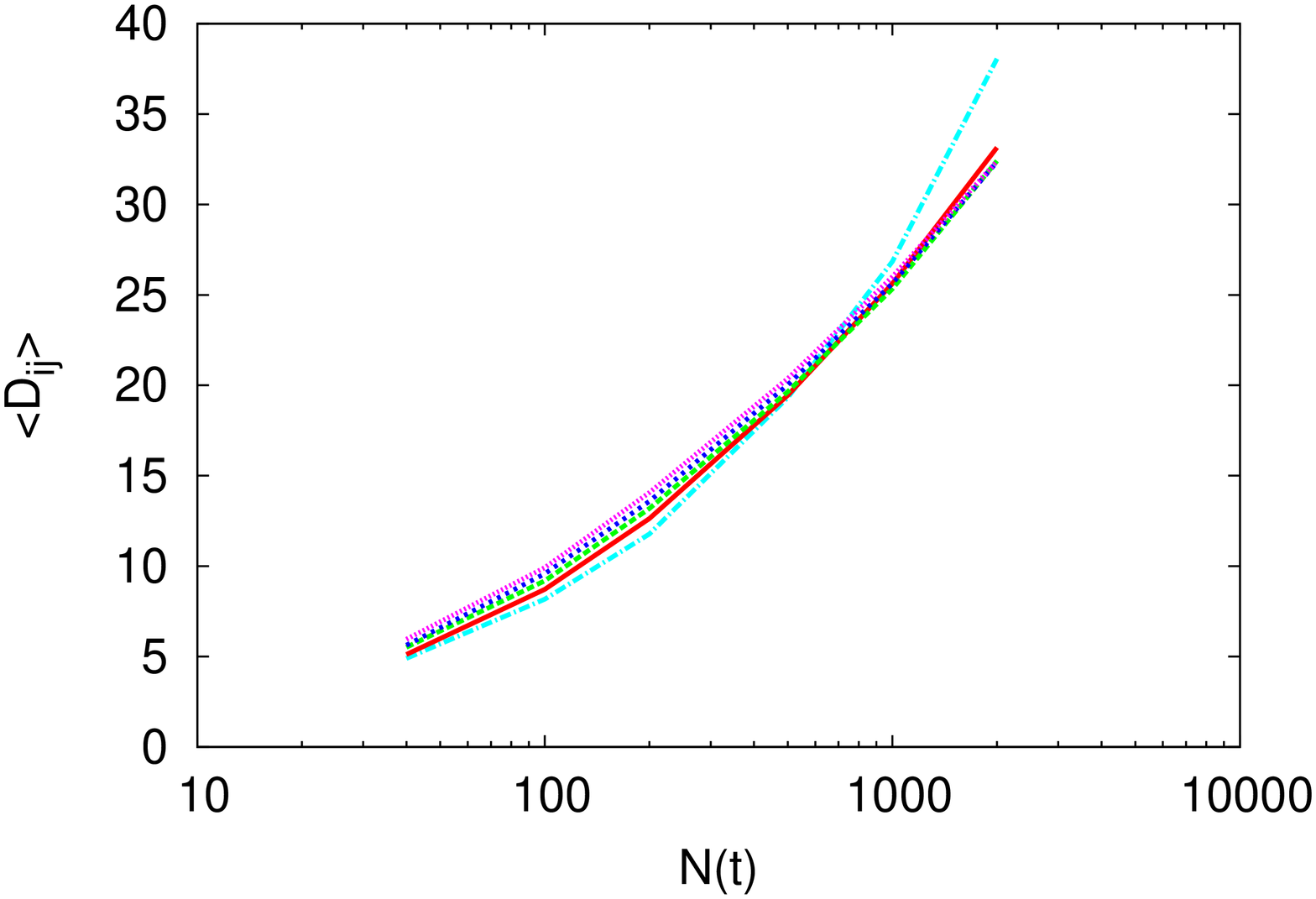}
   \begin{center} (d) Ave. dist. on shortest paths 
   on the nets according to Eden model \end{center}
 \end{minipage} 
\caption{(Color online) Average path length 
$\langle L_{ij} \rangle$ counted by hops (top:(a)(b))
and the distance $\langle D_{ij} \rangle$ measured by 
Euclidean distance (bottom:(c)(d))
in the growing network according to Eden model.
The left:(a)(c) and right:(b)(d)
show the results for the minimum hop paths 
and the shortest distance paths between two nodes, respectively. 
These results are averaged over 100 samples.}
\label{fig_eden_path}
\end{figure}

\subsection{Efficiency of path on the growing networks}
Next, 
we investigate the communication or transportation 
efficiency in the growing networks. 
Obviously, shorter links and paths are better with both 
less construction and maintenance costs. 
Figure \ref{fig_pl} shows the rank plot of link lengths. 
The order of shorter links is 
(c) Eden model $<$ (b) IP model $<$ (a) DLA model
 $<$ (d) Spatial growing model \cite{Hayashi14} 
without the constraint in the surface growth, 
however the difference between (b) and (a) is very small. 
This order of link lengths is reasonable because 
Eden model tends to be compact while IP model tends to 
percolate making some holes as porous structure 
and DLA model tends to spread widely along dendritic edges. 
The tops of Figs. \ref{fig_dla_path}, 
\ref{fig_invasion_path}, and \ref{fig_eden_path}
show the average path length $\langle L_{ij} \rangle$ 
counted by hops between two nodes on the paths of 
(left) minimum hop and (right) shortest distance. 
We obtain 
in the tree-like networks (cyan line)
monotone increasing curves, 
while in the onion-like networks (other color lines)
convex curves with up-down, whose parts in decreasing 
the number of hops on both paths are related to the 
increasing of $\langle k \rangle$ with more links 
as shown at the middle column of Fig. \ref{fig_dla_r-k-R}. 
It is remarkable that 
almost constant path length even for the growing size 
is better than $O(\log N)$ of the SW property.

The bottoms of Figs. \ref{fig_dla_path}, 
\ref{fig_invasion_path}, and \ref{fig_eden_path}
show the average path distance $\langle D_{ij} \rangle$ 
measured by the sum of link lengths 
as Euclidean distances on the paths of 
(left) minimum hop and (right) shortest distance. 
We remark that 
evident difference does not appear in 
$\langle D_{ij} \rangle$ for varying $\delta$ 
in each of Figs. \ref{fig_dla_path}, 
\ref{fig_invasion_path}, and \ref{fig_eden_path} 
except the tree-like networks for $\delta = 0.1$ (cyan line).

The orders of efficient shorter paths are 
Fig. \ref{fig_eden_path}: Eden model 
$<$ Fig. \ref{fig_invasion_path}: IP model 
$<$ Fig. \ref{fig_dla_path}: DLA model
in the measures of 
both $\langle L_{ij} \rangle$ and $\langle D_{ij} \rangle$, 
however the difference among them is small
especially in $\langle L_{ij} \rangle$ for 
$\delta = 0.3 \sim 0.9$. 
In addition, 
all of them in Figs. \ref{fig_dla_path}, 
\ref{fig_invasion_path}, and \ref{fig_eden_path} 
slightly deviate from straight lines of $O(\log N)$ in semi-log plot 
as the SW property, 
although the SW property is obtained 
in the spatial growing networks \cite{Hayashi14} 
without the constraint in the surface growth. 
Thus, 
our proposed networks are efficient because the 
increasing rates of the average path lengths and distances 
are suppressed in the growing size.
Note that the length $\langle L_{ij} \rangle$ 
on the minimum hop path (left) is shorter 
than that on the shortest distance path (right), 
while the distance $\langle D_{ij} \rangle$ 
on the minimum hop path (left) is longer 
than that on the shortest distance path (right), 
in each figure.

\subsection{Resilient connectivity against sequential attacks}
Through this paper, 
we discuss dynamics of network configuration itself at the most 
basic infrastructure for communication or transportation systems, 
in which temporal and/or fixed 
(corresponding to wireless and/or wired) connections are possible 
depending on the time-scale for changing the connection structure 
in a network. 
The quick change results in ad hoc networks, 
while the slow change is treated as an incremental 
modification of network. 
Both cases and the mixed one are not excluded, 
however we have assumed that each node or link is persisted 
once it is added unless removed by failures or attacks 
to simplify the discussion.
While other dynamics of information flows, rumor spreading, 
opinion formation, synchronization, or logistics on a network, 
is significant for applications in 
wireless, sensor, mobile communication systems or 
autonomous transportation systems, in which 
operation protocols for birth and death of communication or transportation 
request, routing, avoidance of congestion, 
task allocation, queuing, awareness of location, 
monitoring of system's states or conditions, and so on \cite{Barbeau07}, 
are necessary.
We have pointed out the issues of rethinking 
packet generation according to population, decentralized 
routing strategy, and link hierarchy among long and short ranges 
with high and low transfer speeds \cite{Hayashi12} 
in the state-of-the-art network technologies.
In addition, 
although several models have been proposed for the 
coevolution of network formation and opinion spreading 
\cite{Holme06,Zanette06} 
and for the coupled dynamics with network evolution and packet flows 
 \cite{Tero10,Hayashi12a,Noh08,Noh09,Xuan10}, 
they are beyond our current scope because of more complex 
dynamics between network generation and information flow 
in various protocols, device technologies, users, and 
situations of utilization.

\begin{figure}[htp]
 \begin{minipage}{0.31\textwidth} 
   \centering
   \includegraphics[height=50mm,angle=-90]{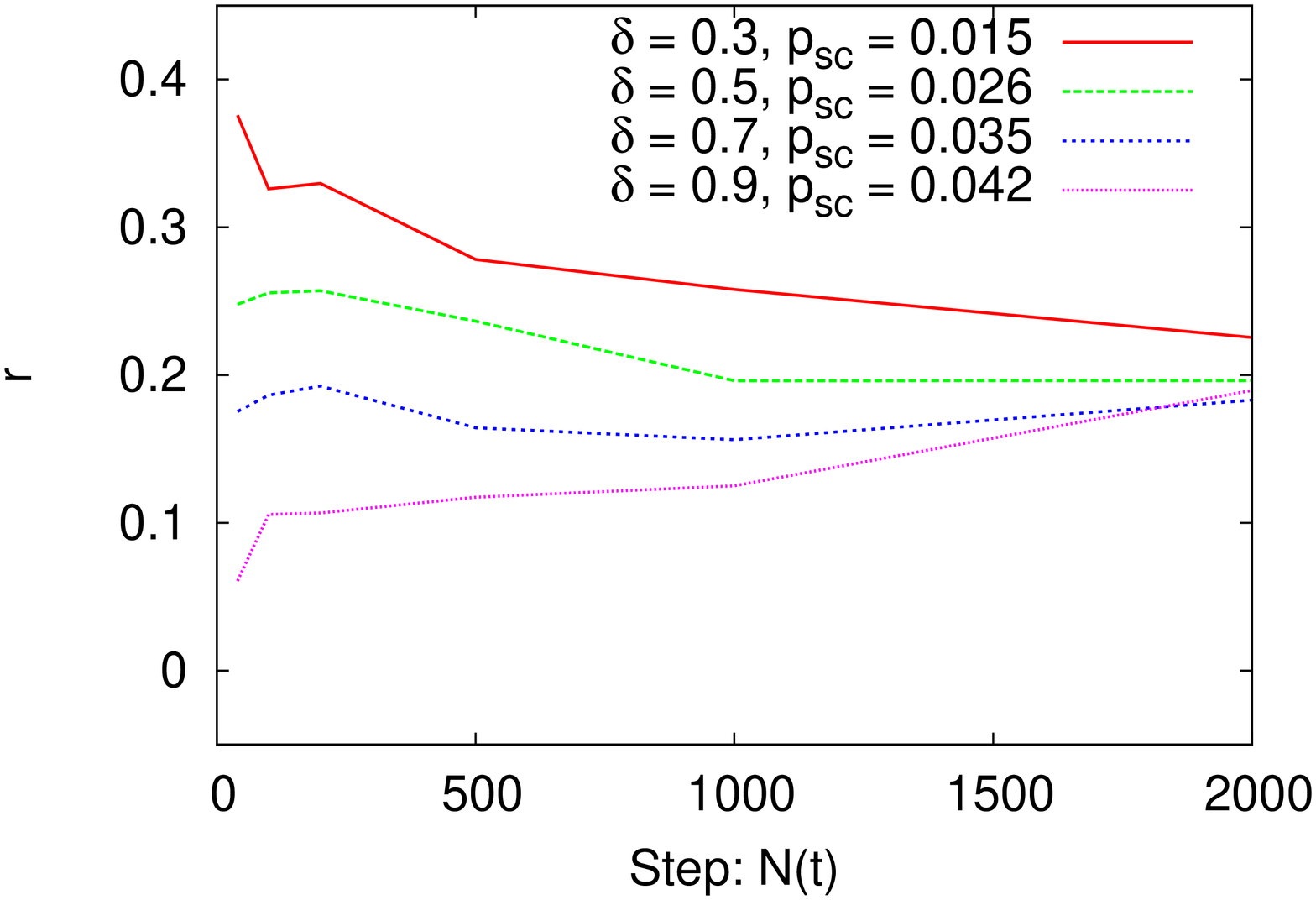}
 \end{minipage} 
 \hfill 
 \begin{minipage}{0.31\textwidth} 
   \centering
   \includegraphics[height=50mm,angle=-90]{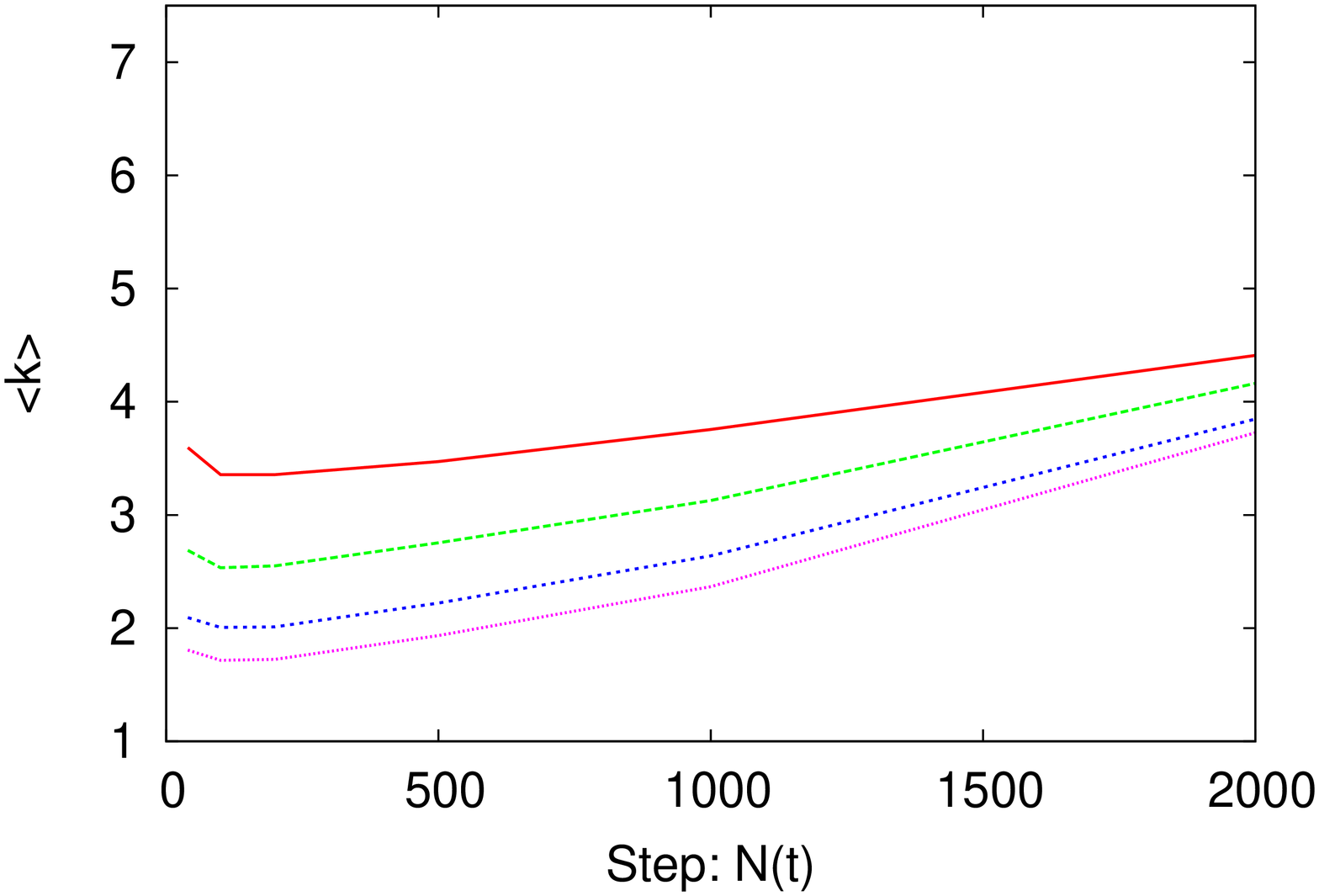}
 \end{minipage} 
 \hfill 
 \begin{minipage}{0.31\textwidth} 
   \centering
   \includegraphics[height=50mm,angle=-90]{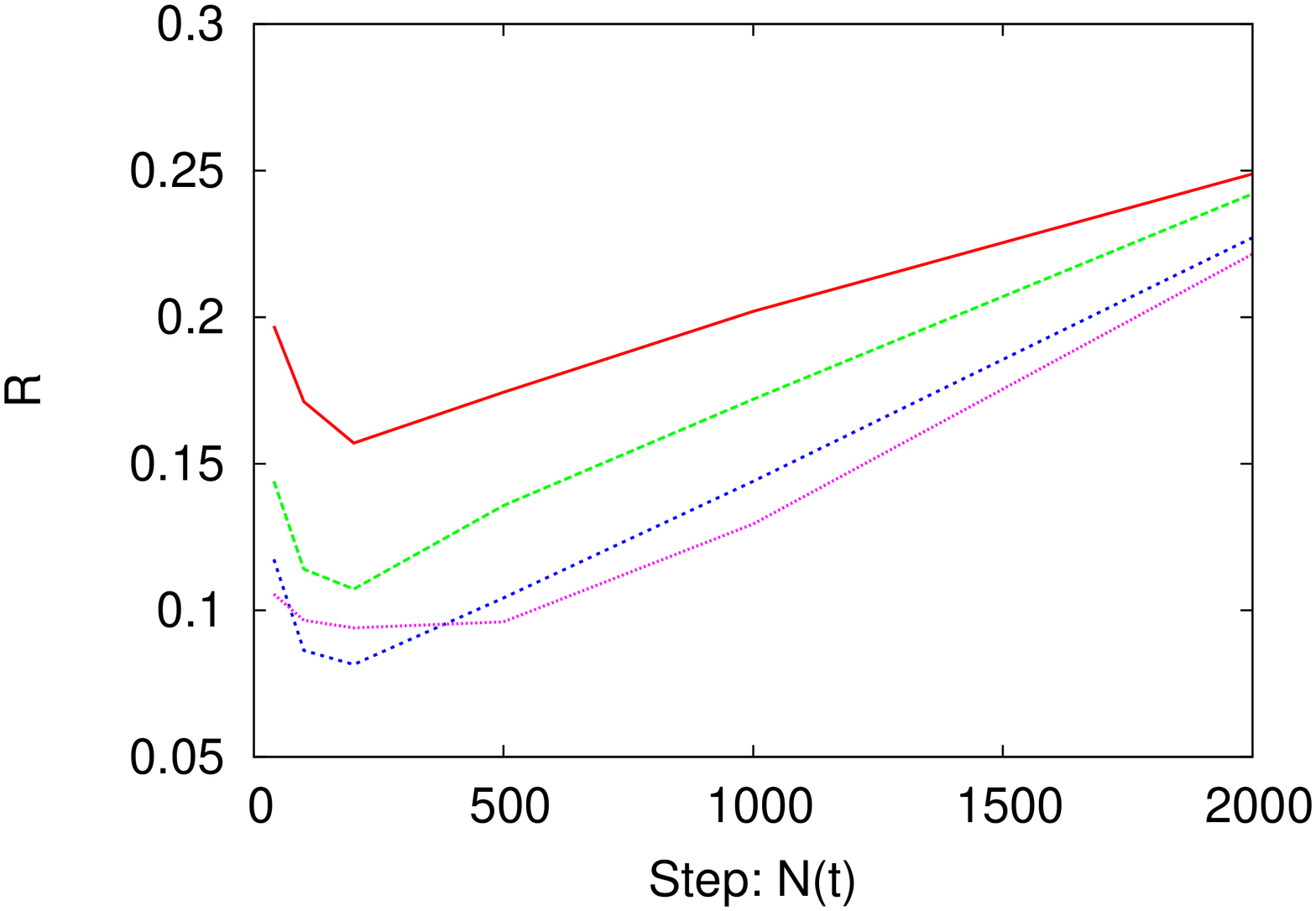}
 \end{minipage} 
 \hfill 
 \begin{minipage}{0.31\textwidth} 
   \centering
   \includegraphics[height=50mm,angle=-90]{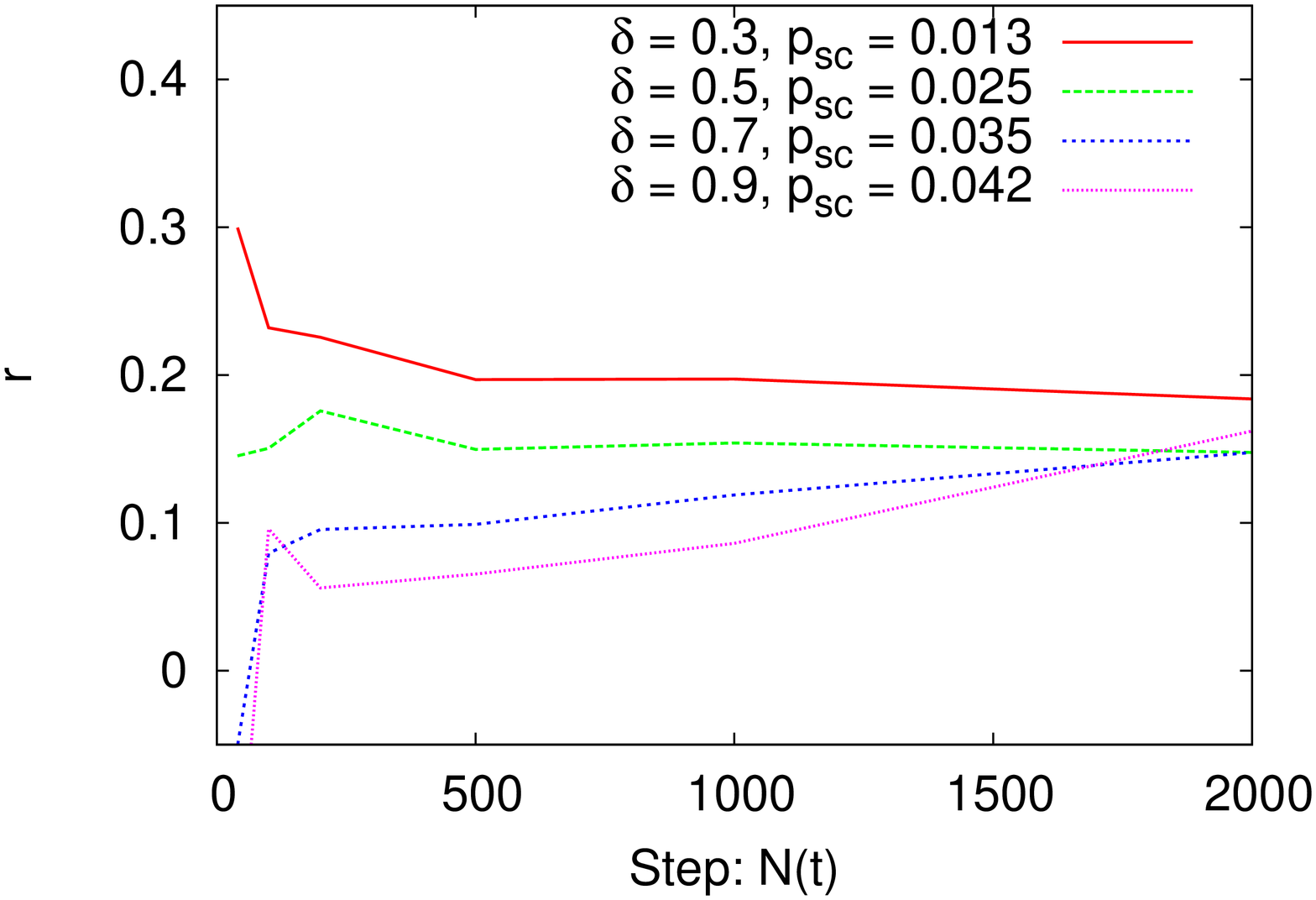}
 \end{minipage} 
 \hfill 
 \begin{minipage}{0.31\textwidth} 
   \centering
   \includegraphics[height=50mm,angle=-90]{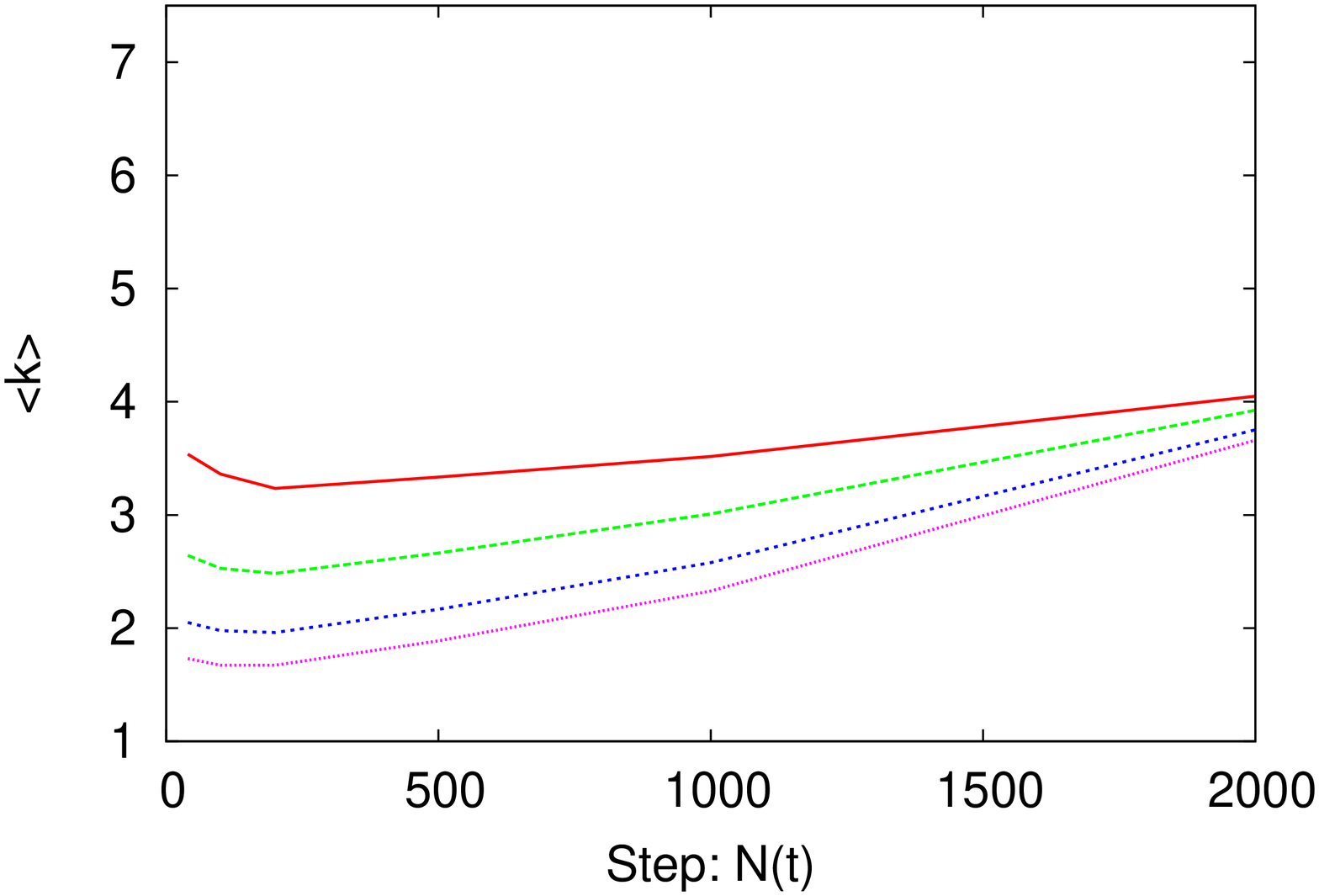}
 \end{minipage} 
 \hfill 
 \begin{minipage}{0.31\textwidth} 
   \centering
   \includegraphics[height=50mm,angle=-90]{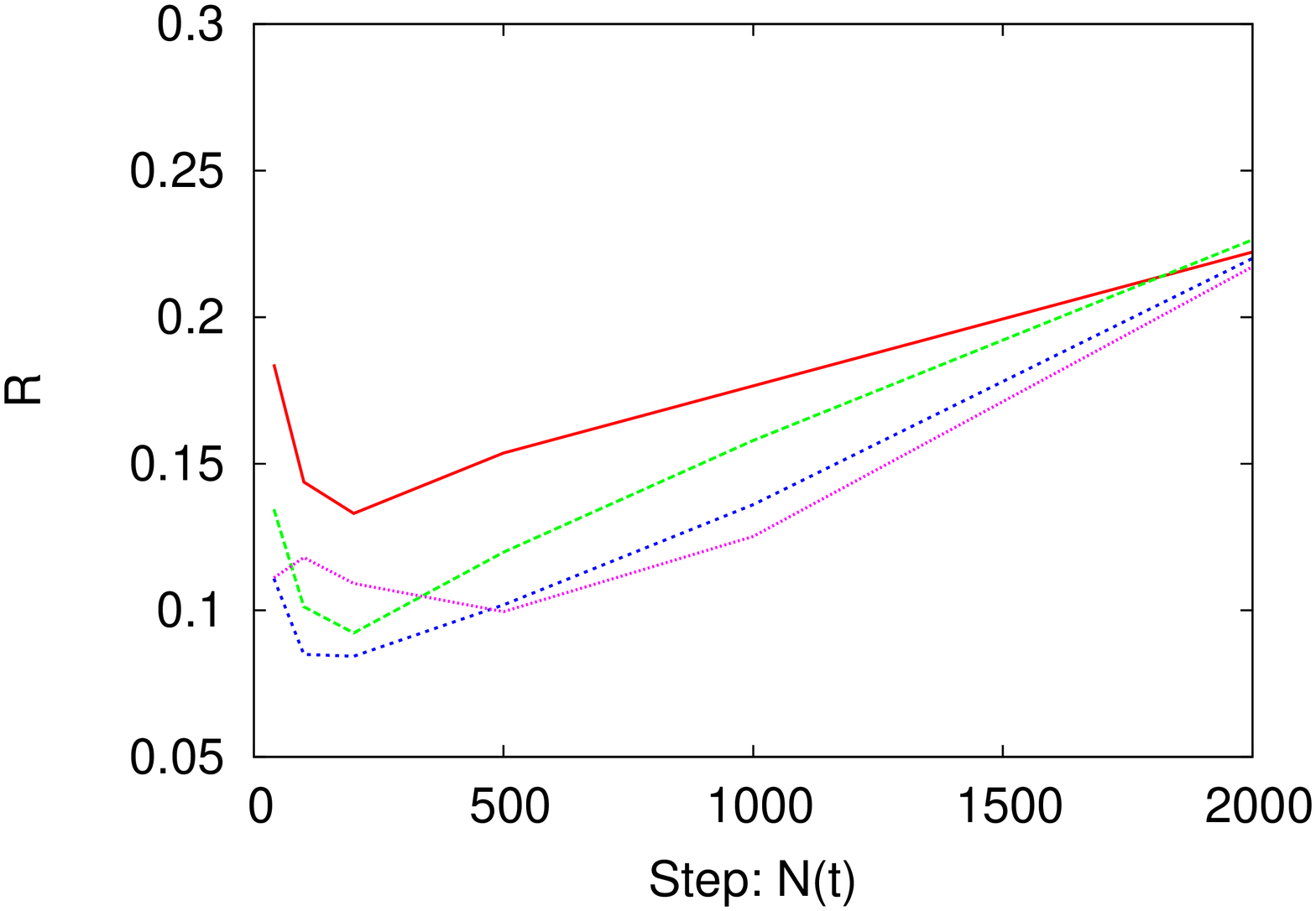}
 \end{minipage} 
 \hfill 
\begin{minipage}[htb]{0.31\textwidth} 
   \centering
   \includegraphics[height=50mm,angle=-90]{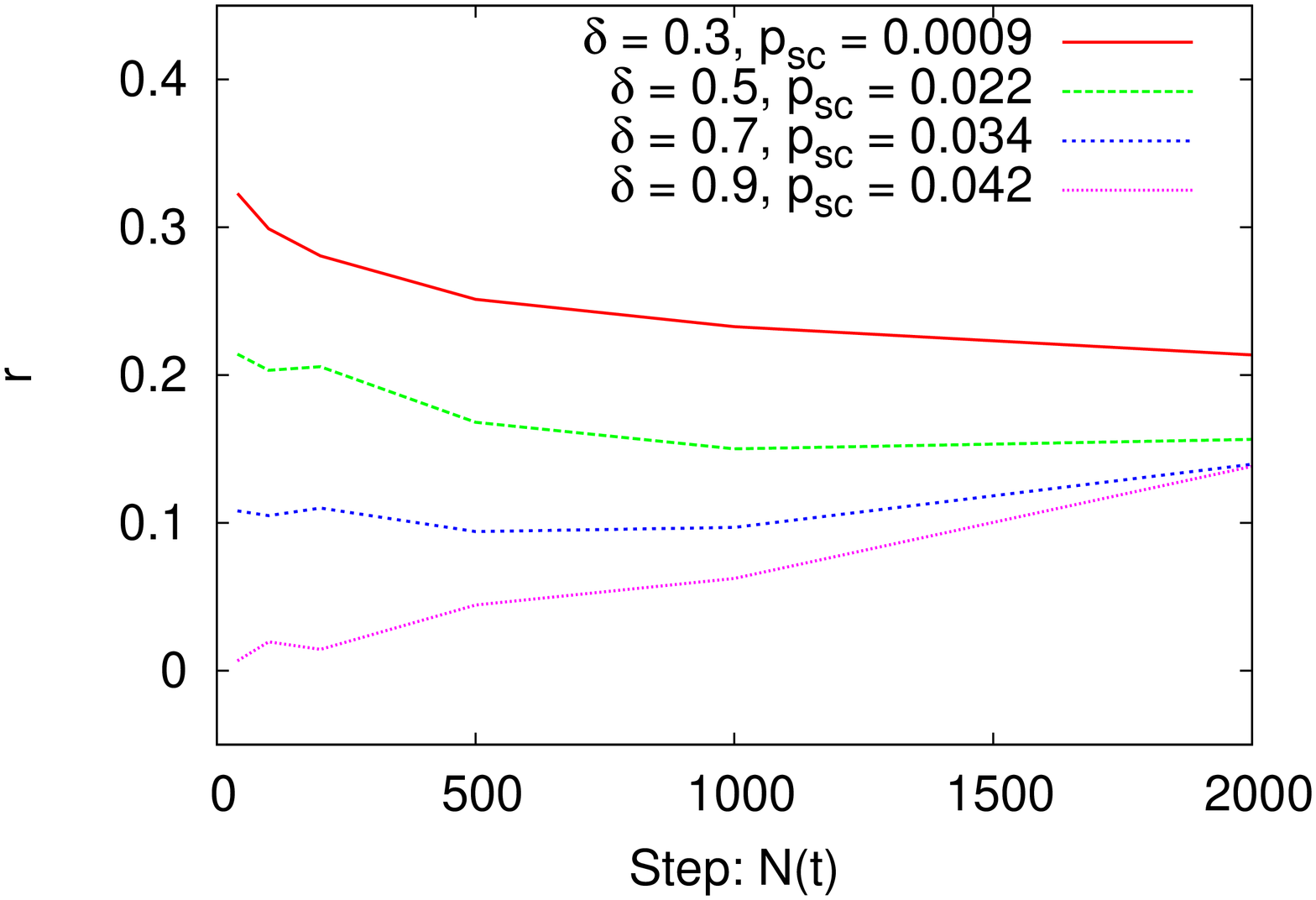}
 \end{minipage} 
 \hfill 
 \begin{minipage}[htb]{0.31\textwidth} 
   \centering
   \includegraphics[height=50mm,angle=-90]{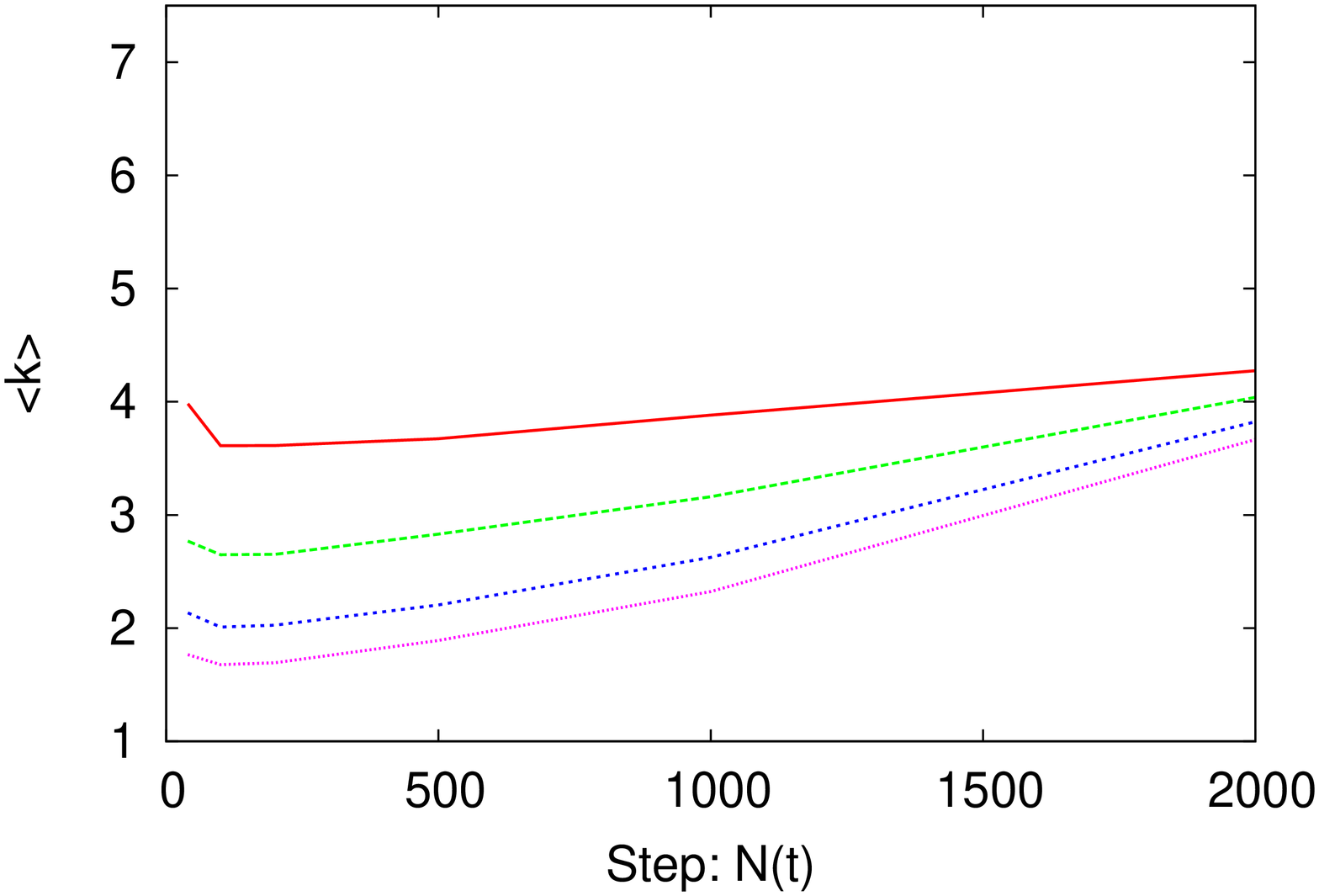}
 \end{minipage} 
 \hfill 
 \begin{minipage}[htb]{0.31\textwidth} 
   \centering
   \includegraphics[height=50mm,angle=-90]{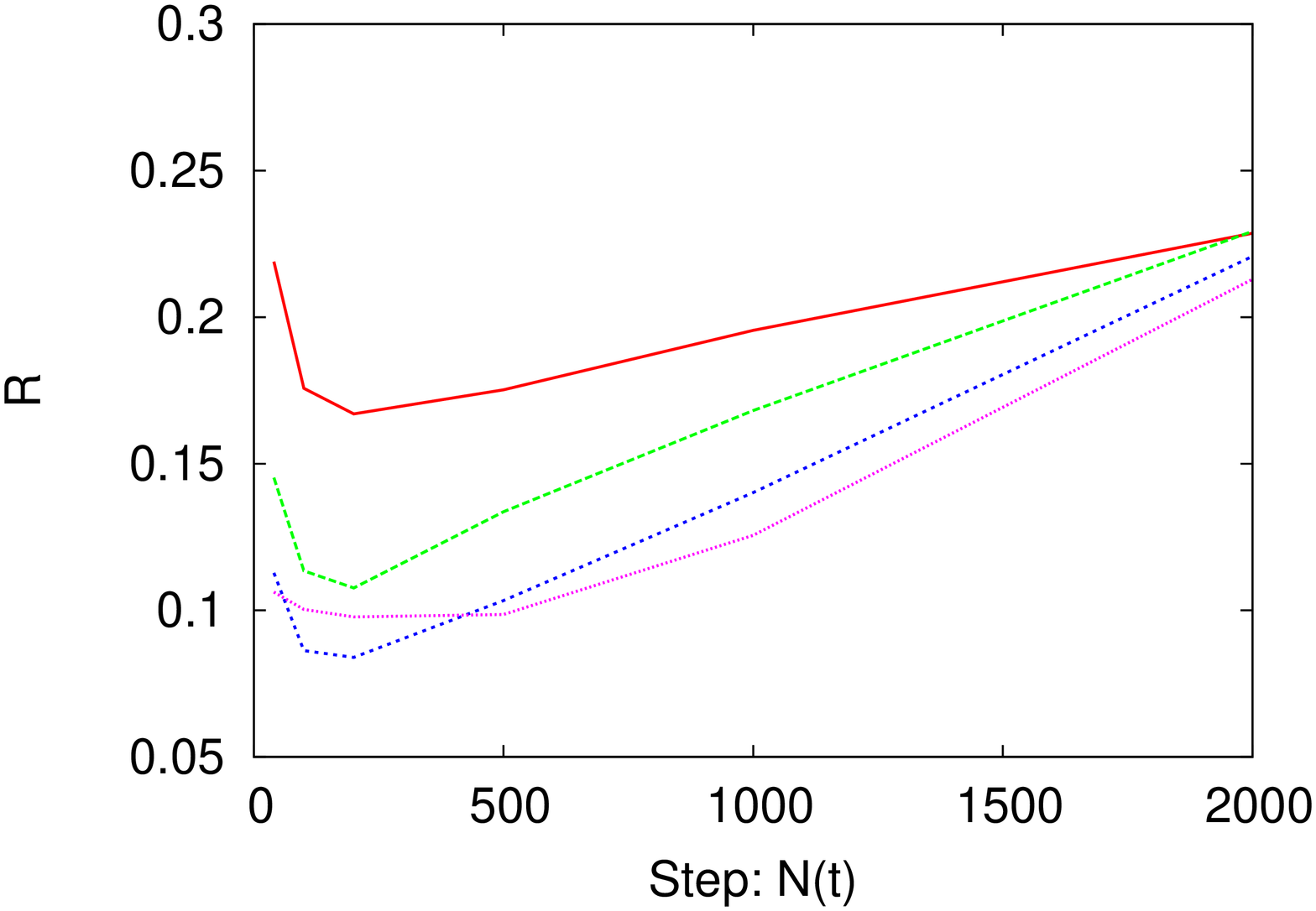}
 \end{minipage} 
\caption{(Color online) Time-courses of assortativity $r$, 
average degree $\langle k \rangle$, robustness index $R$ 
against the malicious attacks 
in the networks damaged by sequential attacks at every 10 time steps.
These results are averaged over 100 samples
according to (top) DLA, (middle) IP, 
(bottom) Eden models.} 
\label{fig_attack_r-k-R}
\end{figure}

On the other hand, 
we consider an interaction of network generation 
with a change of environment, especially in sequential attacks.
We show that 
our proposed networks have resilient connectivity 
even in the previously described growing scheme of constant 
$\delta$, $p_{sc}$, and $IT$, 
without any remedial measures for insistent attacks.
Figure \ref{fig_attack_r-k-R} shows the time-courses of 
$r$, $\langle k \rangle$, and $R$ against the further attacks 
in the networks damaged by sequential attacks.
We set the same values of $\delta$, $p_{sc}$, and $IT$ in the 
previous subsection, 
and the largest degree node with its links 
is removed through the recalculation of degrees 
at every $10$ time steps.
At the left of Fig. \ref{fig_attack_r-k-R} 
top, middle, and bottom for DLA, IP, and Eden models, 
respectively, 
the damaged networks by the sequential attacks 
have positive degree-degree correlations around 
$r \approx 0.2$ somehow or other.
At the middle and right of Fig. \ref{fig_attack_r-k-R}, 
the robustness is recovered with slowly increasing 
$\langle k \rangle$, 
since $R$ is decreasing in the first stage of small $N(t)$ but 
increasing after a time.
We obtain similar behavior in these spatial growth 
models as shown in Fig. \ref{fig_attack_r-k-R} from top to bottom, 
and the case of $\delta = 0.3$ (red line) 
with the stronger effect of the copying is better than 
other cases of $\delta = 0.5$, $0.7$, and $0.9$ (green, blue, 
and magenta lines). 
In particular, in the case of $\delta = 0.9$ (magenta line), 
$R$ does not increase until $N(t) \approx 500$ due to the 
insistent attacks. 
All values of $r$, $\langle k \rangle$, and $R$ 
in Fig. \ref{fig_attack_r-k-R} are smaller, 
as compared with the results for the pure growing networks 
in Fig. \ref{fig_dla_r-k-R}. 
Thus, even the damaged networks maintain 
the robust onion-like topology in the growing scheme.
We remark that in the sequential attacks 
the robustness is protected locally by the copying, 
and enhanced globally in connecting isolated clusters by 
adding shortcut links between randomly chosen nodes.  
Remember the local proxy and complementary functions in subsection 2.1.
When more severe attacks in shorter interval at every $5$ time steps 
are given, around the robustness index 
$R \approx 0.15 \sim 0.2$ against the further attacks, 
the networks no longer have an onion-like topological 
structure because of 
the assortativity $r \approx 0.0$ or $r < 0.0$ in this 
growing scheme of constant $\delta$, $p_{sc}$, and $IT$ without 
any remedial measures.

\section{Conclusion} \label{sec4}
We have proposed a spatial design method of robust and efficient 
networks with typical surface growth patterns.
It is self-organized through the simultaneous progress of the 
copying and adding shortcut links \cite{Hayashi14} 
taking into account linking 
homophily to make an onion-like topological structure with 
positive degree-degree correlations. 
For the copying, 
the selection of node is not uniformly at random but 
limited to the neighbors of the perimeter of connected cluster 
in the growing network on a space.
In spite of the constraint on the surface growth, 
we have obtained that 
the robustness against random failures and malicious attacks 
is nearly optimal of slightly weaker than 
the entirely rewired version \cite{Wu11}.
Moreover, the efficiency of path measured in the hop count 
is better than $O(\log N)$ of the SW property 
but one measured in the Euclidean distance 
slightly degraded from the SW property. 
In particular, 
the growing network becomes more robust and efficient 
in time-course. 
However, we may have to consider a resource allocation problem
for the increasing of $\langle k \rangle$ required with more links. 
On the other hand, 
we have also found that there are no huge hubs bounded by 
an exponential tail of degree distribution, 
and that large degree nodes are 
spontaneously interspersed on a space.

Moreover, we have shown that the robust onion-like topological structure 
remains even in the random growth damaged by the sequential attacks. 
When a node is removed by serious disasters or insistent attacks, 
more effective strategies than a constant random growth 
can be considered to repair the damaged parts. 
If a new node is added near the removed node as the proxy 
of it instead of the random location on
a neighbor site on the perimeter in the surface growth, 
the network is probably more effectively healed over.
If the parameters of $\delta$, $p_{sc}$, and $IT$ are regulated 
according to the damages, 
the network is recovered to the original level of performance 
for the communication or the transportation.
These trials will require further study with healing and 
recovering processes, e.g. we should consider 
which parts of network have priorities to prevent spreading 
damages in topological and spatial importance.
In a realistic situation such as disaster or battle-field, 
resources for the strategies are limited and should be 
allocated effectively in the priorities.

In summary, 
our growing network models suggest that robust and efficient 
onion-like topological structure can emerge 
even when the positions of nodes are limited 
in the contact area of the spatial network. 
In the cooperative growing mechanism, 
the local proxy and complementary robust functions by the copying 
and adding shortcut links 
will be useful for temporal evolution of 
resilient network and a repair strategy in 
large-scale disasters or system crises.
However, it is an issue in general what type of constraint 
for locating nodes on a space 
is an obstacle to the emergence of onion-like topology. 
It may be related to limited facilities from geographical 
and economical viewpoints.

\section*{Acknowledgment}
The author would like to thank Mitsugu Matsushita 
for his valuable comments about fractal growth models on a space
and to anonymous reviewers for their suggestions to improve the 
readability of manuscript. 
This research is supported in part by
a Grant-in-Aid for Scientific Research in Japan, No. 25330100.

\bibliographystyle{elsart-num}
\bibliography{fractal-bibfile}

\begin{thebibliography}{10}
\expandafter\ifx\csname url\endcsname\relax
  \def\url#1{\texttt{#1}}\fi
\expandafter\ifx\csname urlprefix\endcsname\relax\def\urlprefix{}\fi

\bibitem{Dressler07}
F.~Dressler, Self-Organization in Sensor and Actor Networks, John Wiley \& Sons,
  2007.

\bibitem{Barabasi99}
A.-L. Babar\'{a}si, R.~Albert, Science 286 (1999) 509--512.

\bibitem{Sole02}
R.~V. Sole, R.~Pastor-Satorras, E.~Smith, T.~B. Kepler, 
Advances in Complex Systems 5~(1) (2002) 43--54.

\bibitem{Satorras03}
R.~Pastor-Satorras, E.~Smith, R.~V. Sole, 
Journal of Theoretical Biology 222~(2) (2003) 199--210.

\bibitem{Helbing97}
D.~Helbing, J.~Keltsch, P.~Moln\'{a}r, Nature 388 (1997) 47--50.

\bibitem{Tero10}
A.~Tero, S.~Takagi, T.~Saigusa, K.~Ito, D.~P. Bebber, M.~D. Fricker, K.~Yumiki,
  R.~Kobayashi, T.~Nakagaki, 
Science 327~(5964) (2010) 439--442.

\bibitem{Hayashi12a}
Y.~Hayashi, Y.~Megumo, Physica A 391 (2012) 872--879.

\bibitem{Doye05}
J.~P.~K. Doye, C.~P. Massen, Physical Review E 71 (2005) 016128.

\bibitem{Zhou05}
T.~Zhou, G.~Yan, B.-H. Wang, Physical Review E 71 (2005) 046141.

\bibitem{Nagel08}
W.~Nagel, J.~Mecke, J.~Ohser, V.~Weiss, 
Image Analysis \& Stereology 27~(2) (2008) 73--78.

\bibitem{Hayashi09a}
Y.~Hayashi, Advances in Complex Systems 12~(1) (2009) 73--86.

\bibitem{Hayashi09b}
Y.~Hayashi, Physica A 388~(1) (2007)
  991--998, corrigendum (2011).
\newline\urlprefix\url{http://www.sciencedirect.com/science/article/pii/S0378437112002592}

\bibitem{Krapivsky10}
P.~L. Krapivsky, S.~Redner, E.~Ben-Naim, A Kinetic View of Statistical Physics,
  Cambridge University Press, 2010.

\bibitem{Alava05}
M.~J. Alava, S.~Dorogovtsev, Physical Review E 71 (2005) 036107.

\bibitem{Watts98}
D.~J. Watts, S.~H. Strogatz, Nature 393 (1998) 440--442.

\bibitem{Albert00}
R.~Albert, H.~Jeong, A.-L. Barab\'{a}si, Nature 406 (2000) 36--44.

\bibitem{Schneider11}
C.~M. Schneider, A.~A. Moreira, J.~S.~Andrade Jr., S.~Havlin, H.~J.
  Herrmann, PNAS 810~(10) (2011) 3838--3841.

\bibitem{Herrmann11}
H.~J. Herrmann, C.~M. Schneider, A.~A. Moreira, J.~S.~Andrade Jr., S.~Havlin, 
Journal of Statistical Mechanics  (2011) P01027.

\bibitem{Tanizawa12}
T.~Tanizawa, S.~Havlin, H.~E. Physical Review E 85 (2012) 046109.

\bibitem{Filho15}
C.~I. N.~S. Filho, A.~A. Moreira, R.~F.~S. Andrade, H.~J. Herrmann,
  J.~S.~Andrade Jr., Scientific Report 5~(9082).
\newline\urlprefix\url{www.nature.com/articles/srep09082}

\bibitem{Wu11}
Z.-X. Wu, P.~Holme, Physical Review E 81 (2011) 026116.

\bibitem{Newman01}
M.E.J.~Newman, S.H.~Strogatz, D.J.~Watts, 
Physical Review E 64 (2001) 026118.

\bibitem{Hayashi14}
Y.~Hayashi, IEEE Xplore Digital Library Proc. of 2014 IEEE 8th Int. Conf. on
  SASO: Self-Adaptive and Self-Organizing Systems (2014) 50--59,
  http://arxiv.org/abs/1411.7719.

\bibitem{Zolli13}
A.~Zolli, A.~M. Healy, Resilience: Why Things Bounce Back, Simon \& Schuster,
  2013.

\bibitem{Schneider13}
C.~M. Schneider, N.~Yazdani, N.A.M.~Ara\'{u}jo, S.~Havlin, H.J.~Herrmann, 
Scientific Reports 3 (2013) 1969. 
http://www.nature.com/articles/srep01969


\bibitem{Kertesz14}
M.~Stippinger, J.~Kert\'{e}sz, 
Physica A 416 (2014) 431--487.


\bibitem{Yang13}
X.-H. Yanga, S.-L. Lou, G.~Chen, S.-Y. Chen, W.~Huang, 
Physica A 392 (2013) 3531--3536.

\bibitem{Colman13}
E.~R. Colman, G.~J. Rodgers, Physica A 392 (2013) 5501--5510.

\bibitem{Newman00}
M.~E. Newman, C.~Moore, D.~J. Watts, Physical Review Letters 84 (2000) 3201.

\bibitem{Hayashi06}
Y.~Hayashi, J.~Matsukubo, Physica A 380 (2007) 552--562.

\bibitem{Hayashi10}
Y.~Hayashi, Y.~Ono, Physical Review E 82 (2010) 016108.

\bibitem{Callaway01}
D.~S. Callaway, J.~E. Hopcroft, J.~M. Kleinberg, M.~E.~J. Newman, S.~H.
  Strogatz, Physical Review E 64 (2001) 041902.

\bibitem{Meakin98}
P.~Meakin, Fractals, scaling and growth far from equilibrium, Cambridge
  University Press, 1998.

\bibitem{Ben-Jacob97}
E.~Ben-Jacob, Contemporary Physics 38~(3) (1997) 205--241.

\bibitem{Wilkinson83}
D.~Wilkinson, J.~F. Willemsen, 
Journal of Physics A: Mathematical and General 16 (1983) 3365--3376.

\bibitem{Stanley94}
H.~E. Stanley, A.~Coniglio, S.~Havlin, J.~Lee, S.~Schwarser, M.~Wolf, 
Physica A 205 (1994) 254--271.

\bibitem{Witten83}
T.~A. Witten, L.~M. Sander, Physical Review B 27~(9) (1983) 36--44.

\bibitem{Eden61}
M.~Eden, A two-dimensional growth process, in: J.~Neyman (Ed.), Proceedings of
  the 4tn Berkeley Symposium on Mathematical Statistics and Probability,
  Vol.~IV of IV, 1961, pp. 223--239.

\bibitem{Brunet07}
R.~Xulvi-Brunet, I.~M. Sokolov, Physical Review E 75 (2007) 46117.

\bibitem{Pajek}
http://vlado.fmf.uni-lj.si/pub/networks/pajek/

\bibitem{Holme02}
P.~Holme, B.~J. Kim, C.~N. Yoon, S.~K. Han, Physical Review E 65 (2002) 056109.

\bibitem{Newman03a}
M.~E. Newman, Physical Review Letters 89~(20) (2003) 208701.

\bibitem{Newman10}
M.~E. Newman, Networks -An Introduction-, Vol.~68, Oxford University Press, New
  York, NY, 2010.

\bibitem{Barbeau07}
M.~Barbeau, E.~Kranakis, Principles of Ad-hoc Networking, John Wiley \& Sons,
  2007.

\bibitem{Hayashi12}
Y.~Hayashi, Rethinking of communication requests, routing, and navigation
  hierarchy on complex networks -for a biologically inspired efficient search
  on a geographical space-, in: I.~Bilogrevic, A.~Rezazadeh, L.~Momeni (Eds.),
  Networks -Emerging Topics in Computer Science, iConcept Press, 2012, Chapter~4,
  pp. 67--88.
\newline\urlprefix\url{https://www.iconceptpress.com/book/networks--emerging-topics-in-computer-science/11000032/}

\bibitem{Holme06}
P.~Holme, M.~E. Newman, Physical Review E 74 (2006) 056108.

\bibitem{Zanette06}
S.~Gil, D.~H. Zanette, Physics Letters A 356~(2) (2006) 89--94.

\bibitem{Noh08}
S.-W. Kim, J.~D. Noh, Physical Review Letters 100 (2008) 118702.

\bibitem{Noh09}
S.-W. Kim, J.~D. Noh, Physical Review E 80 (2009) 026119.

\bibitem{Xuan10}
Q.~Xuan, F.~Du, T.-J. Wu, G.~Chen, Physical Review E 80 (2010) 046116.

\end{thebibliography}



\end{document}